\documentclass[11pt,a4paper]{amsart}
\usepackage[margin=3cm]{geometry}

\usepackage{amsmath,amsfonts,amssymb,amscd,amsthm}
\usepackage[usenames,dvipsnames,svgnames]{xcolor}

\usepackage[utf8]{inputenc}
\usepackage{graphicx}
\usepackage{caption}
\usepackage{subcaption}
\usepackage[numbers]{natbib}
\usepackage{doi}
\usepackage{autonum}
\usepackage{makecell}
\usepackage{hyperref}

\def\H{\boldsymbol{H}}
\def\n{\boldsymbol{n}}
\def\0{\boldsymbol{0}}
\def\vv{\boldsymbol{v}}
\def\g{\boldsymbol{g}}
\def\E{\boldsymbol{E}}  
\def\J{\boldsymbol{J}}
\def\B{\boldsymbol{B}}
\def\f{\boldsymbol{f}}
\def\F{\boldsymbol{F}}
\def\R{\boldsymbol{R}}
\def\bphi{\boldsymbol{\phi}}
\def\x{\boldsymbol{x}}
\def\z{\boldsymbol{z}}

\def\Hcurl{\mathcal{H}(\textbf{{curl}}, \Omega)}
\def\Hcurlo{\mathcal{H}_0(\textbf{{curl}}, \Omega)}
\def\esp{{\mathcal{V}}}

\def\nedesp{\mathcal{ND}}
\def\geophy{K}

\def\Q{\mathcal{Q}}
\def\Am{\boldsymbol{\mathcal{A}}}

\def\Km{\boldsymbol{\mathcal{K}}}

\def\Cm{\boldsymbol{\mathcal{C}}}
\def\Mm{\boldsymbol{\mathcal{M}}}
\def\Jac{\boldsymbol{\mathcal{J}}}

\def\FEMPAR{{\texttt{FEMPAR}}}
\begin{document}

\title[]{Simulation of High Temperature Superconductors and experimental validation}

\author[M. Olm]{Marc Olm$^\dag$}

\author[S. Badia]{Santiago Badia$^{\ddag,\dag}$}

\author[A. F. Mart\'in]{Alberto F. Mart\'in$^{\dag,\ddag}$}

\thanks{$\dag$ Centre Internacional de M\`etodes Num\`erics en Enginyeria, Parc Mediterrani de la Tecnologia, Esteve Terrades 5, E-08860 Castelldefels, Spain E-mail: {\tt molm@cimne.upc.edu}\\
  $\ddag$ Universitat Polit\`ecnica de Catalunya, Jordi Girona1-3, Edifici C1, E-08034 Barcelona $\&$ Centre Internacional de M\`etodes Num\`erics en Enginyeria, Parc Mediterrani de la Tecnologia, Esteve Terrades 5, E-08860 Castelldefels, Spain E-mail: {\tt sbadia@cimne.upc.edu} (SB), {\tt amartin@cimne.upc.edu} (AM). \\ MO gratefully acknowledges the support received from the Catalan Government through the FI-AGAUR grant. SB gratefully acknowledges the support received from the Catalan Government through the ICREA Acad\`emia Research Program. This work has partially been funded by the European Union through the FP7 project FORTISSIMO, under the grant agreement 609029. The authors thankfully acknowledge the computer resources at Marenostrum IV and the technical support provided by BSC under the RES (Spanish Supercomputing Network).
}
 
\date{\today}

\begin{abstract}
  In this work, we present a parallel, fully-distributed finite element numerical framework to simulate the low-frequency electromagnetic response of superconducting devices, which allows to efficiently exploit HPC platforms. We select the so-called $H$-formulation, which uses the magnetic field as a state variable. N\'ed\'elec elements (of arbitrary order) are required for an accurate approximation of the $H$-formulation for modelling electromagnetic fields along interfaces between regions with high contrast medium properties. An $h$-adaptive mesh refinement technique customized for N\'ed\'elec elements leads to a structured fine mesh in areas of interest whereas a smart coarsening is obtained in other regions. The composition of a tailored, robust, parallel nonlinear solver completes the exposition of the developed tools to tackle the problem. First, a comparison against experimental data is performed to show the availability of the finite element approximation to model the physical phenomena. Then, a selected state-of-the-art 3D benchmark is reproduced, focusing on the parallel performance of the algorithms.   
\end{abstract}

\maketitle

%\noindent{\bf 2010 Mathematics Subject Classification:} 35Q30; 65N30; 76N10.

\tableofcontents

% citet and citep
\section{Introduction}
High Temperature Superconductor (HTS) devices possess a number of unique properties that make them attractive for its use in a wide range of engineering applications. In order to design and optimize devices using superconducting tapes or bulks, computational tools are a powerful technique to simulate its electromagnetic behaviour by solving the system of partial differential equations (PDEs) that governs the problem. In this context, Finite Element (FE) methods are commonly used because they can handle complicated geometries whilst providing a rigorous mathematical framework. However, the electromagnetic modelling of superconductors at low frequencies is an extremely challenging simulation process that stresses many aspects of a numerical code such as multiphysics modelling, multiscale modelling, highly nonlinear behaviour, and a large number of time steps involved. Hence an appropriate definition of the formulation, the FE method, and the solver will play a crucial role in order to obtain meaningful results in a reasonable amount of time. 

Many formulations exist for the eddy-current problem \cite{BIRO1999391}. These formulations can be mainly classified into three kinds, named after the variables used in the system of PDEs that one aims to solve: the $A$-$V$ formulation \cite{grilli_numerical_2004, lousberg_numerical_2009, campbell_direct_2009}, which is based on the magnetic vector potential, the $T$-$\Phi$ formulation \cite{grilli_finite_2005, amemiya_magnetic_2006, stenvall_programming_2010}, which is based on the current vector potential, and the $H$-formulation \cite{pecher_3d-modelling_2003, 7422024, grilli_development_2013, zhang_3d_2012}, which is based directly on the magnetic field. Additionally, the mixed $H$-$\varphi$-$\Psi$ formulation \cite{Stenvall_hformulation_2014, lahtinen_homology_2015} of the FE method uses cohomology basis functions in the dielectric region and allows to treat the air as an exact zero conductivity region. An alternative approach to the FE method is the variational method, which is valid for any electric field-current density relation and exists for several formulations: the $H$-formulation \cite{Bossavit_Numerical_1994, elliott_finite-element_2007, badia_electromagnetics}, the effective magnetization $T$-formulation \cite{pardo_3d_2017} and the $J$-$\phi$ formulation \cite{PRIGOZHIN1996190, pardo_electro_2015, prigozhin_solution_1998, navau_alternating_2008}. Other accurate results have been achieved with circuit models, such as those in \cite{van_nugteren_high_2016} for calculating AC losses. In this work, the authors have selected the most common and widespread formulation, which is the $H$-formulation. The $H$-formulation provides the direct solution to the magnetic vector field, and has the advantage of dealing with boundary conditions in the model in a simple way. External magnetic fields can be applied directly by setting boundary values of the magnetic field, while currents in the superconductor device can be injected through Amp\`ere's law. We stress, however, that the techniques presented here are also applicable with minor adaptations to other formulations as well.

In this work, the FE discretization of the magnetic field relies on the curl-conforming edge (or N\'ed\'elec) element \cite{nedelec_mixed_1980} (of arbitrary order). Edge elements are preferred over grad-conforming Lagrangian ones, since they facilitate the modelling of the field near singularities by allowing normal fields components to jump across interfaces between two different media with highly contrasting properties \cite{mur_edge_1994}. In general, Lagrangian (nodal) elements with a weak imposition of the divergence constraint can converge to singular solutions for homogeneous problems (see, e.g., \cite{badia_nodal-based_2012}), but these methods are not robust for heterogeneous problems like the ones in HTS modelling. 

The high complexity of the problem at hand (i.e., modelling of superconducting devices with surrounding air or dielectric material regions) certainly requires customized solutions for every step in the simulation pipeline, namely mesh generation, discretization of the PDE system at hand and solution of the nonlinear system arising from discretization. With regard to mesh generation, computational cost may become rapidly expensive as we have to mesh not only the region of interest (superconducting device and immediate surrounding) but also the entire dielectric region. Common practice in the $H$-formulation FE modelling is to choose the latter region to be large enough such that interferences between the external applied magnetic field and the magnetic field generated by the superconducting device are avoided. This may imply that a large number of Degrees of Freedom (DoFs) are used on the mesh cells covering the dielectric region, while only a small portion of these DoFs might be actually needed for the accurate approximation of the magnetic field on this region. The most immediate approach found in the literature to tackle this issue consists on the usage of {\em conforming} unstructured meshes with variable size cells, where the mesh cells are coarsened as the boundary of the dielectric domain is approached. The usage of a more fitted (to the superconductor) dielectric domain has been used as well in \cite{granados_h-formulation_2016}. In such a case, one has to take into account self-generated magnetic fields while imposing external magnetic field boundary conditions. Finally, a more involved approach is the employment of cohomology basis functions, which allows to significantly reduce the number of DoFs in the dielectric region \cite{Stenvall_hformulation_2014, lahtinen_homology_2015}. Alternatively, methods taking the current density as state variable may reduce the number of DoFs, since only the sample volume is taken into account. For 2D problems, this has been done by the variational method in J formulation \cite{PRIGOZHIN1996190}, integral methods \cite{brandt_1995, grilli_integral} and circuit methods \cite{nugteren_2016}. In this work, an Adaptive Mesh Refinement (AMR) strategy for N\'ed\'elec elements has been implemented. Using AMR, we can introduce an aggressive coarsening of the mesh in the dielectric region, whereas a fine mesh is achieved in the superconducting device. On the other hand, one can start with a very coarse (possibly unstructured) {\em conforming} mesh that represents the geometry at hand, drastically reducing the mesh generation computational cost. The AMR strategy is based on octree-based meshes, which can be compactly represented and efficiently manipulated in high-end distributed-memory computers. In this work, the {\tt p4est} \cite{BursteddeWilcoxGhattas11} MPI library is used for such purpose. For the solution of the nonlinear problem at every time step, we use the Newton-Raphson solver. At each nonlinear iteration, the resulting linear system is solved by means of preconditioned Krylov subspace iterative solvers \cite{Saad_book}. An efficient preconditioner is crucial for their robustness and (parallel/algorithmic) scalability. In this work, we ground on the so-called Balancing Domain Decomposition by Constraints (BDDC) preconditioning approach \cite{dohrmann_2003, badia_highly_2014, badia_multilevel_2016}. For the problem at hand, we propose a curl-conforming BDDC preconditioner equipped with the coarse space presented in \cite{toselli_dual-primal_2006} for FETI-DP methods, and the approach in \cite{badia_physics_based_2017} to deal with high jumps of material properties.

Research on the simulation of the HTS problem has typically focused on the application side, and has considered moderate scale test cases with commercial software implementations of linear (at most quadratic) edge elements; see, e.g., \cite{zhang_3d_2012, hong_numerical_2006, ainslie_numerical_2016}. On the other hand, 3D problems are of high interest in HTS modelling \cite{lousberg_numerical_2009, grilli_finite_2005, zhang_3d_2012, pardo_3d_magnetization_2017, farinon_2d_2014}, but far from being at the maturity level as one can find in 2D, due to their high computational complexity and the poor (parallel) scalability of commercial software. Indeed, for large-scale FE 3D simulations, the efficient exploitation of HPC resources becomes a must for providing a reasonable time-to-solution. In this context, the current work goes one step beyond by proposing a {\em parallel}, {\em fully-distributed} simulation software pipeline for the electromagnetic behaviour of HTS based on state-of-the-art numerical techniques for every building block.  

The proposed algorithms are available at \FEMPAR{} \cite{badia-fempar,fempar-web-page}, a general-purpose, scientific FE software for the fast  solution of multiphysics problems governed by PDEs on high-end computing environments, from the Desktop/Laptop to HPC clusters and supercomputers. Therefore, this work also aims to introduce \FEMPAR{} to the HTS modelling community as a new and powerful HPC tool for their simulations. In order to show its applicability, a validation with the Hall probe mapping experiment \cite{granados_h-formulation_2016} is performed, obtaining a good agreement between the simulation results and the experimental data. In order to show the benefit of the proposed {\em fully-parallel} simulation software, a selected state-of-the-art 3D benchmark \cite{kapolka_3d_2017} is reproduced with excellent time-to-solution reductions on a massively parallel supercomputer.

The outline of the article is as follows. The problem is defined in Sect.~\ref{sec-formulation} and some notation is introduced.  The FE approximation of the problem is developed in Sect.~\ref{sec-fe_approx}. In Sect.~\ref{sec-mesh_refinement}, we present advanced mesh generation techniques customized for our model problem. In Sect.~\ref{sec-solver}, we present a customized nonlinear parallel solver suitable for the problem at hand. We present a detailed set of numerical experiments in Sect.~\ref{sec-results}, which include a validation phase against experimental data and the reproduction of a selected benchmark, together with a strong scaling analysis. Finally, some conclusions are drawn in Sect.~\ref{sec-conclusions}.  

\section{The system of equations}\label{sec-formulation}
\subsection{Notation}\label{sec-notation}
In this section, we introduce the problem to be solved and its particularities. Let $\Omega\subset\mathbb{R}^d$ be a bounded domain with $d=2,3$ the space dimension. Let us denote by $L^2(\Omega)$ the space of square integrable functions. Furthermore, we will make use of the space
\begin{align}
%\Hs  &:= \{ v \in L^2(\Omega)  \hbox{ }| \hbox{ } \nabla v \in L^2(\Omega)^d \} \\
\Hcurl &:= \{ \vv \in L^2(\Omega)^d  \hbox{ }| \hbox{ } \nabla \times \vv \in L^2(\Omega)^d \},
\end{align}
and its subspace
\begin{align}
%\Hso  &:= \{ v \in \Hs \hbox{ }| \hbox{ } v=0 \hbox{ in } \partial\Omega) \} \\
\Hcurlo &:= \{ \vv \in \Hcurl \hbox{ }| \hbox{ } \n \times \vv = \0 \hbox{ in } \partial\Omega \},
\end{align}
where $\n$ denotes the outward unit normal to the boundary of the domain $\Omega$. In the sequel, bold characters will be used to describe vector functions or tensors, while regular ones will determine scalar functions or values. No difference is intended by using upper-case or lower-case letters for functions. Calligraphic letters are used to describe functional spaces and bold calligraphic letters will denote bilinear operators.  
\subsection{Maxwell equations in electromagnetics}
Let us first state the Maxwell equations, which physically describe magnetostatics. Let us consider $\Omega\subset\mathbb{R}^d$ to be a simply connected nonconvex polyhedral domain with a connected Lipschitz continuous boundary $\partial\Omega$. The differential Maxwell equations read
\begin{align}
 \nabla \times \E &= - \frac{\partial \B}{\partial t}, && \text{Maxwell-Faraday equation} \label{eq-maxwell1} \\
 \nabla \times \B &= \mu_0 \J, && \text{Amp\`ere's circuital law} \label{eq-maxwell2} \\ 
 \nabla \cdot \B  &= 0, && \text{Gauss's law for magnetism} \label{eq-maxwell3} \\
 \nabla \cdot \E  &= 4 \pi \sigma,  && \text{Gauss' law} \label{eq-maxwell4}
\end{align}
in $\Omega  \times (0,T]$, where $\E$ is the electric field, $\B$ is the magnetic field, $\J$ is the electric current density, $\mu_0>0$ is the magnetic permeability of the vacuum, and $\sigma$ is the electric charge density. This form of the equations is valid for negligible displacement current. Furthermore, we add the constitutive law that specifies the (possibly nonlinear) relationship between the electric field and the current density in a material by 
\begin{align} 
\E = \rho \J, && \text{Ohm's law} \label{eq-conslaw}
\end{align}
$\rho>0$ being the material resistivity tensor (inverse of conductivity). In this work, we restrict ourselves to non-magnetic media since we consider the constitutive law $\B = \mu_0 \H$ for magnetic fields. 

\subsection{The $H$-formulation}\label{sec-Hformulation}
After some trivial manipulation of Eqs. (\ref{eq-maxwell1})-(\ref{eq-maxwell2}) and the substitution of the constitutive law in Eq. (\ref{eq-conslaw}), one can obtain the so-called $H$-formulation for the magnetic field $\H$. The proposed formulation reads: seek a magnetic field $\H$ solution of
\begin{align}
\frac{\partial \mu_0 \H}{\partial t} + \nabla \times \rho \nabla \times \H  = \f & \qquad \hbox{in } \, \Omega \times (0,T],   \label{eq-dcurl_p}
\end{align}
% (i.e., divergence must vanish in order to ensure the divergence-free condition for $\H$ at the continuous level)
where $\f$ is a solenoidal given source term. Taking the divergence of Eq.~(\ref{eq-dcurl_p}), and given that $\H$ is solenoidal at the initial time, follows that $\H$ is solenoidal for every time. Besides, Eq.~(\ref{eq-dcurl_p}) needs to be supplied with appropriate boundary and initial conditions. The boundary of the domain $\partial \Omega$ is divided into its Dirichlet boundary part, i.e., $\partial \Omega_D$, and its Neumann boundary part, i.e., $\partial \Omega_N$, such that $\partial \Omega_D \cup \partial \Omega_N = \partial \Omega$ and $ \partial \Omega_D \cap \Omega_N = \emptyset$. Then, boundary and initial conditions for the problem at hand read
\begin{align}
\H \times \n = \g && \quad \hbox{on } \, \partial \Omega_D \times (0,T], \\
\n \times (\rho \nabla \times \H) = 0 && \quad \hbox{on } \, \partial \Omega_N \times (0,T],\\
\H(\x,t=0)=\0 && \hbox{in } \Omega.
\end{align}
Note that Dirichlet boundary conditions prescribe the tangent component of the magnetic field on the boundary of the domain, while Neumann boundary conditions prescribe the tangent component of the electric field $\E$ (see Eq. (\ref{eq-conslaw})). Finally, the variational form of the $H$-formulation reads as follows: find $\H \in \Hcurl$ such that 
\begin{align}
( \frac{\partial \mu_0\H}{\partial t}, \vv ) + (\rho \nabla \times \H, \nabla \times \vv) &= (\f, \vv),  & \forall \vv \in \Hcurlo. \label{eq-double_curl}
\end{align}

\subsection{Transmission conditions}\label{sec-transmission}
Natural boundary conditions appear on the formulation after integrating by parts Eq. (\ref{eq-double_curl}):
\begin{align}\label{eq-Neumann_curl}
\int_{\Omega} \left( \nabla \times \rho \nabla \times \H \right) \cdot \vv  &= \int_{\Omega} \left( \rho \nabla \times \H \right)\cdot \left( \nabla \times \vv \right) - \int_{\partial\Omega_N}\left( \rho \nabla \times \H \right) \cdot \left( \n \times \vv \right),
\end{align} 
where we can identify the condition $\n \times (\rho \nabla \times \H)$ to be introduced in the Neumann boundary $\partial \Omega_N$. Consider now two different non-overlapping regions on the domain $\Omega$ corresponding to two different media, namely $\Omega_{1}$ and  $\Omega_{2}$ such that $\Omega = \Omega_1 \cup \Omega_2$ and $\Gamma = \Omega_1 \cap \Omega_2$. Let us denote by $\{ \n_{\Gamma_1}$,$\n_{\Gamma_2} \}$ the unit normal pointing outwards of $\{ \Omega_1, \Omega_2 \}$. Clearly, $\n_{\Gamma_1} = -\n_{\Gamma_2}$ and we state the natural interface conditions (\textit{transmission conditions}) for Eq.~(\ref{eq-dcurl_p}) as
\begin{align}
\n \times ( \rho_1 \nabla \times \H_1 - \rho_2 \nabla \times \H_2) = 0 & \qquad \hbox{on } \Gamma, \label{eq-neumE}
\end{align} 
where $\n$ holds in this case for $\n = \n_{\Gamma_1}$. Note that, with no other sources, Eq. (\ref{eq-neumE}) enforces the continuity of the tangent component of the electric field $\E$ over the interface, i.e., $\n \times (\E_1 - \E_2) = 0$. For the problem at hand, the domain will be composed of an HTS device $\Omega_{\rm hts}$ and a surrounding dielectric region $\Omega_{\rm air}$.

On the other hand, currents in the superconductor device are injected through Amp\`ere's circuital law, Eq. (\ref{eq-maxwell2}), in a closed surface $S$ as (by the Stokes theorem)
\begin{align}\label{eq-Iapp}
\int_{S} (\nabla \times \H)\cdot \n = \oint_{\partial S} \H \cdot \boldsymbol{\tau} = I_{\rm app},
\end{align}
where now $\n$ denotes the unit normal pointing outwards to the surface $S$ defined by a section of $\Omega_{\rm hts}$ (the domain itself in a 2D case). On the other hand, $\boldsymbol{\tau}$ is the unit tangent to the surface boundary. The scalar value $I_{\rm app}$ is the net current enforced in the superconductor in the perpendicular direction to the surface.

The final HTS problem will read as follows: find $\H \in \Hcurl$ such that Eq.~(\ref{eq-double_curl}) holds in $\Omega_{\rm hts}$ and $\Omega_{\rm air}$, together with the transmission condition (\ref{eq-neumE}) and the constraint (\ref{eq-Iapp}).

\subsection{Material modelling}\label{sec-csm}
%For type II superconductors, as HTS, the magnetic flux penetrates within the material for magnetic fields above the first critical field, denoted by $\H_{c1}$. In reversible superconductors with no flux vortex pinning, this flux penetration is roughly uniform, resulting in virtually no macroscopic current density. However, under strong pinning, induced mostly by defects, there appears non-uniform flux penetration, which can be described by flux creep \cite{brandt_susceptibility_1997}, being the Critical State Model (CSM) a particular limit. The CSM, proposed simultaneously by Bean  and London , gives a macroscopic explanation of the irreversible magnetization behavior. 
%The critical state model (CSM), proposed simultaneously by Bean \cite{bean_magnetization_1964} and London \cite{london_alternating_1963}, states the existence of a limiting macroscopic superconducting current density $J_c$ that a hard superconductor can carry. However, in both models, $\E-\J$ relations are non-smooth and can be multi-valued. Therefore, they should be approximated by an analytical and smoother function, e.g., a power law, for computational purposes.  
In previous sections, a suitable $H$-formulation (applicable to general electromagnetics) has been presented. In this section, the general formulation is extended to superconductivity by means of the constitutive law definition that relates current densities and electric fields. We consider a dielectric domain $\Omega_{\rm air}$ large enough for neglecting boundary effects associated to the magnetization of the superconductor. In order to model its non-conducting behaviour, we consider a conductivity (inverse of resistivity) value that ideally tends to 0. However, the dramatic jump of resistivity on the interface introduces a boundary layer on the interface that would require huge computational resources to be captured, whereas we are mainly interested in the superconductor behaviour. Thus, it is common practice to consider a fixed value for the resistivity in $\Omega_{\rm air}$ that, whilst maintaining a large magnitude ratio with regard to the superconducting material, allows the computation to take place with a desired level of precision. On the other hand, a nonlinear electric field-current density relation is used in $\Omega_{\rm hts}$ to describe the penetration of the magnetic flux and induced currents. For describing $\E(\J)$, we will use the power law 
\begin{align}\label{eq-powerlaw}
\E = \frac{E_c}{J_c} \left( \frac{\| \J \| }{J_c} \right)^{n} \J, 
\end{align} 
where $E_c$ is the critical electric field, $J_c$ is the critical current density, and $n$ is the exponent of the power law. This kind of expression tends to the analytical Bean's model \cite{bean_magnetization_1964} when $n$ tends to infinity. One can identify Ohm's law (Eq. (\ref{eq-conslaw})) in this $\E$-$\J$ relation, with the following expression for the resistivity parameter in the superconducting region:
\begin{align} 
\rho_{\rm hts}(\H) = \frac{E_c}{J_c} \left( \frac{ \| \nabla \times \H \| }{J_c} \right)^{n}. \label{eq-JcB}
\end{align}  
In turn, $J_c$ may be considered a fixed current density value, or a value dependent on the magnetic field (magnitude and direction). For instance, for the magnetization of type-II superconductors, the Kim's model \cite{kim_magnetization_1963} introduces a dependence on the magnetic field strength, 
\begin{align}
& \quad
J_c(\B) = \frac{J_{c0} B_0}{B_0 + \mu_0 \| \H \| },
\end{align} 
where $J_{c0}$ and $B_0$ are parameters determined by the physical properties of the superconducting material.
\section{Numerical approximation}\label{sec-fe_approx}

\subsection{FE approximation}
In this section, we discuss a conforming FE space with respect to $\Hcurl$. Let $\mathcal{T}_h$ be a partition of $\Omega$ into a set of hexahedra (quadrilaterals in 2D) geometrical cells $\geophy$. Using Ciarlet's definition, a FE is represented by the triplet $ \{ \geophy, \esp, \Sigma \}$, where $\esp$ is the space of functions on $\geophy$ and $\Sigma$ is a set of functionals on $\esp$. These functionals are called DoFs of the FE. Let us first define polynomial spaces that will be needed in forthcoming definitions. The space of polynomials of degree less than or equal to $k>0$ in all the variables $\{x_i\}_{i=1}^d$ is denoted by $\mathcal{Q}_k(K)$. Let us also define the corresponding truncated polynomial space $\mathcal{P}_k(K)$ as the span of the monomials with degree less than or equal to $k$. Below, we define the local space of functions and its corresponding set of DoFs. 

\subsubsection{Edge hexahedral FEs}\label{subsec-hexned}
For this sort of FEs, $\esp_{k}({K})$ is the vector space defined as 
\begin{align}
 \esp_{k}({K}) &:= \{ \Q_{k-1,k}({K}) \times \Q_{k,k-1}({K}) \}, \\
 \esp_{k}({K}) &:= \{ \Q_{k-1,k,k}({K}) \times  \Q_{k,k-1,k}({K}) \times \Q_{k,k,k-1}({K}) \}, 
\end{align}
in the 2D and 3D case, respectively. The set of functionals that form the basis in two dimensions reads  (see \cite{monk_finite_2003}, ch.6) 
\begin{align}
& \frac{1}{\| \mathcal{{E}} \|} \int_{\mathcal{{E}}} ( u_h\cdot \boldsymbol{\tau} ) q  \quad \forall q \in \mathcal{P}_{k-1}(\mathcal{{{E}}}), \text{ for every edge } \mathcal{E} \text{ of } \geophy \label{eq-hexnedge_2D} \\ & \frac{1}{\| \hat{K} \|} \int_{{K}}  u_h \cdot \boldsymbol{q} \quad \forall \boldsymbol{q} \in \Q_{k-1,k-2}({K}) \times \Q_{k-2,k-1}({K}),\label{eq-hexninner_2D}
\end{align}
while in three dimensions the set is defined as 
\begin{align}
&\frac{1}{\| \mathcal{{E}} \|} \int_{\mathcal{{E}}} ( u_h\cdot \boldsymbol{\tau} ) q  \qquad \forall q \in \Q_{k-1}(\mathcal{{{E}}}), \text{ for every edge } \mathcal{E} \text{ of } \geophy \label{eq-hexnedge} \\
&\frac{1}{\| \mathcal{{F}} \|} \int_{\mathcal{{F}}} ( u_h\times \n )\cdot \boldsymbol{q} \qquad \forall \boldsymbol{q} \in \Q_{k-2,k-1}(\mathcal{{F}})\times \Q_{k-1,k-2}(\mathcal{{F}}), \text{ for every face } \mathcal{F} \text{ of } \geophy \label{eq-hexneface} \\
&\frac{1}{\| {K} \|} \int_{{K}}  u_h \cdot \boldsymbol{q} \qquad \forall \boldsymbol{q} \in \Q_{k-1,k-2,k-2}({K})\times \Q_{k-2,k-1,k-2}({K}) \times \Q_{k-2,k-2,k-1}({K}),\label{eq-hexninner}
\end{align}
where $\boldsymbol{\tau}$ is the unit vector along the edge and $\n$ the unit normal to the face. 
Note that in the case of the lowest order elements, i.e., $k=1$, only DoFs associated to edges appear. For higher order elements, i.e., $k\geq 2$, we find all kinds of DoFs in both dimensions.

 Note that  local spaces $\esp_k({K})$ lie between the full polynomial spaces of order $k-1$ and $k$, i.e., $\Q_{k-1}^d({K}) \subset \esp_k({K}) \subset \Q_{k}^d({K})$. This kind of elements are called N\'ed\'elec FEs of the first kind \cite{nedelec_mixed_1980}. Another edge FE based on full polynomial spaces, the so-called second kind, was introduced also by N\'ed\'elec in \cite{nedelec_new_mixed_1986}.
 
\subsubsection{Global FE spaces and conformity}\label{subsec-assembly} 
A space is $\mathcal{H}$(\textbf{curl})-conforming if the tangential components at the interface between elements are continuous, i.e., they do not have to satisfy normal continuity over element faces. Any function $\H_h$ can be uniquely determined by the set of DoFs defined for the edge FEs. The discrete global FE-space where the magnetic field solution $\H_h$ lies is defined as 
\begin{align}
\nedesp_k(\Omega) = \{ v_h \in \Hcurl \hbox{ such that } v_h|_K\in \esp_k(K) \, \forall \, K \in \mathcal{T}_h \},
\end{align} 
where $\esp_k$ has been defined for the hexahedral edge FE in Sect. \ref{subsec-hexned}. % Local FE space information is transferred from the reference element to physical elements through the so-called covariant Piola mapping ${\funmap}_\geophy(\hat{v}) \doteq \hat{\funmap}_\geophy (\hat{v}) \circ \geomap^{-1}_\geophy$, where    
%\begin{align}\label{eq-Piola_map}
%\hat{\funmap}_K(\vv) \doteq \jacobian^{-T} \vv.  
%\end{align}
%The Piola mapping, which preserves tangential traces of vector fields \cite{rognes_efficient_2009}, is the key to achieve a curl-conforming global space. 
It can be checked that, by enforcing the continuity of DoF values on edges/faces for all the elements that contain them, one already enforces continuity of the tangent component across elements \cite{monk_finite_2003}.

\subsection{Time discretization}\label{subsec:time_int}
Let us consider a partition of the time interval $[0,T]$ into $N$ time slabs. We denote the $n$-th time slab by $\Delta t^n = (t^{n-1},t^n],$ for $n=1,\ldots,N$. We also denote each time slab size by $| \Delta t^n |$. Time integration is performed with a $\theta$-method, even though the use of other time integrators is straightforward. For the sake of clarity, we use the Backward-Euler time integration in the presentation of the method. The already discretized in time problem reads: Given $\H(t^0)=\0$, find at every time step $n=1,\ldots,N$ the solution $\H_h^n \in \nedesp_k$ such that   
\begin{align}
& \frac{\mu_0}{| \Delta t^n |} (\H_h^n,\vv_h) + (\rho({\H_h^n)}\nabla \times \H_h^n,\nabla \times \vv_h)  \\ 
\nonumber & \qquad =  (\f_h^n,\vv_h) + \frac{\mu_0}{| \Delta t^n |} (\H_h^{n-1},\vv_h), &&  \forall \vv_h \in \nedesp_k \\
\end{align}
where $\f_h^n$ is the discrete version of the source term $\f$ evaluated at time $t^n$. Note that non-homogeneous initial conditions are readily imposed by considering $\H^0_h = \H_h(t^0)$, where $\H_h(t^0)$ is the interpolation of the initial value onto the FE space $\nedesp_k$.
  
\subsection{Dirichlet Boundary Conditions}\label{subsec-dirichlet}
The $H$-formulation is preferred over other formulations, among other reasons, due to its straightforward manner of handling magnetic fields and currents. External magnetic fields can be applied directly by setting boundary values of the magnetic field on Dirichlet boundaries, while currents in the superconductor device can be injected through Amp\`ere's law (see Eq.~(\ref{eq-maxwell2})) through constraints. 
Dirichlet boundary conditions will be strongly imposed in the resulting system (usual implementation in FE codes), hence we need the DoF values over the boundary to be fixed. $\H_h$ DoFs are obtained by means of moments defined over edges, faces, and cells. As a result, one has to use the corresponding N\'ed\'elec FE interpolator, which consists in evaluating the moments (e.g., Eqs.~(\ref{eq-hexnedge_2D}) and (\ref{eq-hexninner_2D}) or (\ref{eq-hexnedge}), (\ref{eq-hexneface}) and (\ref{eq-hexninner})) for the continuous {boundary} function. 
%in the reference FE element, using the mapping (\ref{eq-Piola_map}) from the physical to the reference space (see \cite[p. 134]{monk_finite_2003}). 

\section{Mesh generation by hierarchical AMR}\label{sec-mesh_refinement}
Uniform refinement is not a choice for problems that exhibit localized phenomena and multiscale features, since the size of the resulting system may rapidly become prohibitive.
The purpose of any mesh adaptive method is to achieve a high degree of accuracy in areas of the domain of particular interest while saving computational efforts in other areas. To this end, the mesh is refined in regions of the domain that present a complex behaviour of the solution. AMR is mostly used with \emph{a posteriori} error estimate that determines the accuracy of the solution in every mesh cell and requires to know the solution of the PDE at every iteration (dynamic AMR). However, the areas of interest in the problem at hand are located in an {\em a priori} known region of the domain, the superconductor region, therefore we can leverage AMR with a refined static mesh approach. Let us clarify this approach with the problem at hand.

Superconducting regions of the domain and immediate surroundings show high values for the gradients of the solution (i.e., high variation of the solution in small regions of the domain), while a smoother solution is expected in a vast majority area of the dielectric region. Besides, quantities of interest (e.g., AC losses) are computed on $\Omega_{\rm air}$, hence we are mainly interested in taking control over the degree of accuracy that can be achieved there. On the other hand, the ratio between the volume of the superconductor material and the dielectric region is very low and thus saving efforts in $\Omega_{\rm air}$ is crucial to obtain results in a reasonable amount of time. At this point, we should stress that we restrict ourselves to $h$-adaptivity, i.e., only the mesh is adapted, in contrast to the so-called $hp$-adaptivity, where the polynomial order $p$ of the FEs  (see Sect. \ref{sec-fe_approx}) may also be adapted, and thus vary among mesh cells. 

%For PDE problems posed on geometrically complex domains, octree-based meshes can be combined with special numerical discretization techniques such as embedded (or immersed) boundary FE methods \cite{BADIA2017, badia_verdugo_2017}. This latter possibility is though not explored in this paper. We restrict ourselves to HTS modelling problems on simple boxed domains.

\subsection{Hierarchical AMR on octree-based meshes}
\begin{figure}[t!]
    \centering
    \begin{subfigure}[t]{0.25\textwidth}
        \includegraphics[width=\textwidth]{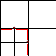}
        \caption{\emph{Hanging} entities marked in red.}
        \label{fig-hanging_node}
    \end{subfigure}
    \hspace{0.5cm}
    \begin{subfigure}[t]{0.25\textwidth}
        \includegraphics[width=\textwidth]{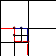}
        \caption{Not permitted \emph{hanging} entities marked in blue.}
        \label{fig-2lev_hang_nodes}
    \end{subfigure}
       \hspace{0.5cm}
        \begin{subfigure}[t]{0.25\textwidth}
        \includegraphics[width=\textwidth]{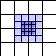}
        \caption{Centered refinement pattern. $\ell\in\{2,3,4\}$ from clearest to darkest.}
        \label{fig-refmesh}
    \end{subfigure}
    
    \caption{$h$-adaptive refined (single octree) {\em non-conforming} meshes.}
    \label{fig-nc_meshes}
\end{figure}

In this work, we explore hierarchically refined octree-based hexahedral meshes \cite{Tu_2005}. The mesh generation process in this context starts with a (possibly unstructured) {\em conforming} coarse mesh. This mesh can, e.g. be as simple as a single quadrilateral or hexahedron. Hierarchical AMR is a multi-step process in which at each single level, some cells of the input mesh to the level are marked for refinement. For the problem at hand, the criteria underlying which cells are marked for refinement is purely geometric. A cell marked for refinement is partitioned into four (2D) or eight (3D) children cells by bisecting all cell edges $\mathcal{E}_\geophy$. The resulting triangulation $\mathcal{T}_h$ can be thought as a collection of quads (2D) or octrees (3D) where the cells of the starting coarse mesh are the roots of these trees, and children cells branch off their parent cells. The leaf cells in this hierarchy form the mesh in the usual meaning, i.e., $\mathcal{T}_h$. Thus, for every cell $K \in \mathcal{T}_h$ we can define $\ell(K)$ as the level of $K$ in the aforementioned hierarchy, where $\ell(K)=0$ for the root cells, and $\ell(K)=\ell({\rm parent}(K))+1$ for any other cell. For the sake of clarity, Fig.~\ref{fig-refmesh} depicts cells at different levels of refinement for an initial single cell mesh. Octree-based meshes can be very compactly represented, and efficiently manipulated in high-end distributed-memory computers \cite{BursteddeWilcoxGhattas11}. Besides, they provide multi-resolution capability by local adaptation, i.e., octree-based mesh cells (actually the leaves in the hierarchy) can be at different levels of refinement. They are thus potentially {\em non-conforming}, e.g., we may have a mesh corner {\em hanging} in the middle of an edge or face, and {\em hanging} edges or faces enclosed as a subpart of coarser geometric entities (see Fig.~\ref{fig-hanging_node}). The fact that the mesh is {\em non-conforming} introduces additional complexity in the implementation of conforming FEs, specially in parallel codes for distributed-memory computers. This degree of complexity is nevertheless significantly reduced by enforcing the so-called 2:1 balance ratio, i.e., geometrically neighbouring cells may differ by only a single level of refinement. In this sense, in Figs.~\ref{fig-hanging_node} and \ref{fig-2lev_hang_nodes}, allowed \emph{hanging} geometric entities are depicted in red, whereas in Fig.~\ref{fig-2lev_hang_nodes}, not allowed ones are shown in blue. Clearly, the latter mesh is the result of a refinement process that does not accomplish the 2:1 balance, thus not permitted in our AMR approach. Note that in order to enforce the 2:1 balance in the situation depicted in Fig. \ref{fig-2lev_hang_nodes}, one would need to apply additional refinement to some cells with lower values for $\ell(K)$ until the 2:1 restriction is satisfied. This restriction is widely used in the AMR literature as a reasonable trade-off between performance gain and complexity of implementation \cite{Tu_2005, BursteddeWilcoxGhattas11}. Using octree-based hexahedral meshes we can introduce an aggressive coarsening of the mesh in the dielectric region, whereas a fine mesh is achieved in the superconducting device, see Fig. \ref{fig-h_adaptive_mesh}. In the present work, an extension of the AMR algorithm to support N\'ed\'elec elements has been implemented.

\subsection{Conformity of the Global FE space} 
In order to preserve the conformity of the global FE space $\nedesp_k$ (thus to guarantee the continuity of the tangent component of the FE solution), we cannot allow an arbitrary value for DoFs placed on top of \emph{hanging} geometric entities. We will denote this sort of DoFs by \emph{hanging} DoFs. During the assignment of DoFs to geometric entities, we can identify two different kinds of them: free DoFs and fixed DoFs. Free DoFs are independent DoFs which will appear in the final system, whereas fixed DoFs, which can be either Dirichlet or \emph{hanging}, are the ones that are constrained. On the one hand, Dirichlet DoFs are left out of the system (see Sect. \ref{subsec-dirichlet}) since their values are known by imposition of strong boundary conditions. On the other hand,  there are generally two possibilities in order to handle constraints for {\em hanging} DoFs (just like in the case of Dirichlet DoFs). The first one is to keep them in the global system, and enforce additional constraints that they must satisfy. Our preferred solution, also implemented in the case of nodal (Lagrangian) elements elements in \cite{coupling_kus_2017}, relies on the elimination of the \emph{hanging} DoFs from the global system, so as the continuity is already enforced at the basis functions level.

\section{Nonlinear transient solver}\label{sec-solver}
HTS modelling requires not only a reliable constitutive model but also a robust and efficient nonlinear solver. Superconductor phenomena may occur in a very short time period and thus abrupt changes of behaviour are found in a very small time scale. Furthermore, the nonlinearity associated to the constitutive law $\E-\J$ presents extreme parameters $(n \sim 30-100)$. Spatial scales, time scales, and nonlinearity make superconductivity modelling a very challenging task from a computational point of view. The nature of the nonlinearity is very stiff, since the exponent  in (\ref{eq-powerlaw}) usually takes high values, which makes the Lipschitz continuity constant of the nonlinear PDE operator at hand large. For this purpose, a specific nonlinear transient solver is proposed in this section. %The time integration is performed with the Backward Euler technique (Sect. \ref{subsec:time_int}). Thus, at every time step one aims to solve a nonlinear problem. A suitable formulation for solving the electromagnetic problem has been presented in Sect. \ref{sec-Hformulation}, whereas in this section we present the nonlinear solver employed. 
\subsection{Algebraic form}\label{sec-operators}
For the sake of clarity in forthcoming sections, let us write the problem in algebraic form.  The magnetic vector field $\H_h$ is expanded by means of the vector shape functions $\{ \bphi^i \}^{N_H}_{i=1}$ related to the curl-conforming edge element. Let us define the following element-wise matrices in the element $K$: 
\begin{align}
\Mm_K &:= \sum_{i=1}^{N_H} \sum_{j=1}^{N_H} \int_{K} \bphi^i \cdot \bphi^j \\
\Km_K &:= \sum_{i=1}^{N_H} \sum_{j=1}^{N_H} \int_{K}\rho(\H_h) (\nabla \times \bphi^j)\cdot(\nabla \times \bphi^i)
\end{align}
Consider also the right-hand side discrete vector $\F_K^i  := \int_{K} \f_K \bphi^i$. Then, the usual assembly is performed to obtain global matrices and arrays.

Once the operators have been introduced in algebraic form, the problem for a single time step $t^n$ ($n>0$) in algebraic form reads: 
\begin{align}\label{eq-alg}
\begin{bmatrix}
\frac{\mu_0}{| \Delta t^n | }\Mm + \Km(\H_h^n) & \Cm^t \\
\Cm & \0
\end{bmatrix} \begin{bmatrix} \H_h^n \\ \lambda^n \end{bmatrix} = 
\begin{bmatrix} \F_h(t^n) + \frac{\mu_0}{| \Delta t^k | }\Mm \H_h^{n-1} \\ {I}_{\rm app} \end{bmatrix},
\end{align}
where $\H_h$ is the discrete function containing the DoF values for the magnetic field. $\lambda^n$ is the Lagrange multiplier introduced to enforce the current constraint, and $\Cm$ is the one-row matrix that enforces the current value $I_{\rm app}$ over a given closed surface $S$ through Eq.~(\ref{eq-Iapp}), i.e., the application of a constraint over the solution 
\[
\Cm \H_h^n = \sum_{i=1}^{N_H} \int_{S} \left( \nabla \times \bphi^i \right) \cdot \n {\H_h^{n_i}} = {I}_{\rm app}.
\]

\subsection{Adaptive time stepping}\label{sec-time}
Time scales may be very small in this problem due to the applied fields frequency. However, the process of magnetization of the superconductor allows to identify different needs in different periods of the process. Although restrictions in the time step size are severe in some periods of time (when $\| \J \|$ becomes larger than $J_c$ in some region), the time step can be relaxed in monotone magnetization curves. The same effect occurs for the validation model that will be considered in Sect. \ref{sec-vali}, where an injected current is kept constant for a period of time before proceeding to the following current load increment. A simple adaptive time stepping will be used to accelerate convergence. The time step size is updated with the nonlinear solver convergence history for the last converged time step as 
\begin{align}
| \Delta  t^n | = \frac{ \kappa }{\# \rm{iters}} | \Delta  t^{n-1} |, 
\end{align}
where $\kappa$ stands for a selected growing ratio. Usually, one may select $\kappa$ as the ``ideal'' number of iterations to convergence sought in the nonlinear algorithm. Finally, the trial time step size is restricted to upper and lower bound values
\begin{align} 
| \Delta t^n |  = \left\{ \begin{array}{ll}
  \Delta t_{\rm min} & \mbox{if $| \Delta t^n | \leq \Delta t_{\rm min} $},\\
| \Delta t^n | & \mbox{if $\Delta t_{\rm min} \leq | \Delta t^n | \leq \Delta t_{\rm max} $}, \\ 
  \Delta t_{\rm max} & \mbox{if $\Delta t_{\rm max} \leq  | \Delta t^n | $}.\end{array} \right. 
\end{align}
\subsection{Linearization}\label{sec:solver}
The problem and its residual are stated in an algebraic form as
\begin{align}
\Am(\x) \x = \boldsymbol{b}, \qquad \R = \Am(\x) \x - \boldsymbol{b} = \0,
\end{align}
where the full vector of unknowns $\x$ and the right-hand side $\boldsymbol{b}$ have been presented in (\ref{eq-alg}). 
 It is essential to build a robust nonlinear solver together with an effective adaptive time stepping technique.
Note that the resistivity takes a constant value $\rho_{\rm air}$ in the air region $\Omega_{\rm air}$ hence the problem is linear in this part of the domain. However, a highly nonlinear problem is found in the superconductor region $\Omega_{\rm hts}$. Therefore, our strategies focus on the linearization of the problem associated to the extreme nonlinearity given by the resistivity $\rho_{\rm hts}(\H)$. For that purpose, we will make use of a composition of nonlinear solvers. Our nonlinear solver is the composition of a Newton-Raphson (NR) method with an exact derivation of the Jacobian and a cubic backtracking (BT) line search algorithm (see \cite{brune_composing_2013}). 

By means of the NR method, we obtain (for the time step $t^n$) the direction of the solution update at the iterate $k$, i.e., $\delta \x^{n,k}=\x^{n,k+1}-\x^{n,k}$, solving the linearized problem for the current linearization point ($ \H_h^{n,k}, \lambda^{n,k}$)
%\begin{align}\label{eq-linearizedsystem}
%\frac{\partial \R^{n,k}}{\partial ( \H_h^{n,k}, \lambda^{n,k} )} \delta( \H_h^{n,k}, \lambda^{n,k} ) = \Jac %\delta \x^{n,k} = -\R^{n,k}.
%\end{align}
\begin{align}\label{eq-linearizedsystem}
\Jac( \x^{n,k} ) \delta \x^{n,k} = -\R(\x^{n,k}).
\end{align}
Later, the BT technique tries to minimize the residual of the iterate $\x^{n,k+1} = \x^{n,k} + \beta \delta \x^{n,k}$ with the found direction $\delta\x^{n,k}$ by means of the step length $\beta$, i.e.,  
\begin{align}
\beta = \underset{0 < \tilde{\beta} \leq 1}{\operatorname{argmin}} \| \R ( \x^{n,k} + \tilde{\beta} \delta \x^{n,k} ) \|^2,
\end{align} 
and the process is repeated until a convergence criteria is attained. It is not our intention to define nor the BT technique neither the basic NR algorithm, which can be found in \cite{brune_composing_2013}, but we will introduce the expression of the application of the Jacobian operator $\Jac$ that is specific to our formulation. To this end, let us first define the discrete residual of the resulting algebraic system evaluated at the point ($ \H_h^{n,k}, \lambda^{n,k}$). For the sake of simplicity, time step and iterate indices $\{n,k\}$ will be omitted in the rest of the section, where expressions always refer to a concrete evaluation point. The component-wise definition of the residual $\R$, of the form $\R(\H_h,\lambda) = [\R^{\H_h}, R^{\lambda}]$, follows
\begin{align} 
\R^{H_h^i}(\H_h, \lambda) &= \sum_{j=1}^{N_H} \int_{\Omega_{\rm hts}} \frac{\mu_0}{| \Delta t |} \bphi^j \cdot \bphi^i   H_h^j + \sum_{j=1}^{N_H} \int_{\Omega_{\rm hts}}  \rho_{\rm hts}(\H_h) (\nabla \times \bphi^j) \cdot (\nabla \times \bphi^i)  H_h^j  \nonumber \\ & \qquad + \int_{S} (\nabla \times \bphi^i)\cdot \n \lambda - \int_{\Omega_{\rm hts}} \f \cdot \bphi^i, \\ 
R^\lambda(\H_h) &= \sum_{j=1}^{N_H} \int_{S} (\nabla \times \bphi^j)\cdot \n H_h^j - I_{\rm app}
\end{align}
for the magnetic field DoFs $\{H^i_h\}_{i=1}^{N_H}$ and the Lagrange multiplier $\lambda$ used to enforce the applied current $I_{\rm app}$ in the closed surface $S$. In this case, $\n$ denotes the unit normal to the surface $\Omega_{\rm hts}$.  The application of the Jacobian to a given direction $\z$, i.e., $\Jac(\x, \z)$, given the linearization point $\x$, i.e., $\Jac(\x) \z$, reads: 
\begin{align}
\Jac(\x) \z = {D} \R(\x, \z) = \Am(\x) \z + D \Am (\x, \z) \x. 
\end{align}
where, e.g., ${D} \R(\x, \z)$ is the G\^{a}teaux derivate of $\R$ at $\x$ in the direction of $\z$. Following the notation proposed in Sect. \ref{sec-operators}, the linearized $\Jac(\x)$ at each iterate $\{n,k\}$ can be stated as
\begin{align}\label{eq-nl_alg}
\Jac(\x) &=\begin{bmatrix}
\frac{\mu_0}{| \Delta t | }\Mm + \Km(\H_h) + \displaystyle\frac{\partial\Km(\H_h)}{\partial \H_h}{\H_h}& \Cm^t \\
\Cm & \0
\end{bmatrix},
\end{align}
where the original operator $\Am$ can be directly identified. The entries for the block corresponding to the magnetic field read:   
\begin{align} 
{\left[ \Jac(\H_h) \right]}_{ij} = \frac{\partial \R_i}{\partial \H_h^j} = & {\left[ \Am(\H_h)\right]}_{ij} + \displaystyle\int_{\Omega_{\rm hts}} \displaystyle\frac{\partial \rho_{\rm hts}(\H_h)}{\partial H_h^j} (\nabla \times \H_h) \cdot (\nabla \times \bphi^i), 
\end{align}
where $\{i,j\}=1,\cdots, N_H$, i.e., the magnetic field number of unknowns. It becomes clear in this expression that the nonlinearity is given by the discrete magnetic field $\H_h$ whereas the Lagrange multiplier enforcing the current is a linear relation.   
 If we go one step further, for the constitutive law presented in Eq. (\ref{eq-conslaw}), and considering $J_c$ independent of the magnetic field, we obtain the expression for the \emph{tangent} resistivity with respect to the magnetic field at the evaluation point $\H_h$: 
\begin{align}
\frac{\partial \rho_{\rm hts}(\H_h)}{\partial H_h^j} = \frac{E_c}{J_c} n \left( \frac{ \| \nabla \times \H_h \|}{J_c} \right) ^{n-2}  \frac{(\nabla \times \H_h)}{J_c} \cdot \frac{(\nabla \times \bphi^j)}{J_c}.
\end{align}

\subsection{Parallel linear solver}\label{subsec:parallel_solver}
The presented nonlinear solver (Sect.~\ref{sec:solver}) can rely on either parallel sparse direct or iterative solvers to solve the linearized problems that arise at every nonlinear iteration. In contrast to sparse direct solvers, iterative solvers can be efficiently implemented in parallel computer codes for distributed-memory computers. However, they have to be equipped with an efficient preconditioner, which is crucial for their robustness and (parallel/algorithmic) scalability. In this work, we ground on the so-called Balancing Domain Decomposition by Constraints (BDDC) preconditioning approach \cite{dohrmann_2003, badia_highly_2014, badia_multilevel_2016} as a preconditioner for Krylov subspace iterative solvers \cite{Saad_book}. Grad-conforming problems, e.g., Laplacian or linear elasticity problems, can efficiently be solved with these algorithms, leading to robust and {\em weakly scalable} linear solvers (see, e.g., \cite{badia_highly_2014, badia_multilevel_2016}).\footnote{A parallel iterative preconditioned solver is said to be weakly scalable if it is able to keep its efficiency (i.e., number of iterations, time-to-solution, memory consumption) as we increase the number of processors while keeping constant the load per processor (i.e., in the solution of larger global problem sizes).} Indeed, they are at the core of state-of-the-art high-performance scientific computing software projects (see, e.g., \cite{fempar-web-page, petsc-web-page}), and they have been shown to be extremely scalable in the solution of elliptic PDE problems. However, the treatment of curl-conforming problems is much more involved. An extension to the curl-conforming case is presented in \cite{toselli_dual-primal_2006} (for a very related solver); see also the deluxe scaling BDDC method in \cite{dohrmann_bddc_2016} or the physics-based approach \cite{badia_physics_based_2017} for problems with jumps of coefficients. %It is well known that 3D weakly scalable algorithms with minimal coarse spaces for problems posed in $\mathcal{H}$(\textbf{curl}) require a change of basis over some DoFs shared by multiple parts, i.e., interface DoFs \cite{toselli_dual-primal_2006}. Nevertheless, the situation is specially involved when {\em hanging} geometrical entities in $h$-adapted meshes. 
In this work, we are interested in strong scaling\footnote{Strong scaling is the ability of a parallel algorithm to reduce time-to-solution with increasing number of processors in the solution of a fixed problem size.}, since we aim at reducing time-to-solution for a fixed problem size. For the sake of robustness, we have implemented  the most conservative coarse space proposed in \cite{toselli_dual-primal_2006}, called Alg. C in this reference, and the approach in \cite{badia_physics_based_2017} to deal with high jumps of material properties. %, which presents lower complexity of implementation in the frame of the BDDC preconditioner, customized for $h$-adaptive methods. %The selection of the algorithm C, in constrast to the algorithm B, makes the algorithm reach a time-to-solution scalability limit for an earlier number of parts. 

\section{Numerical experiments}\label{sec-results}

In this section, we test the $h$-adaptive FE approximation of the $H$-formulation and the parallel nonlinear solver presented in the previous sections. In general, the physical domains in our simulations consist of a superconducting bulk completely surrounded by a dielectric box (see Fig.~\ref{fig-h_adaptive_mesh} for an illustrative example).  Dirichlet boundary conditions are applied on the boundary of the outer domain. Thus, we make sure that this external boundary is far enough from the superconducting device to avoid interior magnetic fields generated by itself to reach the boundary and interfere external applied fields. Unless otherwise stated, the stopping criteria for the nonlinear solver is the reduction of the ratio between the discrete $\ell^2$-norm of the nonlinear residual and right-hand side below $10^{-10}$. At the same time, the stopping criteria for the iterative linear solver applied to each linearized time step is the reduction of the discrete $\ell^2$-norm of the relative residual below $10^{-12}$. The adaptive time stepping algorithm in Sect.~\ref{sec-time} will be used with $\kappa=5$. %First, we present a comparison of the computational model against experimental data, based on the Hall probe mapping experiment in \cite{granados_h-formulation_2016}. Then, the parallel performance of the scheme will be analyzed for the 3D benchmark in \cite{kapolka_3d_2017}. 

\subsection{Software}

All the algorithms described in this work are available at \FEMPAR{}~\cite{badia-fempar,fempar-web-page}, a general purpose, parallel scientific software for the FE simulation of complex multiphysics problems governed by PDEs. It supports several computing and programming environments, such as, e.g.,  multi-threading via OpenMP for the Desktop/Laptop on moderate scale problems, and hybrid MPI/OpenMP for HPC clusters and massively parallel supercomputers. For each programming environment, it offers a set of flexible data structures and algorithms for each step in the simulation pipeline, which can be customized and/or combined in multiple ways in order to satisfy the particular application problem needs; see \cite{badia-fempar} for a deep coverage of the software architecture of \FEMPAR{}. It is distributed as open source software under the terms of the GNU GPLv3 license. \FEMPAR{} is written in Fortran200X following object-oriented design principles. 

\FEMPAR{} supports arbitrary order edge FEs on both hexahedra and tetrahedra, on either structured or unstructured {\em conforming} meshes (i.e., typically generated by an external mesh generator), and mesh generation and adaptation using hierarchically refined octree-based meshes. The serial and MPI-parallel versions of the process described in Sect.~\ref{sec-mesh_refinement} is grounded on {\tt p4est}~\cite{BursteddeWilcoxGhattas11}. {\tt p4est} is an MPI library for efficiently handling (forest of) octrees on distributed-memory processors. Among others, it provides a set of octree manipulation primitives which are essential for our approach: (1) to adapt an octree by refining (or coarsening) its cells; (2) to redistribute the octree cells among the available processors for dynamic  load-balancing (by means of space-filling curves~\cite{BursteddeWilcoxGhattas11}); and (3) to enforce the 2:1 balance ratio. Using a compact representation, {\tt p4est} provides memory-efficient and scalable algorithms for all the aforementioned manipulation primitives {\cite{BursteddeWilcoxGhattas11}. On top of {\tt p4est}, \FEMPAR{} builds a richer representation of the mesh (essentially mesh cells and lower dimensional geometrical entities connectivity information) to support the implementation of adaptive FE methods using \emph{hanging} DoFs constraints (see Sect.~\ref{sec-mesh_refinement} ). Both the mesh, and the rest of data structures used in the simulation are {\em fully-distributed} among the processors involved in the parallel simulation. This implies, e.g., that each processor holds a partial portion of the global mesh cells, and a subset of the DoFs of the global FE space. It is essential to scale FE simulations to large core counts. 

At the linear solver kernel, it offers several alternatives depending on the programming environment at hand. In this work, we used the ones described in the sequel. For small scale problems on, e.g., a Desktop computer, we rely on a parallel multi-threaded sparse direct solver available at Intel MKL PARDISO \cite{intel_pardiso}. On the other hand, for a hybrid OpenMP/MPI environment, linear solvers are based on preconditioned Krylov subspace solvers. \FEMPAR{} hallmark is an abstract OO framework for the implementation of widely applicable highly scalable multilevel DD solvers. The preconditioners which are accommodated within this framework require the solution of linear systems which are local to each subdomain, and the so-called coarse-grid problem, that is crucial for preconditioner efficiency and scalability. These problems are solved using the aforementioned sparse direct solver. Each MPI task in the parallel computation handles the computations to be performed at a single subdomain. Provided that the algorithm lets a high degree of overlapping to be achieved among fine and coarse-grid tasks, an additional MPI task is spawn in order to carry out coarse-grid-related computations; see, e.g., \cite{badia_highly_2014} for additional details. Our BDDC preconditioner implementation can deal with curl-conforming spaces of arbitrary order, tetrahedral/hexahedral meshes and structured/unstructured partitions. 

\subsection{Experimental framework}

The experiments in this section have been performed on the  MareNostrum-IV~\cite{MNIV} (MN-IV) supercomputer, hosted by the Barcelona Supercomputing Center (BSC). MN-IV is equipped with 3456 compute nodes connected together with the Intel OPA HPC network. Each node is equipped with 2x Intel Intel Xeon Platinum 8160  multi-core CPUs, with 24 cores each (i.e., 48 cores per node), and 96 GBytes of RAM. \FEMPAR{} was compiled with the Intel Fortran compiler (v18.0.1) using system-recommended optimization flags, and linked against the Intel MPI Library (v2018.1.163) for message-passing, and the BLAS/LAPACK and PARDISO available on the Intel MKL library for optimized dense linear algebra kernels, and sparse direct solvers, respectively. All floating-point calculations were performed in IEEE double precision.

\subsection{Comparison against experimental data}\label{sec-vali}
In this section, the proposed FE model is validated against experimental data, obtained by means of the Hall scanning magnetometer experiment, exposed in detail in \cite{granados_h-formulation_2016}. Our goal is to compare the experimental data for a 2G tape sample (see Tab.~\ref{tab-params} for properties) with the numerical results obtained with \FEMPAR{}. The problem of a HTS tape magnetized by a current flowing through it has been solved by several authors (see, e.g., \cite{norris_calculation_1971, nibbio_effect_2001, pardo_alter_2004}). Besides, several comparisons between experimental and numerical results can be found, e.g., in \cite{dutoit_measuring_2005, FUKUI2003224, kim_estimation_2010}.

\begin{table}
\centering
\begin{tabular}{lcccr}
\hline 
\textbf{Parameter}        & \thead{Comparison \\ to experimental \\data}  &    3D benchmark     &  Units \\  \hline
air domain size             &  $100\times100$  &   $100\times100\times100$ & ${\rm mm}^d$          \\
HTS width                   & $12$             &   $10$              & mm                 \\ 
HTS thickness               & $110$            &   $1000$            & $\mu$mm            \\
$n$ power law exponent      &  $32$            &     $24$            &                    \\
$\mu_0$                     &  ${4\pi}\cdot 10^{-7}$ & ${4\pi}\cdot 10^{-7}$ & H/m         \\
$\mu_r$                     & 1.0               &  1.0                       & H/m          \\       
$E_0$                       & $10^{-4}$         &  $10^{-4}$                 & V/m          \\
$J_c$                       & $3.38\cdot 10^8$  &  $1.0\cdot 10^8$    & A/$\rm{m}^2$        \\
$I_c$                       & $446.16$          &  $1.0\cdot 10^3$    & A        \\
$\rho_{\rm air}$            &  $1.0$            &      $10^{-2}$     & $\Omega\cdot$m       \\  
\hline
\end{tabular}
\caption{Geometric and electrical parameters of the HTS tape used in different problems. Domain units dependend on  domain dimension $d$.}
\label{tab-params}
\end{table}

In the experiment we consider that the current is applied in a HTS tape by a sequence of step functions, with time intervals of 100~seconds each before proceeding to the following increment. The current applied is gradually increased up to $I_{\rm app}=460$~A, which corresponds to $1.03 \cdot I_c$ (i.e., $I_c=446.16$). See Fig.~\ref{fig-applied_current_profile} for a clear exposition of the injected current. The applied current remains constant during short periods of time so it allows the flux creep effects to pass, and therefore the current distribution is stabilized along the superconductor specimen.% The physical experiment is carried out at a temperature of $77~K$.

We will compare the experimental and numerical profiles of the vertical component of the magnetic field (i.e., $B_y=\mu_0 H_y$) $400~\mu\rm{m}$ above the HTS tape surface (where the active part of the sensor is located). For the computational model, we simplify the superconducting region as a homogenization of the multiple layers typically found in a 2G tape (e.g., the substrate, the silver, and the copper covering), where only a layer of $1$-$2~\mu\rm{m}$ corresponds to superconducting material. As it is shown in this section, such approximation is accurate in the magnetic field computation at $400~\mu\rm{m}$ above the tape surface. Following \cite{granados_h-formulation_2016}, a critical current dependence with the magnetic field, i.e., $J_c(\B)$, is introduced through tabulated values (see Fig.~\ref{fig-JcB}), where effective $J_c$ values can be obtained at each nonlinear iteration from the corresponding linearization point $\B$. A Lift Factor (LF=$J_c/J_{c0}$) is obtained for each component $B_i$ of $\B$, i.e., $LF_i(B_i)$, where $i\in\{1,2\}$ in this 2D simulation (see Fig.~\ref{fig-JcB}). Then, the resulting $J_c$ is obtained as $J_{c0} \cdot \left\Vert LF \right\Vert$, where $J_{c0}$ is the value of $J_c$ in absence of magnetic field, i.e., $J_c({\bf 0})$, detailed in Tab.~\ref{tab-params}.

Fig.~\ref{fig-validation} shows the comparison between the experimental and numerical vertical magnetic field $\mu_0 H_y$. The numerical results are obtained for first order edge FEs. All plots in Fig.~\ref{fig-validation} show profiles along the $x$-axis section for the full HTS tape length and 6 additional $\rm{mm}$ to each side. Therefore, with the reference domain $\Omega=[0,100]\times[0,100]~\rm{mm}^2$, these profiles correspond to the line $y=50.455,~ 38<x<62~\rm{mm}$. Out of these results, some conclusions can be drawn. First and most important, the experimental data is in good agreement with the performed simulations and therefore, the proposed formulation is able to reproduce the physical phenomena. Second, the magnetic field peak values are, in all cases, in excellent agreement. It is important to note that the experimental data does not possess symmetry, while the computed data respects such expected symmetry. This fact can be attributed to small imperfections in the experiment. Out of the validation test, we can identify three phases. A first one, where the computed data reproduces with a high accuracy the experimental data (see Figs.~\ref{fig-50up}, \ref{fig-100up}, and \ref{fig-200up}). A second stage, in which, even though the peak values are in good agreement, the experimental data presents small variations in the superconductor area  (see Figs.~\ref{fig-300up}-\ref{fig-400down}). Finally, a third stage, coinciding with the unloading of the sample, where a slightly higher discrepancy is observed. However, peak values are correctly captured and the model predicts the physical phenomenon throughout the entire simulation. 

\begin{figure}[b!]
    \centering
    \begin{subfigure}[t]{0.38\textwidth}
        \includegraphics[width=\textwidth]{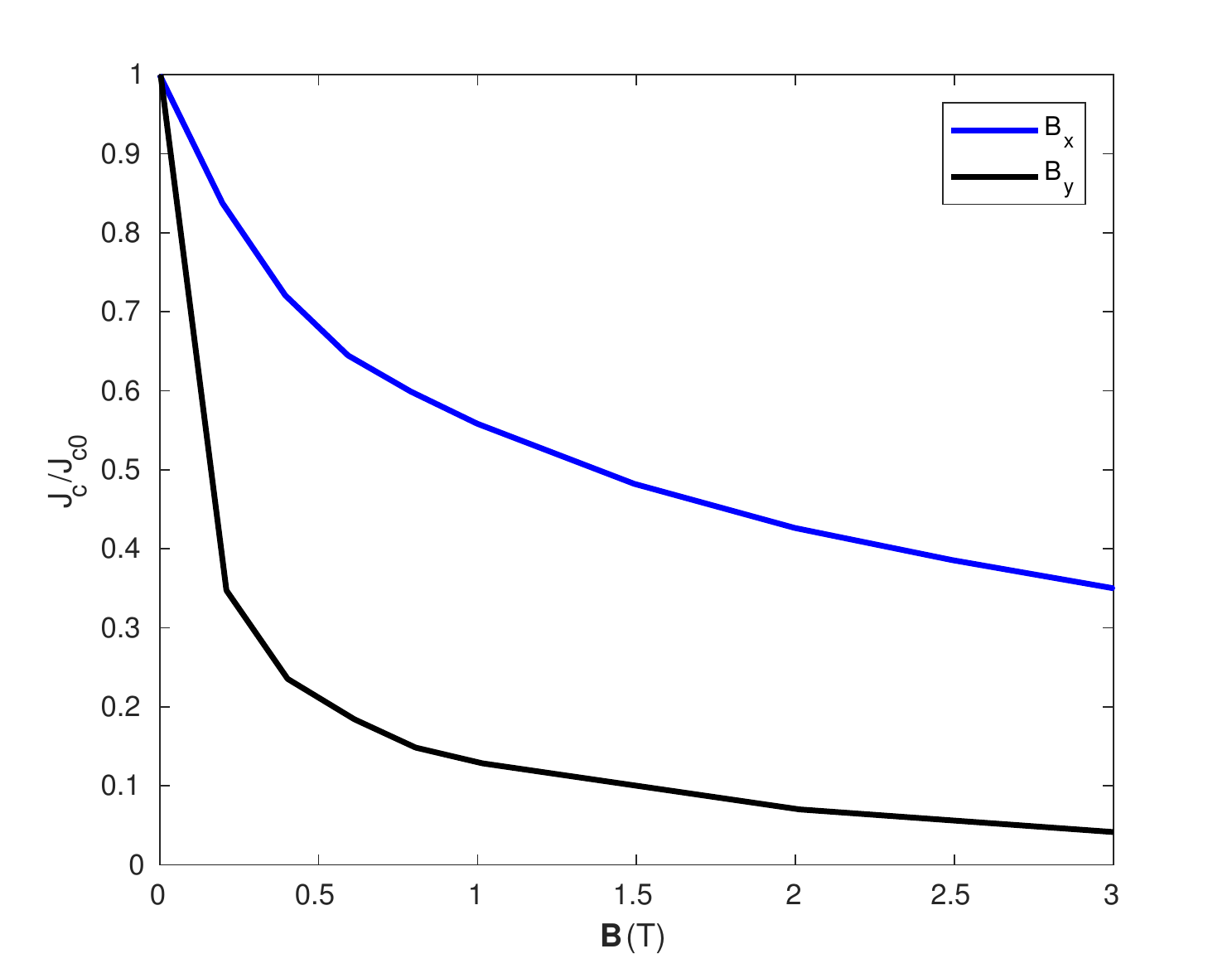}
        \caption{Dependence of $J_c$ on the magnetic field $\B$. Normalized with $J_{c0}$. }
        \label{fig-JcB}
    \end{subfigure}
    \hspace{1cm}
    \begin{subfigure}[t]{0.38\textwidth}
        \includegraphics[width=\textwidth]{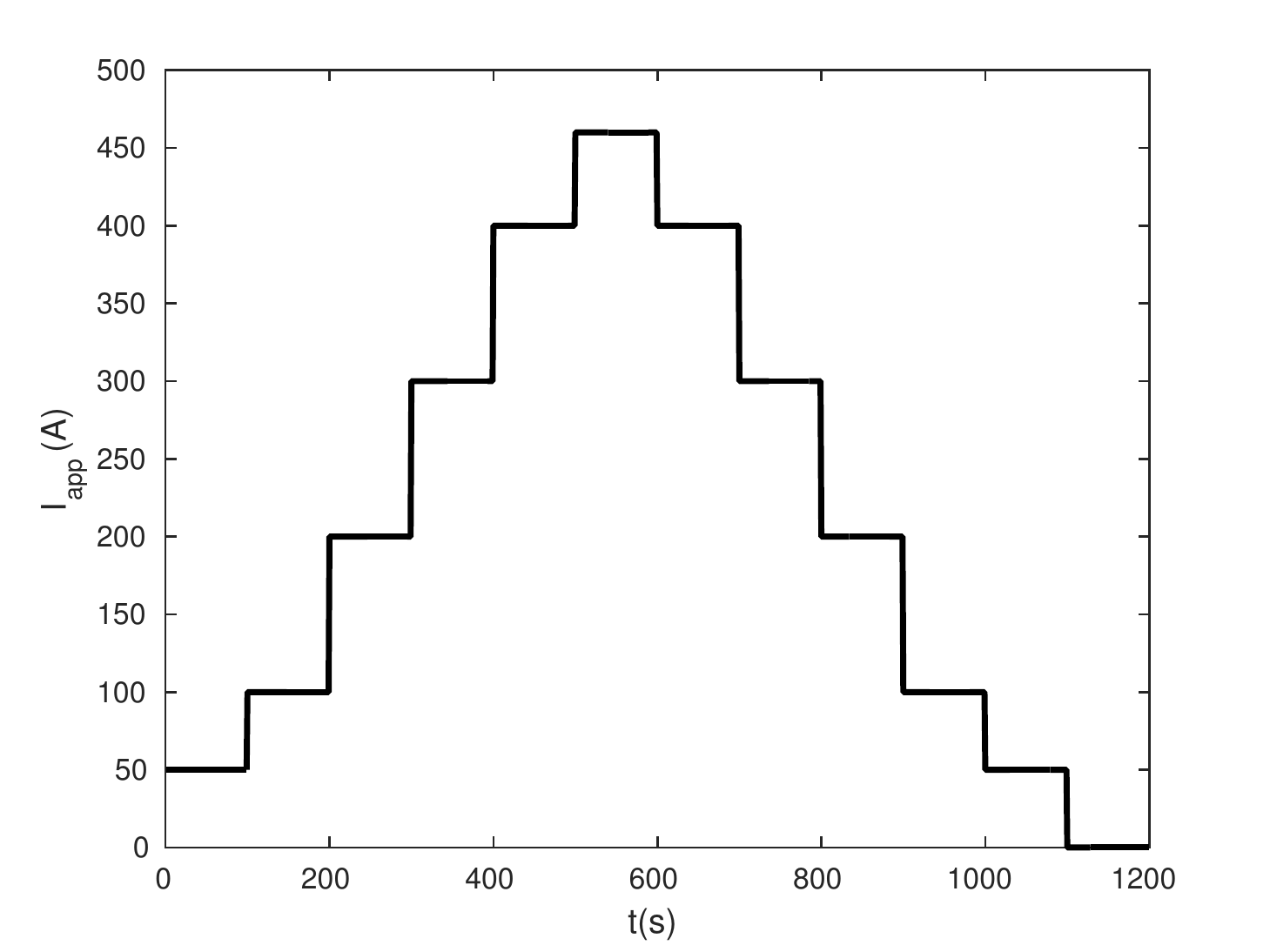}
        \caption{Injected current in the superconductor as a step function. Peak value of $I=460$~A, $I=1.03 I_c$.}
        \label{fig-applied_current_profile}
    \end{subfigure}
    \caption{Validation problem inputs definition.}\label{fig-input_data}
\end{figure}

\begin{figure}[h!]
    \centering
    \begin{subfigure}[b]{0.3\textwidth}
        \includegraphics[width=\textwidth]{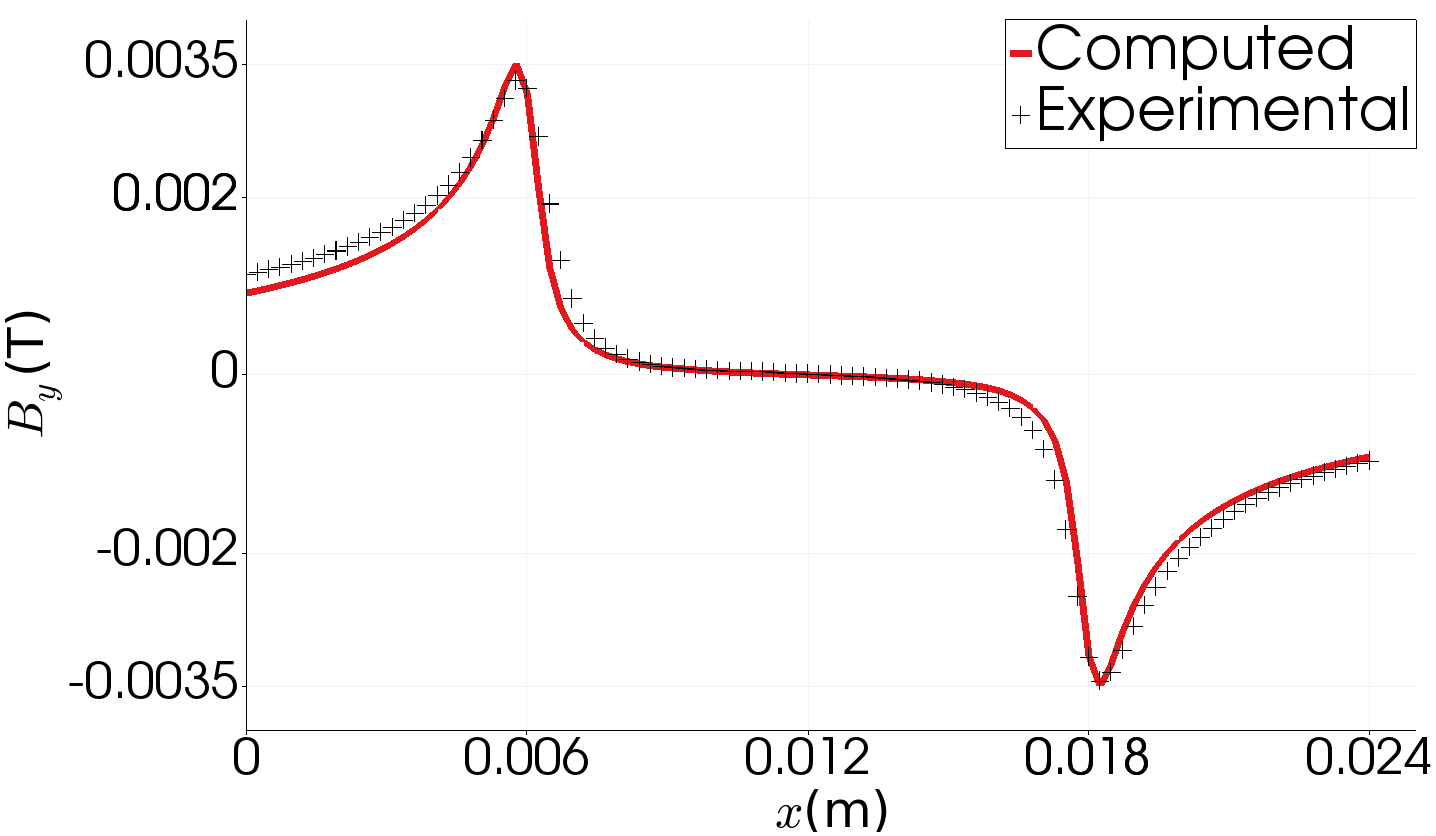}
        \caption{Loading $I_{\rm app}=50$~A.}
        \label{fig-50up}
    \end{subfigure}
    ~ %add desired spacing between images, e. g. ~, \quad, \qquad, \hfill etc. 
    \begin{subfigure}[b]{0.3\textwidth}
        \includegraphics[width=\textwidth]{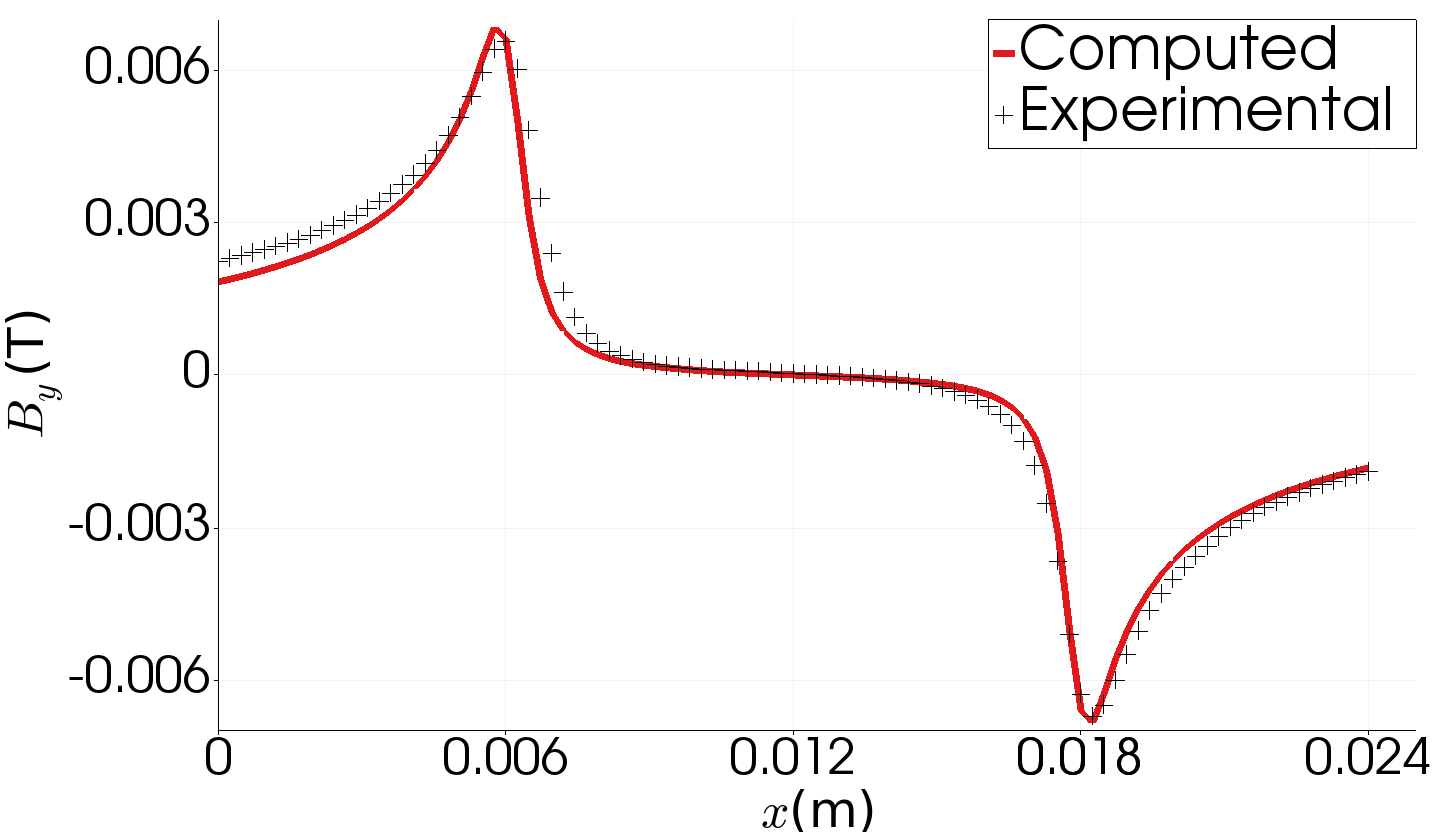}
        \caption{Loading $I_{\rm app}=100$~A.}
        \label{fig-100up}
    \end{subfigure}
        \begin{subfigure}[b]{0.3\textwidth}
        \includegraphics[width=\textwidth]{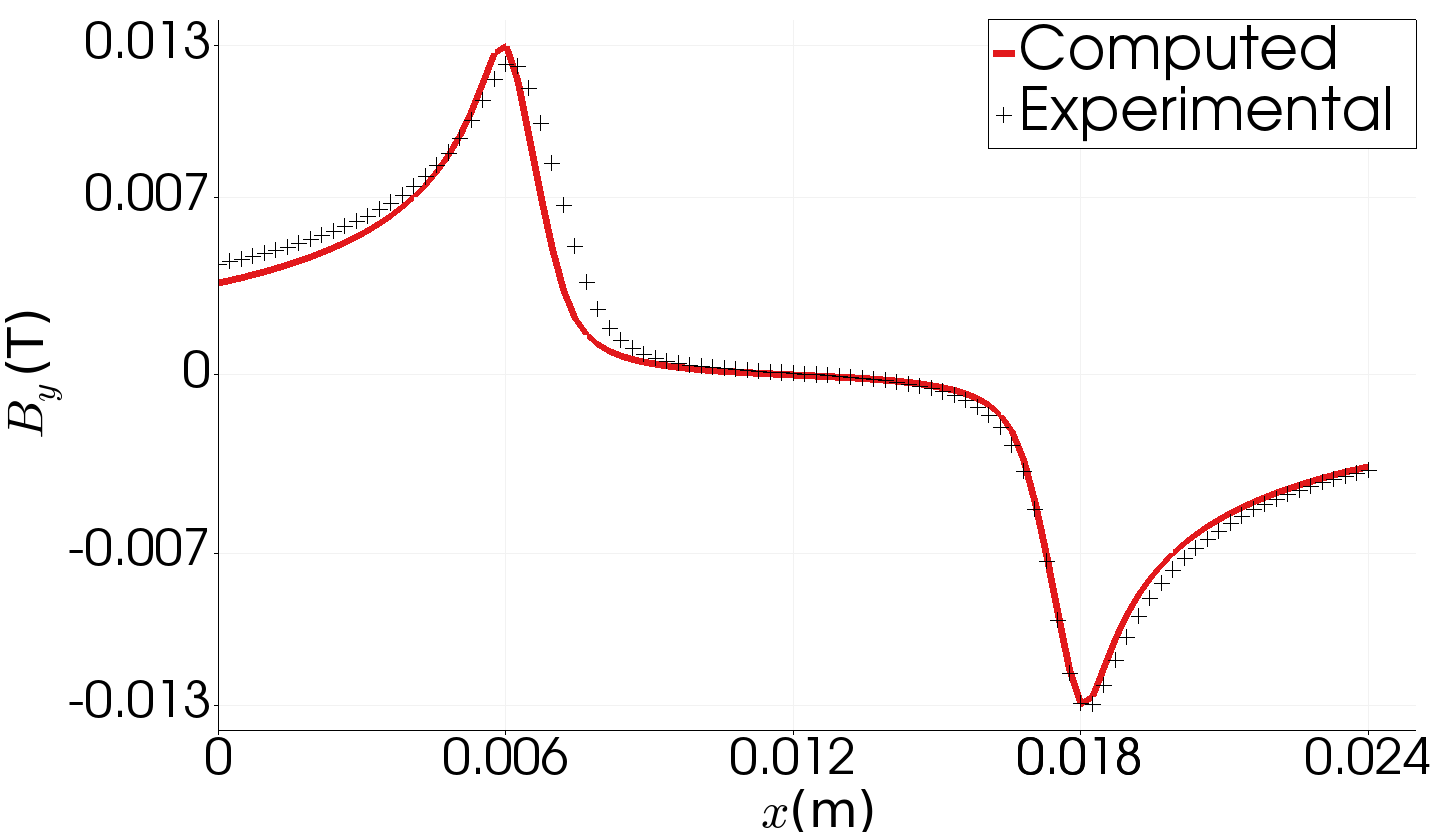}
        \caption{Loading $I_{\rm app}=200$~A.}
        \label{fig-200up}
    \end{subfigure}

    \begin{subfigure}[b]{0.3\textwidth}
        \includegraphics[width=\textwidth]{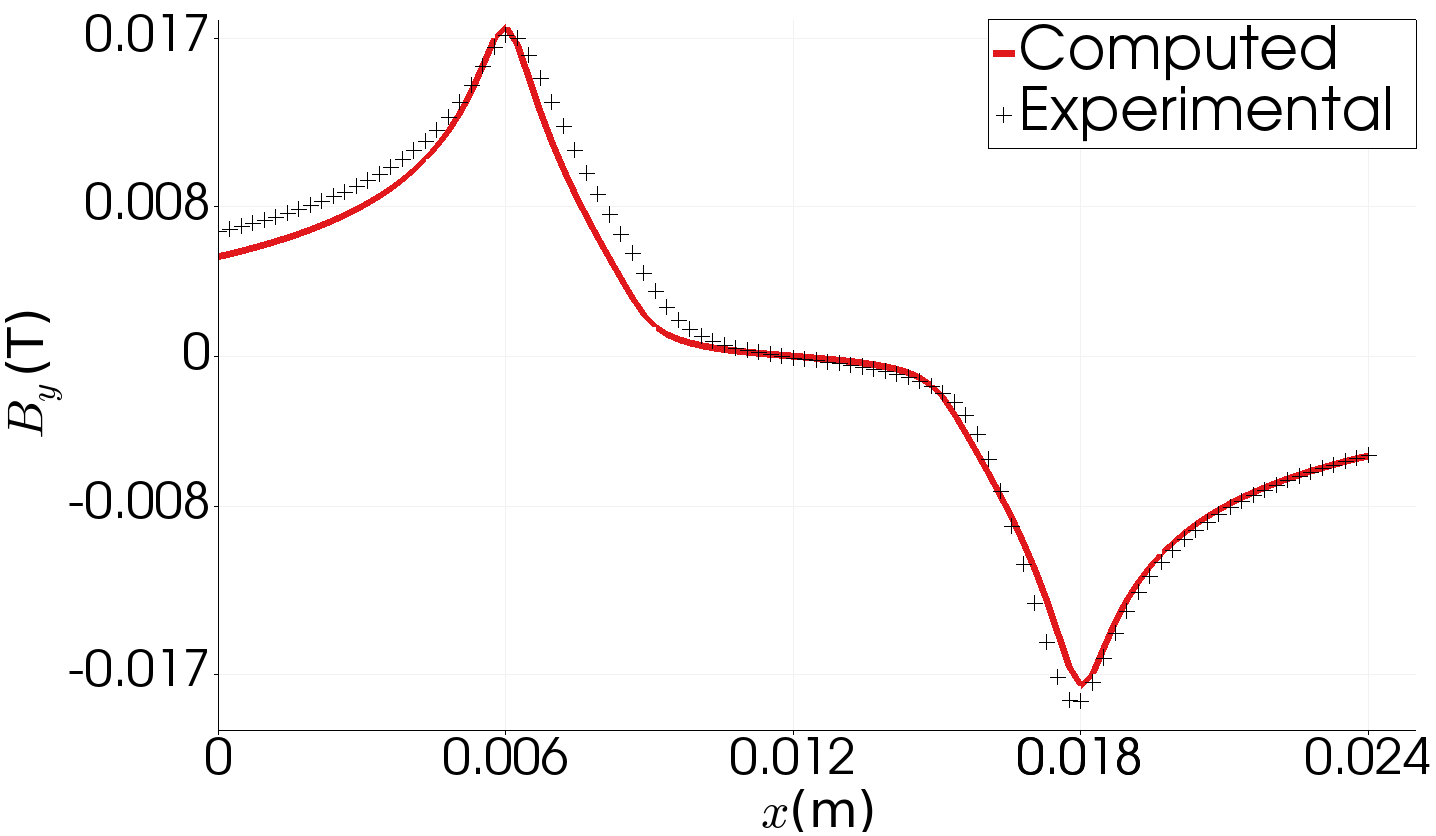}
        \caption{Loading $I_{\rm app}=300$~A.}
        \label{fig-300up}
    \end{subfigure}
        \begin{subfigure}[b]{0.3\textwidth}
        \includegraphics[width=\textwidth]{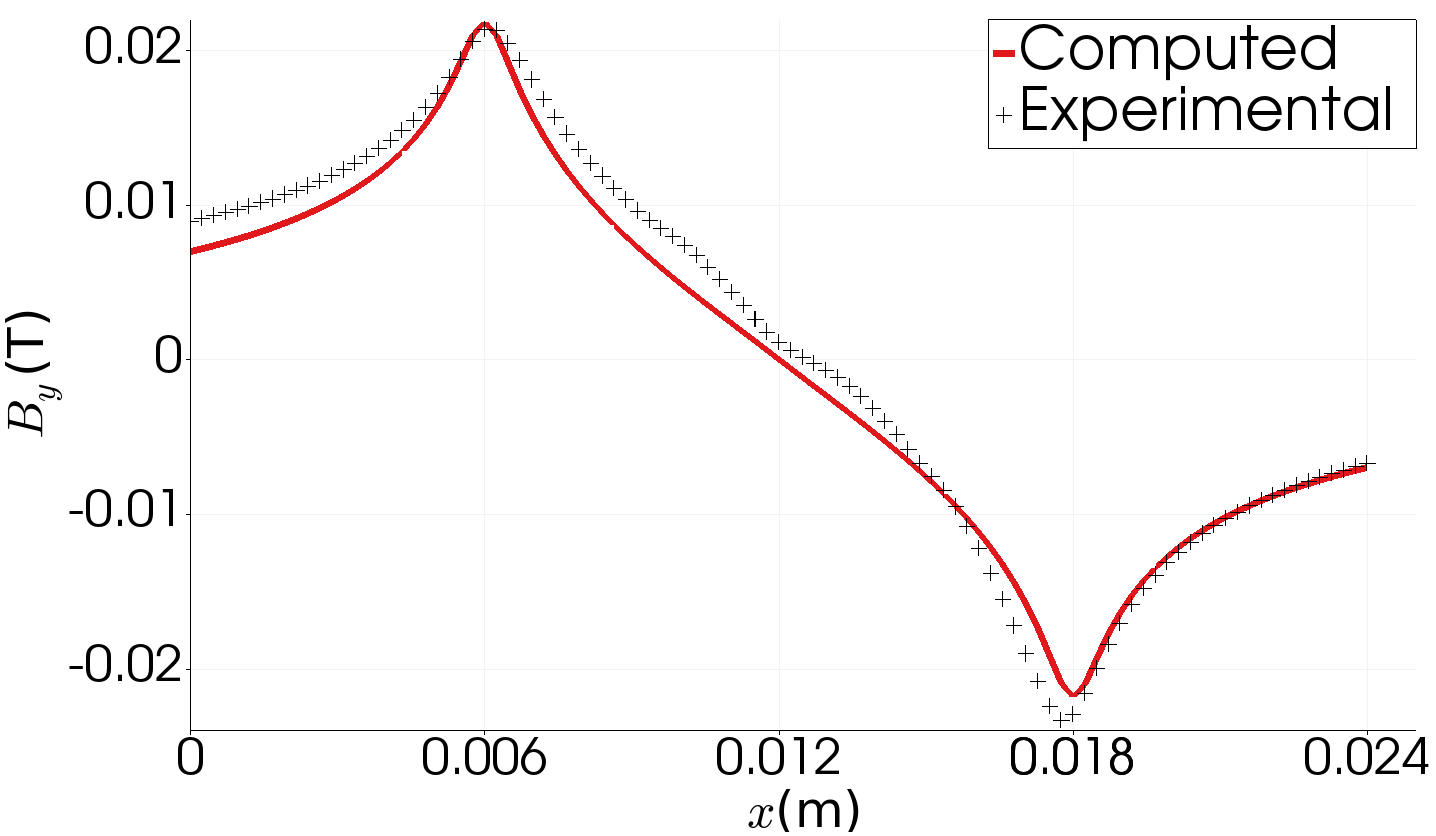}
        \caption{Loading $I_{\rm app}=400$~A.}
        \label{fig-400up}
    \end{subfigure}
        \begin{subfigure}[b]{0.3\textwidth}
        \includegraphics[width=\textwidth]{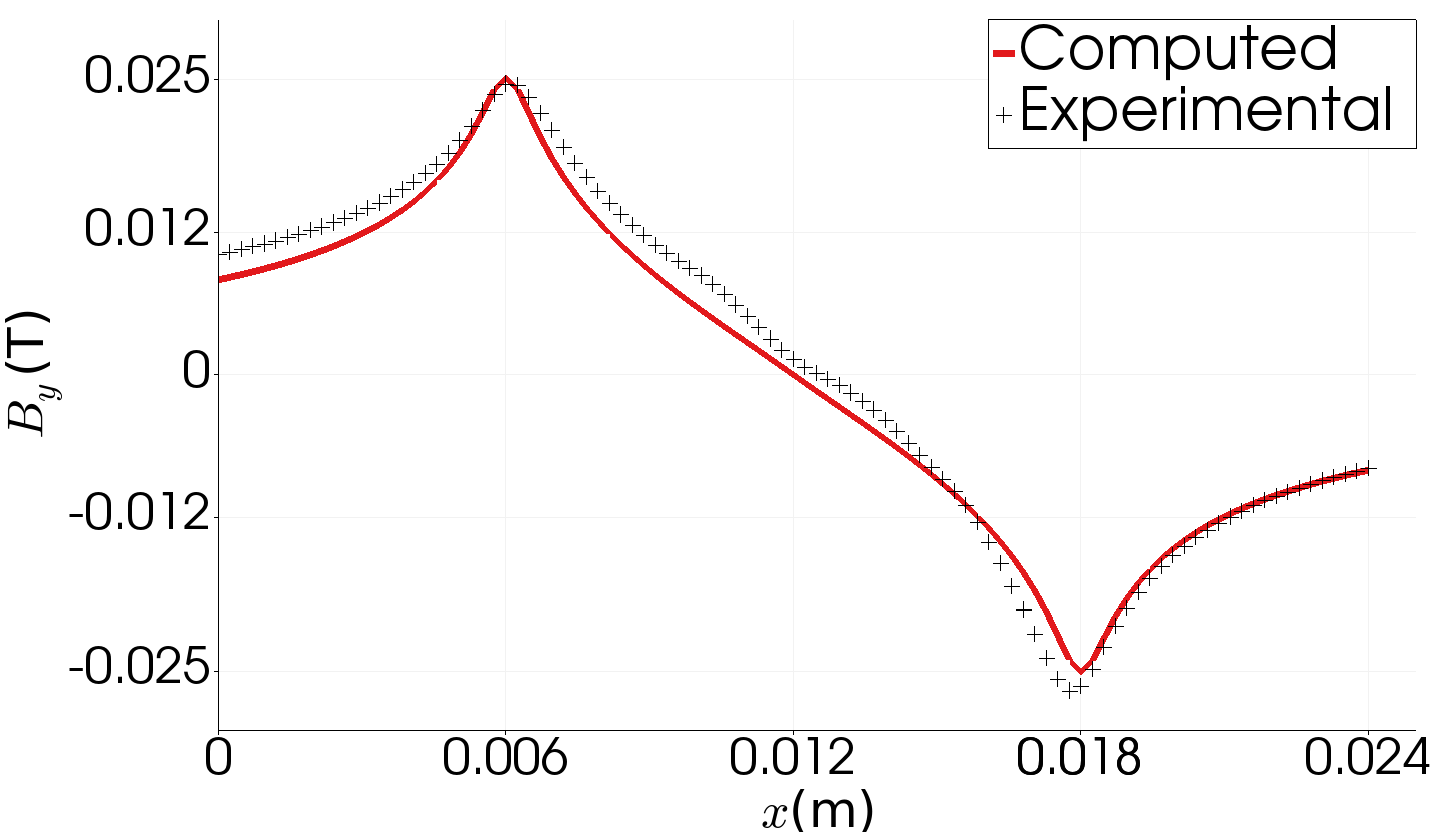}
        \caption{Loading $I_{\rm app}=460$~A.}
        \label{fig-460}
    \end{subfigure}
    
        \begin{subfigure}[b]{0.3\textwidth}
        \includegraphics[width=\textwidth]{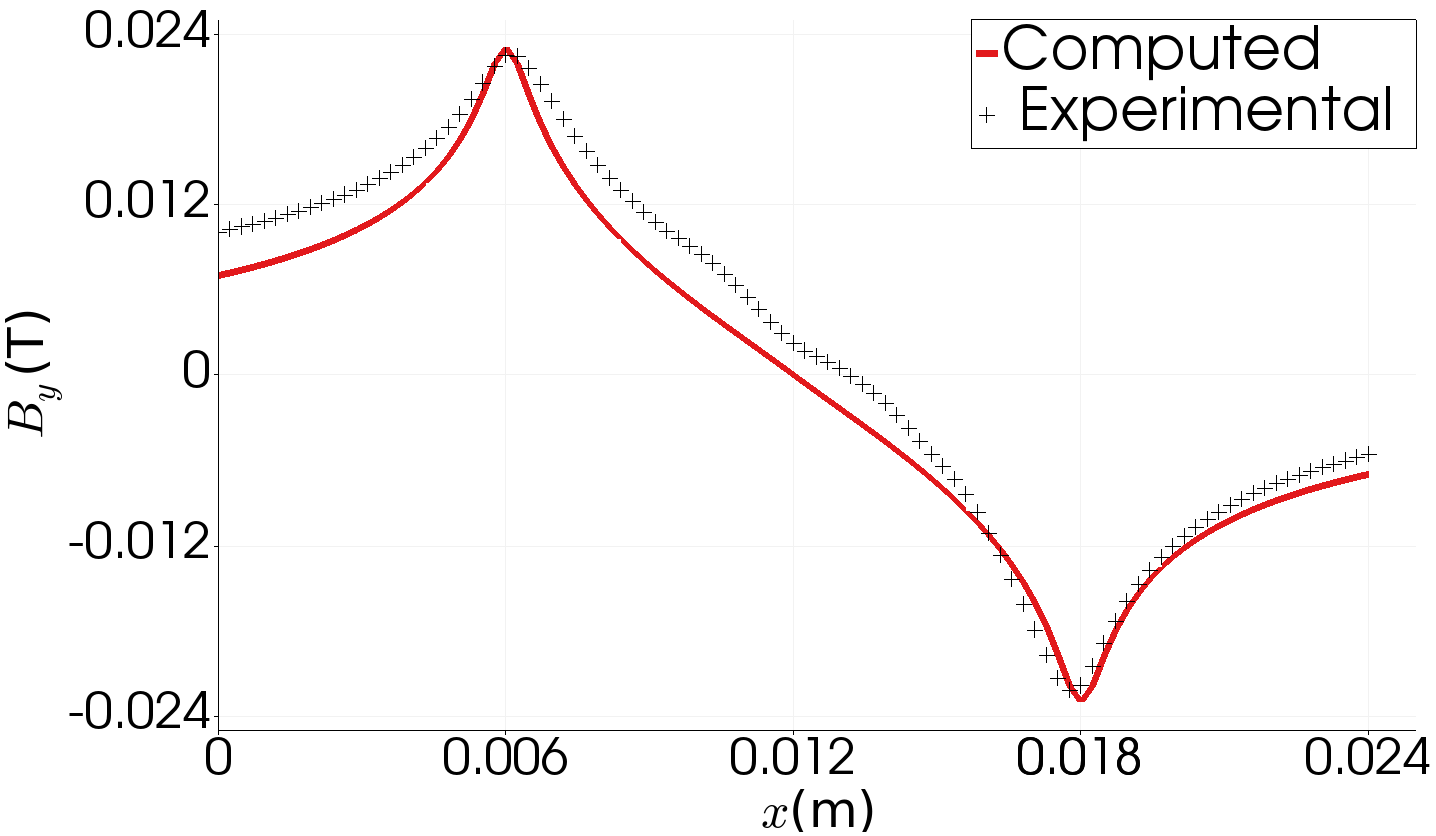}
        \caption{Unloading $I_{\rm app}=400$~A.}
        \label{fig-400down}
    \end{subfigure}
        \begin{subfigure}[b]{0.3\textwidth}
        \includegraphics[width=\textwidth]{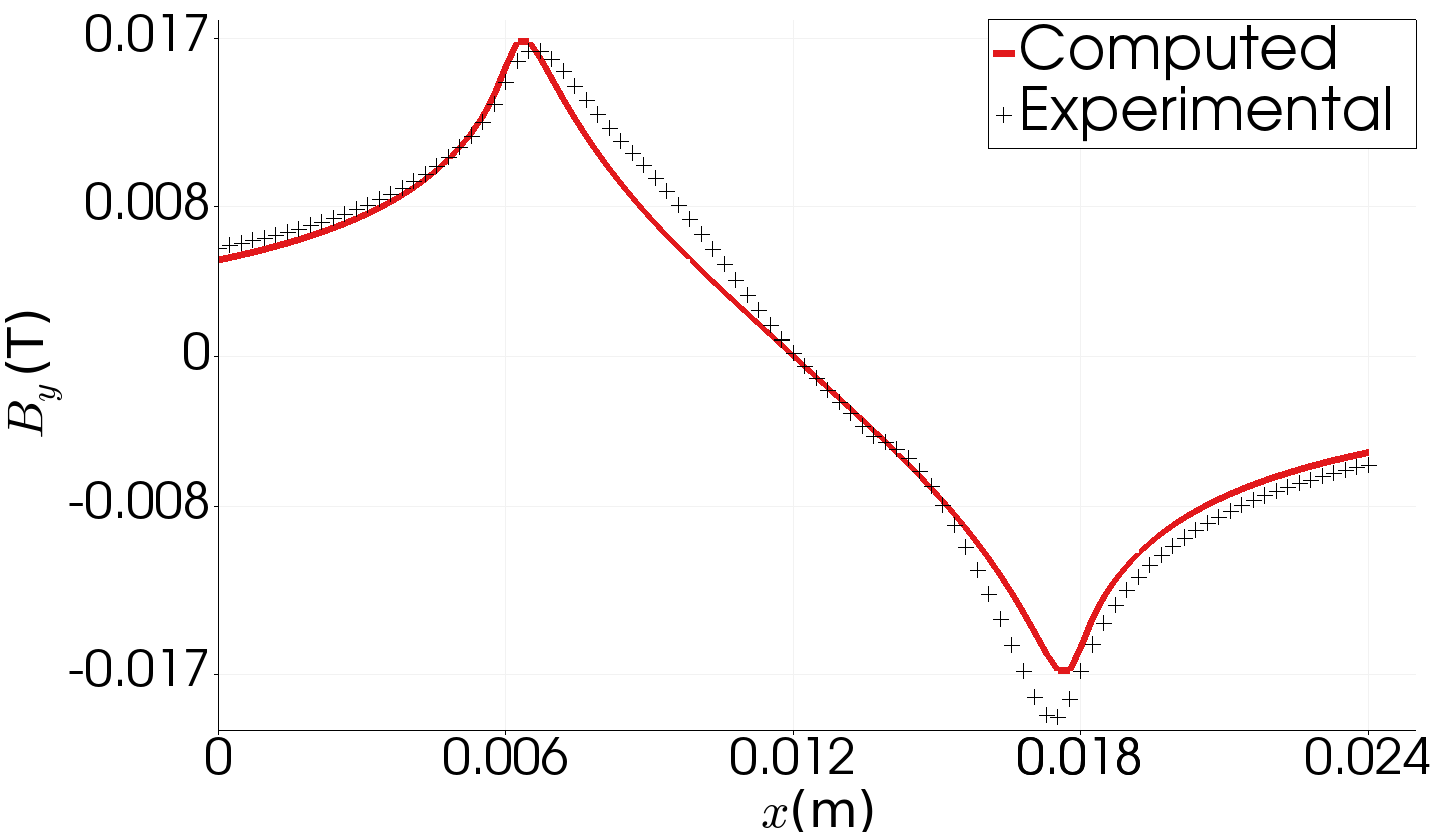}
        \caption{Unloading $I_{\rm app}=300$~A.}
        \label{fig-300down}
    \end{subfigure}
        \begin{subfigure}[b]{0.3\textwidth}
        \includegraphics[width=\textwidth]{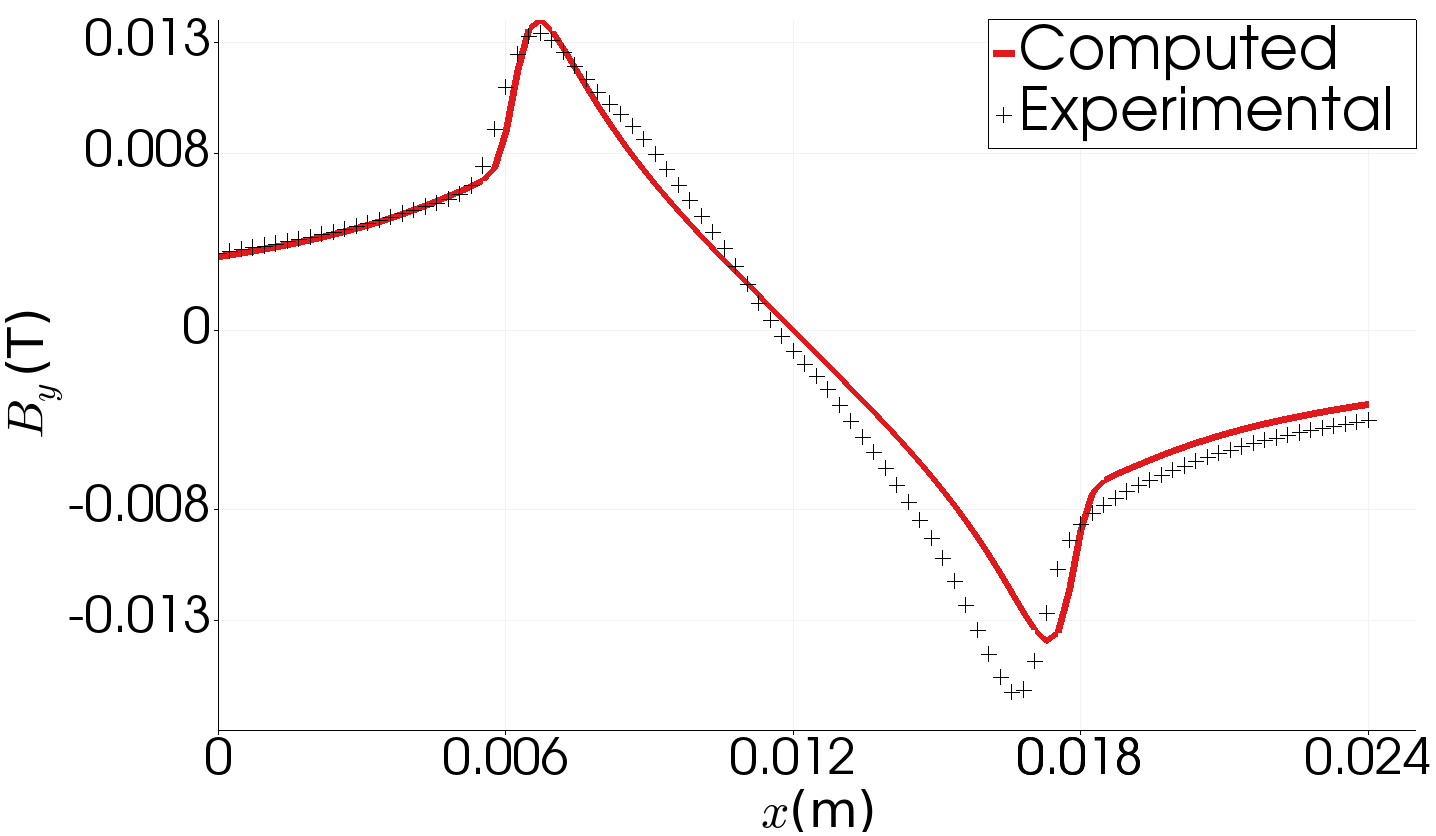}
        \caption{Unloading $I_{\rm app}=200$~A.}
        \label{fig-200down}
    \end{subfigure}
    
        \begin{subfigure}[b]{0.3\textwidth}
        \includegraphics[width=\textwidth]{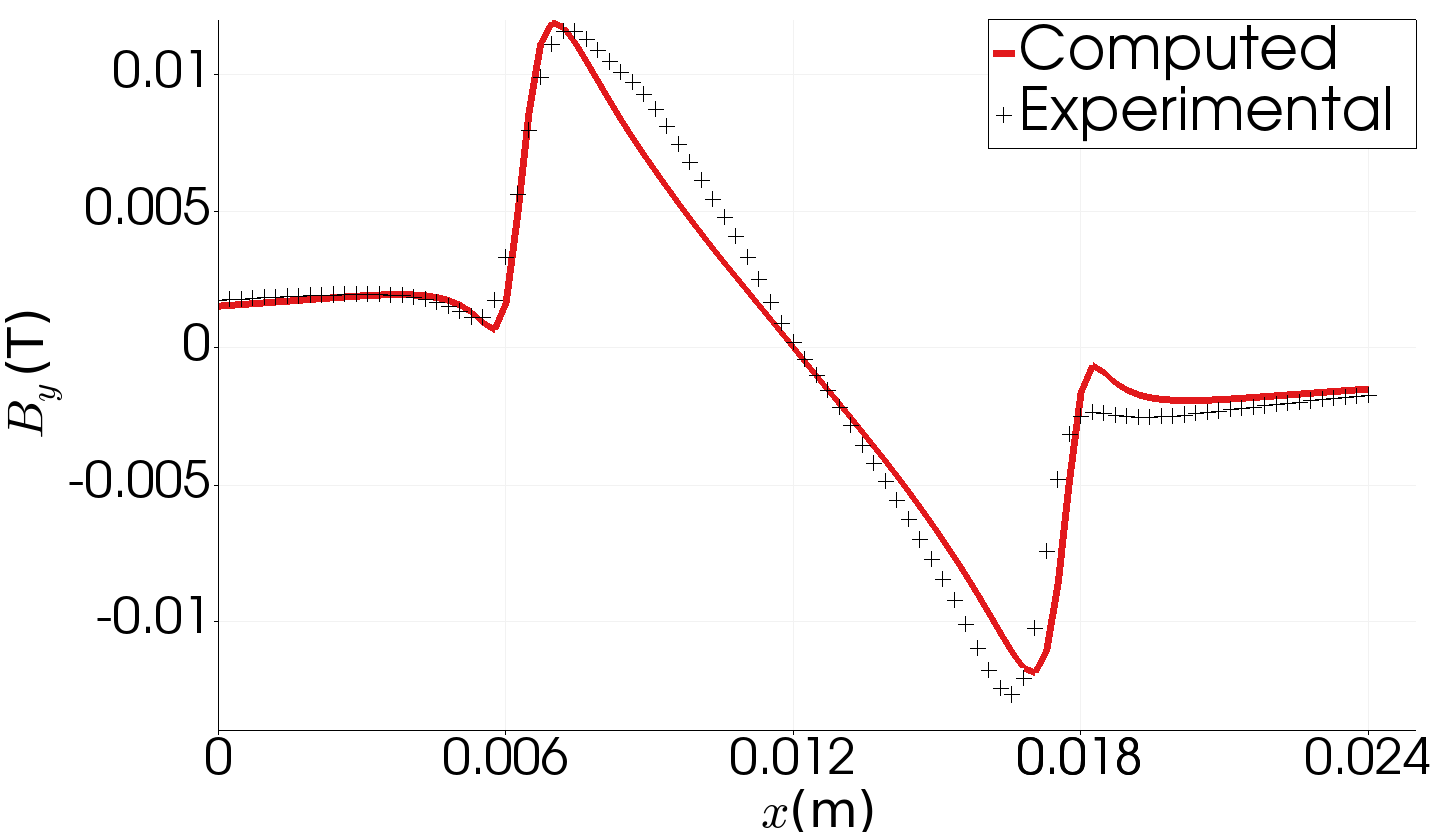}
        \caption{Unloading $I_{\rm app}=100$~A.}
        \label{fig-100down}
    \end{subfigure}
        \begin{subfigure}[b]{0.3\textwidth}
        \includegraphics[width=\textwidth]{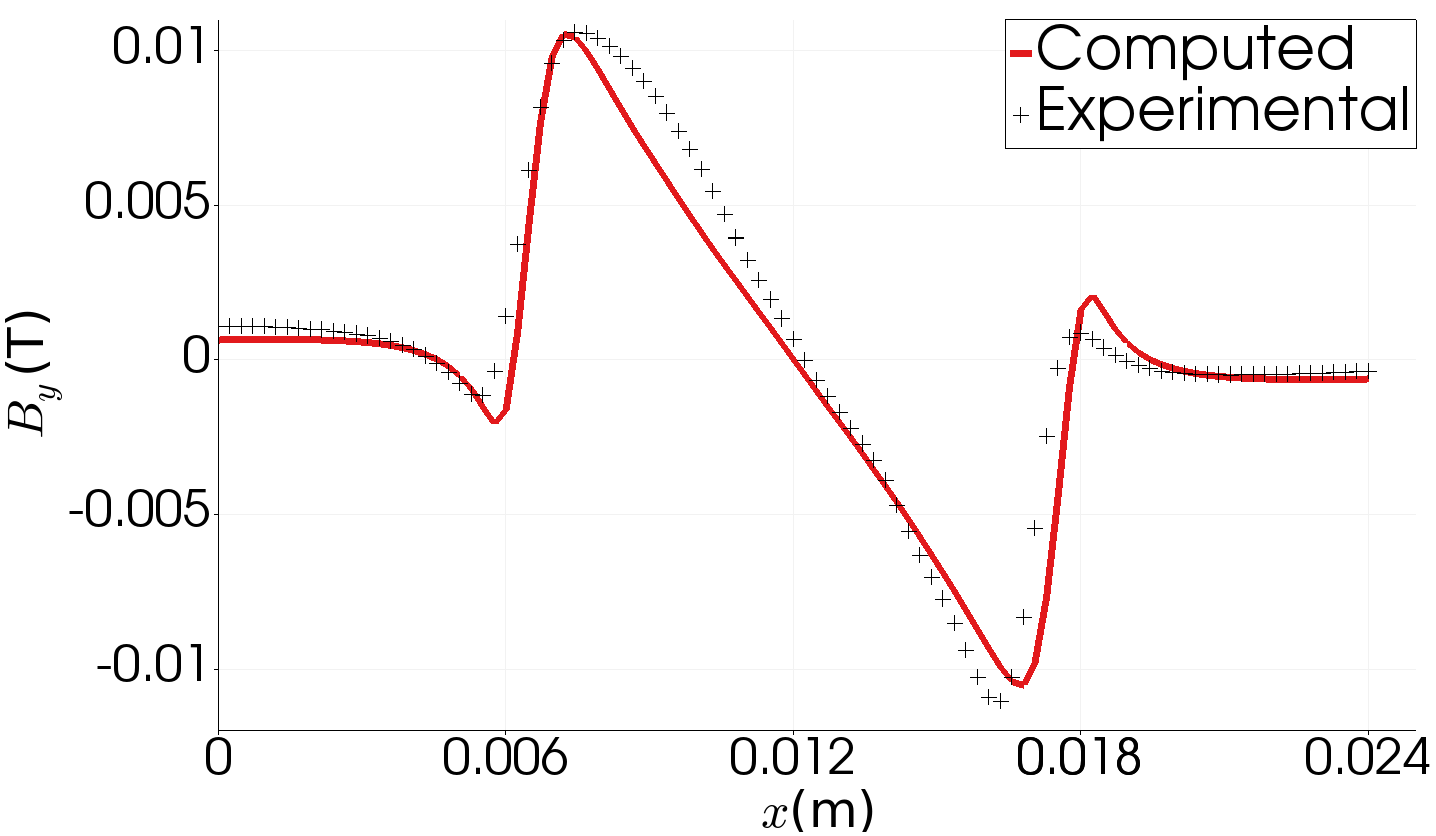}
        \caption{Unloading $I_{\rm app}=50$~A.}
        \label{fig-50down}
    \end{subfigure}
        \begin{subfigure}[b]{0.3\textwidth}
        \includegraphics[width=\textwidth]{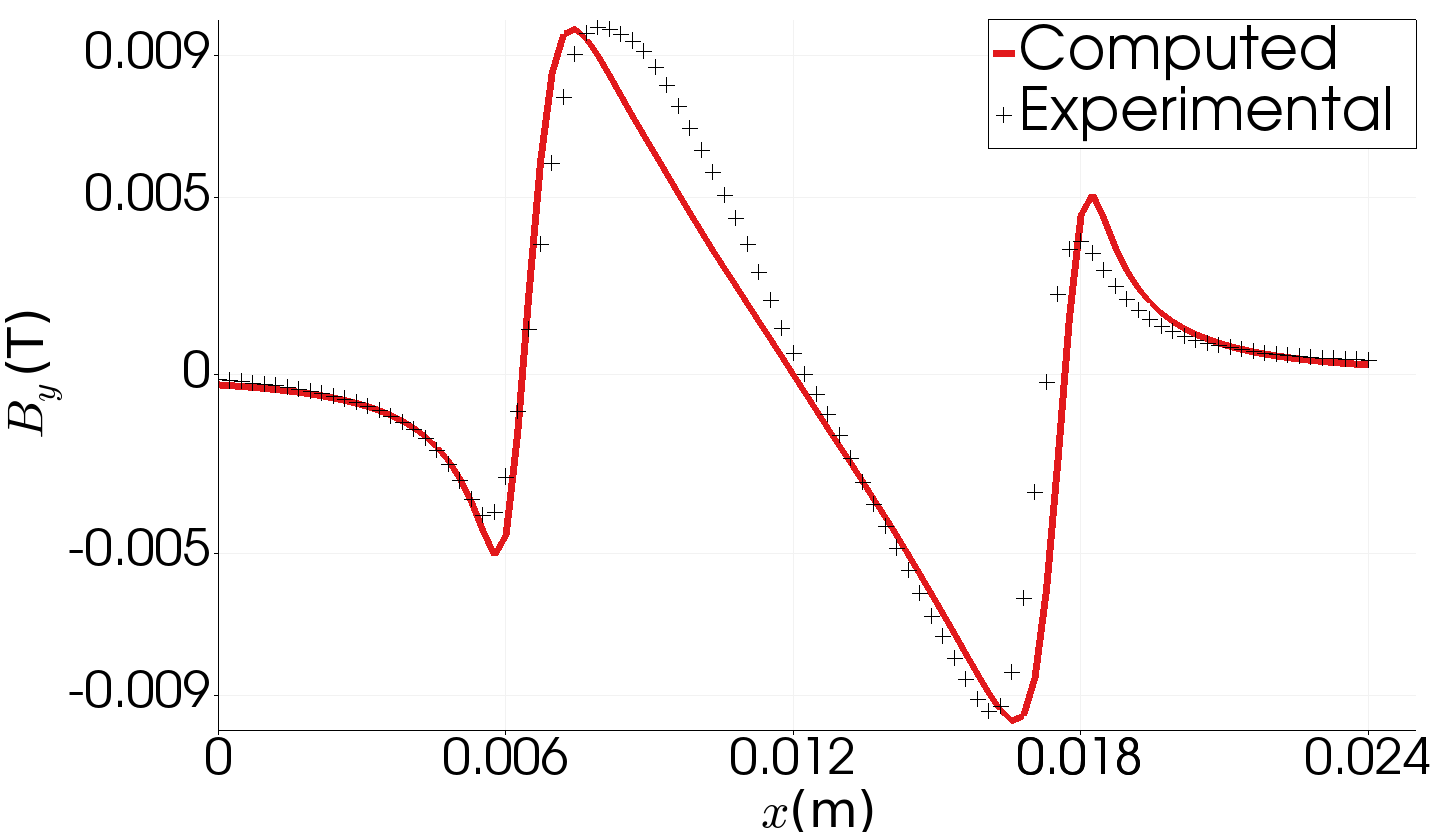}
        \caption{Unloading $I_{\rm app}=0$~A.}
        \label{fig-0down}
    \end{subfigure}
    \caption{Magnetic field $B_y=\mu_0 H_y$ profiles for experimental (Hall probe mapping) and computed data in a full load-unload cycle for the validation test.}\label{fig-validation}
\end{figure}

\subsection{3D benchmark}
In this section, our algorithms are applied to a recently proposed benchmark in \cite{kapolka_3d_2017}, suggested as a stepping stone for future investigations in the 3D modelling of the electromagnetic behaviour of superconductors. The benchmark consists in the magnetization of a superconducting parallelepiped subjected to an AC magnetic field making an angle with the normal to the larger surface. Many authors have previously analyzed the case of a tape under an external field by several methods \cite{PRIGOZHIN1996190, nibbio_effect_2001, brandt_superconductors_1996,  amemiya_numerical_1997,  pardo_the_transverse_2004}. In our case, the scenario is valid to model two different cases: the magnetization of an isotropic superconducting bulk and the case where no current can flow in one direction, i.e., an anisotropic material. This last realistic case models, e.g., stacks of HTS coated conductors used as permanent magnets, where the current cannot flow along a direction due to the presence of high-resistivity layers in the stack.

Consider a box domain $\Omega=[0,100]\times[0,100]\times[0,100]$~${\rm mm}^3$ where the superconductor fills the volume $\Omega_{\rm hts} = [45,55]\times[45,55]\times[49.5,50.5]~{\rm mm}^3$. The air domain is defined as $\Omega_{\rm air} = \Omega \setminus \Omega_{\rm hts}$. There is no source term, i.e., $\f=\0$, and Dirichlet-type boundary conditions are imposed over the entire boundary as a time-dependent magnetic field. Initial conditions are simply $\H(t=0)=\0$~T. On the other hand, no net current flow condition is imposed in the superconductor. The parallelepiped is subjected to a uniform external sinusoidal magnetic field in the $xz$ plane at an angle $\alpha=\pi/6$ with respect to the $x$-axis, an amplitude of $\B_{ext}=200$~mT, and frequency $\omega=$50~Hz. The superconductor behaviour is modelled with the resistivity nonlinear law (see Eq.~(\ref{eq-powerlaw})) with an exponent $n=24$. In addition to an isotropic resistivity, the situation where no current can flow along the $z$-direction is also considered by assigning an anisotropic resistivity value $\rho_z=\rho_{\rm air}$, whereas the resistivity is defined by means of the power law definition in the remaining directions. For the sake of brevity, the two situations are referred to as bulk and stack, respectively. A detailed exposition of the parameters being used can be found in Tab.~\ref{tab-params}. Note the slight reduction of the resistivity parameter in the dielectric region, which allows to enhance convergence performance without any impact in the computed quantities, as will be shown by results. Simulations are performed for a full cycle and an additional quarter of a cycle to take into account the initial magnetization of the HTS device. Maximum and minimum allowed time step sizes are set to $\frac{5T/4}{2\cdot10^2}$ and $\frac{5T/4}{2\cdot10^5}$, with $T=\omega^{-1}$. Therefore, the full simulation is performed with the lower bound of 200 time steps. Through this section, we will make use of several different meshes, described in Tab.~\ref{tab-meshes}. First, aiming to show the accuracy of high-order N\'ed\'elec FEs, we will employ a very coarse mesh consisting of $12 \times 12 \times 2$ cells on the superconducting domain. Finer meshes are obtained through $r$ isotropic refinements (i.e., dividing every single cell into 8 cells). Thus, these meshes are defined by the original mesh and the levels of refinement $r$ in Tab.~\ref{tab-meshes}. Besides, we will make use of two different meshes for the study of the computational times: a coarser mesh, consisting of $52\times52\times6$ cells in the superconducting region and 28624 cells in the whole computational domain, and a finer mesh consisting of $102\times102\times10$ cells in the superconducting region and 147904 in total. This study is performed for first order N\'ed\'elec FEs. Nonlinear iterations are stopped when the Euclidean norm of the residual is below $10^{-4}$. Further details of the benchmark can be found in \cite{kapolka_3d_2017}.

\begin{table}
\centering
\begin{tabular}{lrrrrr}
\hline
    & \thead{ \#cells \\ on $\Omega_{\rm hts}$} &   &  \thead{ \#DoFs \\ on $\Omega_{\rm hts}$} &  \thead{ \#cells \\ on $\Omega$} & \thead{ \#DoFs \\ on $\Omega$} \\ \hline      
Mesh $1$        &  $12\times12\times2$ &  288      & 1,254    &  1,464    & 6,386    \\
Mesh $1^{r=1}$ &  $24\times24\times4$  & 2,304     & 8,500    &  11,712   & 42,948   \\
Mesh $1^{r=2}$ &  $48\times48\times8$  & 18,432    & 61,544   &  93,696   & 312,008  \\
Mesh 2  &  $52\times52\times6$         & 16,224    & 55,438   &  28,624   &  101,368  \\ 
Mesh 3  &  $102\times102\times10$      & 104,040   & 337,222  &  147,904  &  495,190  \\ \hline 
\end{tabular}
\caption{Summary of meshes used to reproduce the benchmark. The superindex $r$ denotes the number of uniform refinements applied to every cell of the original mesh.}
\label{tab-meshes}
\end{table}

\begin{figure}[t!]
    \centering
    \begin{subfigure}[t]{0.49\textwidth}
        \includegraphics[width=\textwidth]{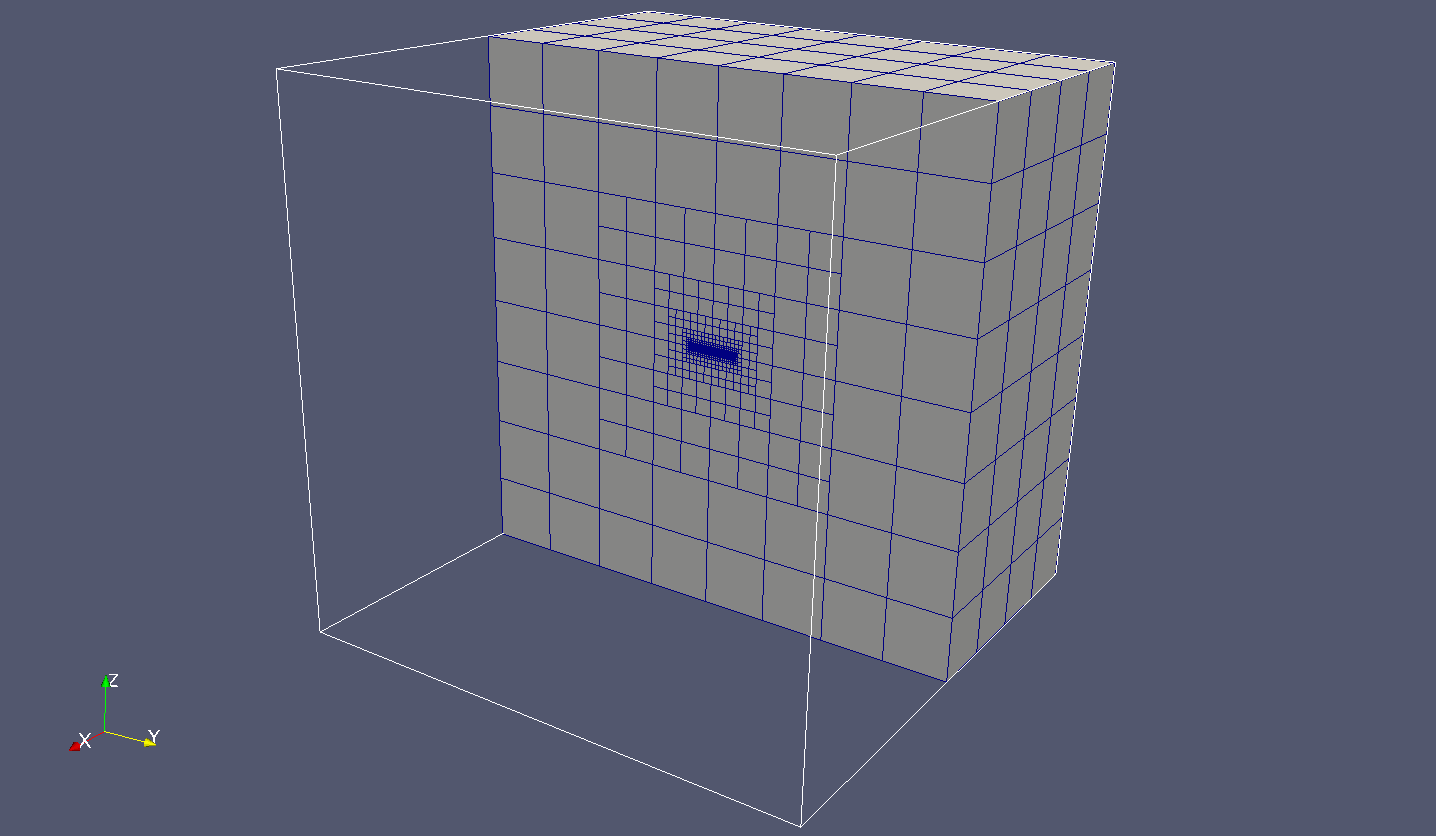}
        \caption{Half domain in cross-section perpendicular to $x$-axis. }
    \end{subfigure}
        \begin{subfigure}[t]{0.49\textwidth}
        \includegraphics[width=\textwidth]{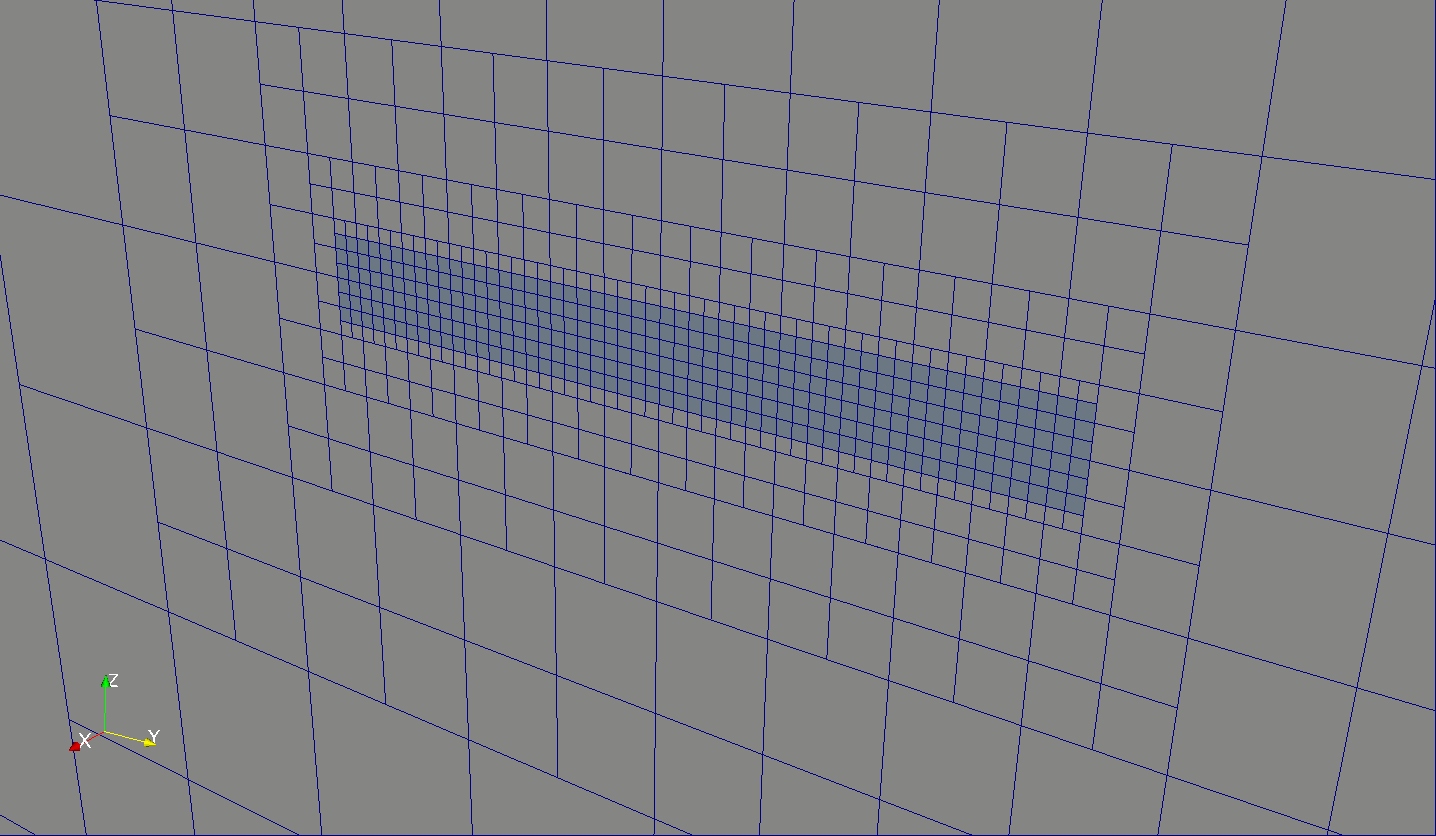}
        \caption{Zoom to centered region. HTS domain depicted in blue. }
    \end{subfigure}
    
        \begin{subfigure}[t]{0.49\textwidth}
        \includegraphics[width=\textwidth]{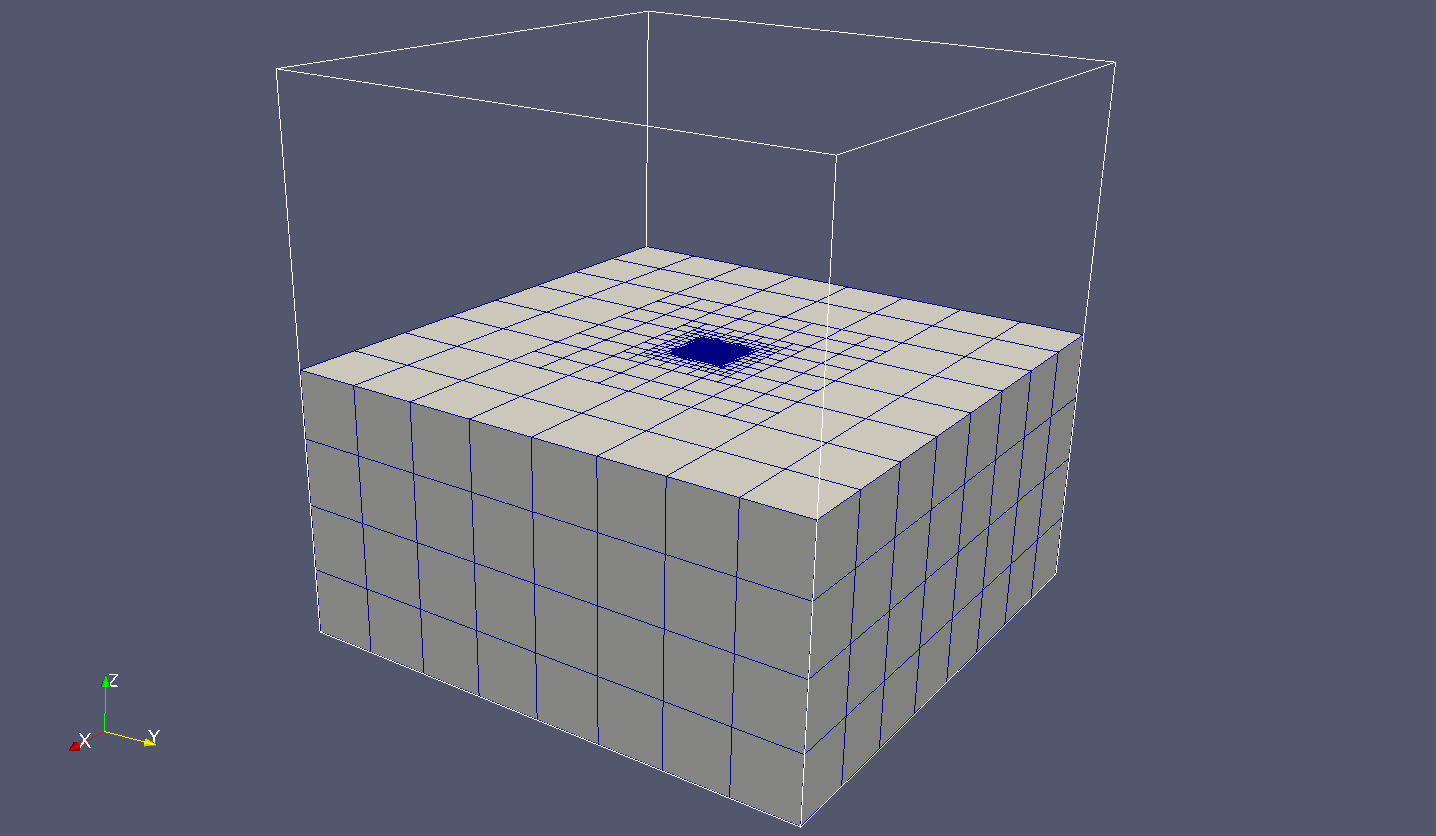}
        \caption{Half domain in cross-section perpendicular to $z$-axis. }
    \end{subfigure}
        \begin{subfigure}[t]{0.49\textwidth}
        \includegraphics[width=\textwidth]{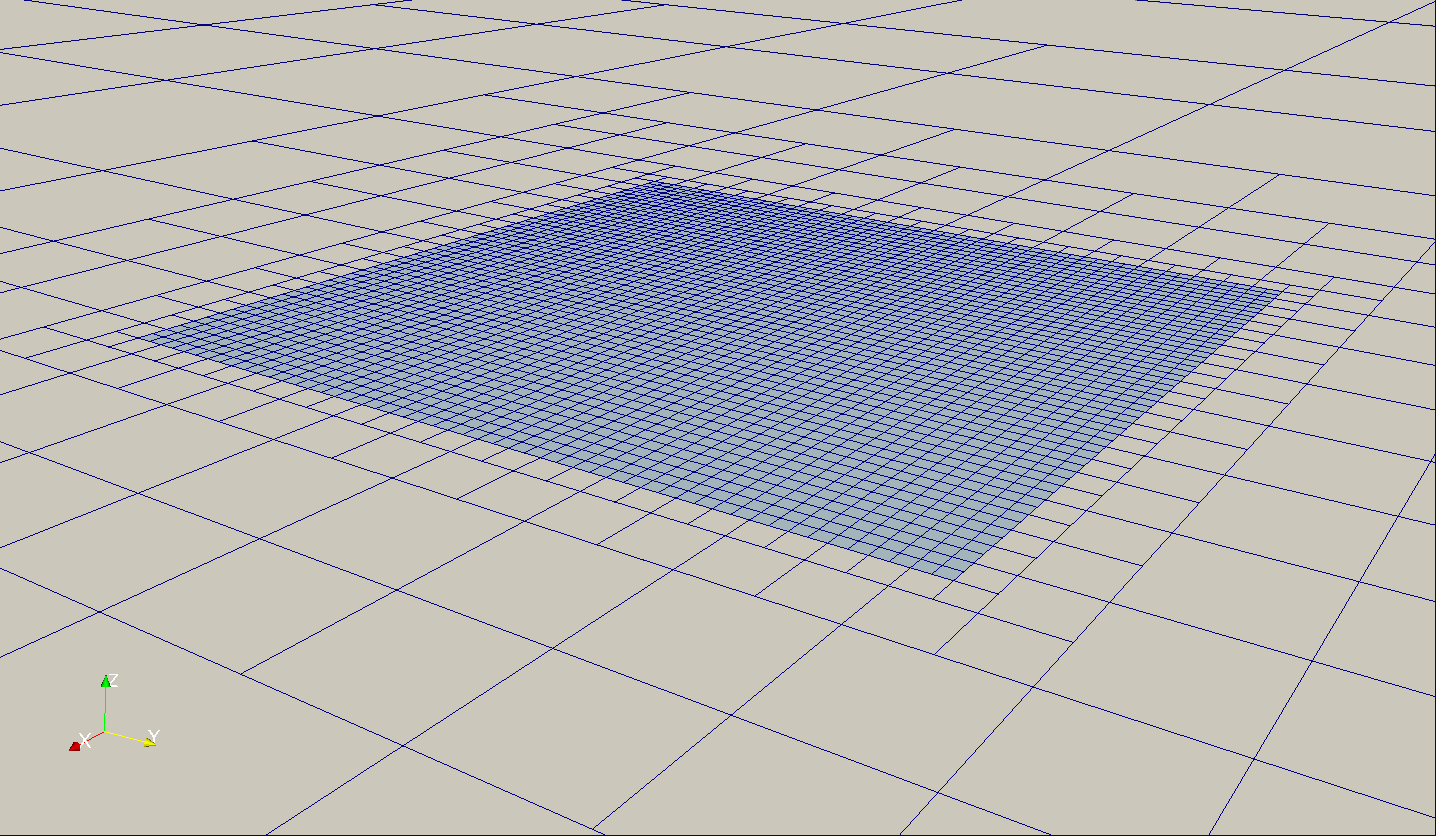}
        \caption{Zoom to centered region. HTS domain depicted in blue. }
    \end{subfigure}
    \caption{Illustration of Mesh 2 (see Tab.~\ref{tab-meshes}). Refinement pattern is common to all meshes. Adaptive refinement technique results in a structured mesh for the HTS device, while a smart coarsening following the 2:1 ratio can be observed in the dielectric domain surrounding the superconducting region. }
    \label{fig-h_adaptive_mesh}
\end{figure}

Fig.~\ref{fig-J_distr} shows the pattern of computed current density distributions within the superconducting device for the bulk and the stack. All the current density plots are taken at $t=5$~ms, when the sinusoidal function reaches its first peak value. On the other hand, Fig.~\ref{fig-magnetization_loops} shows the computed magnetization loops in the bulk and the stack, respectively. Up to three different curves are shown for each magnetization plot: the projection of the magnetization on directions $x$, $z$ and $\alpha$ (i.e., the direction of the applied magnetic field). Shown data is normalized with the magnitude $J_c \cdot b$, with $b$ the length of the side of the base of the parallelepiped. Finally, Fig.~\ref{fig-AC_loss} shows the computed instantaneous power loss in both cases.

\begin{figure}[h!]
    \centering
    \begin{subfigure}[b]{0.49\textwidth}
        \includegraphics[width=\textwidth]{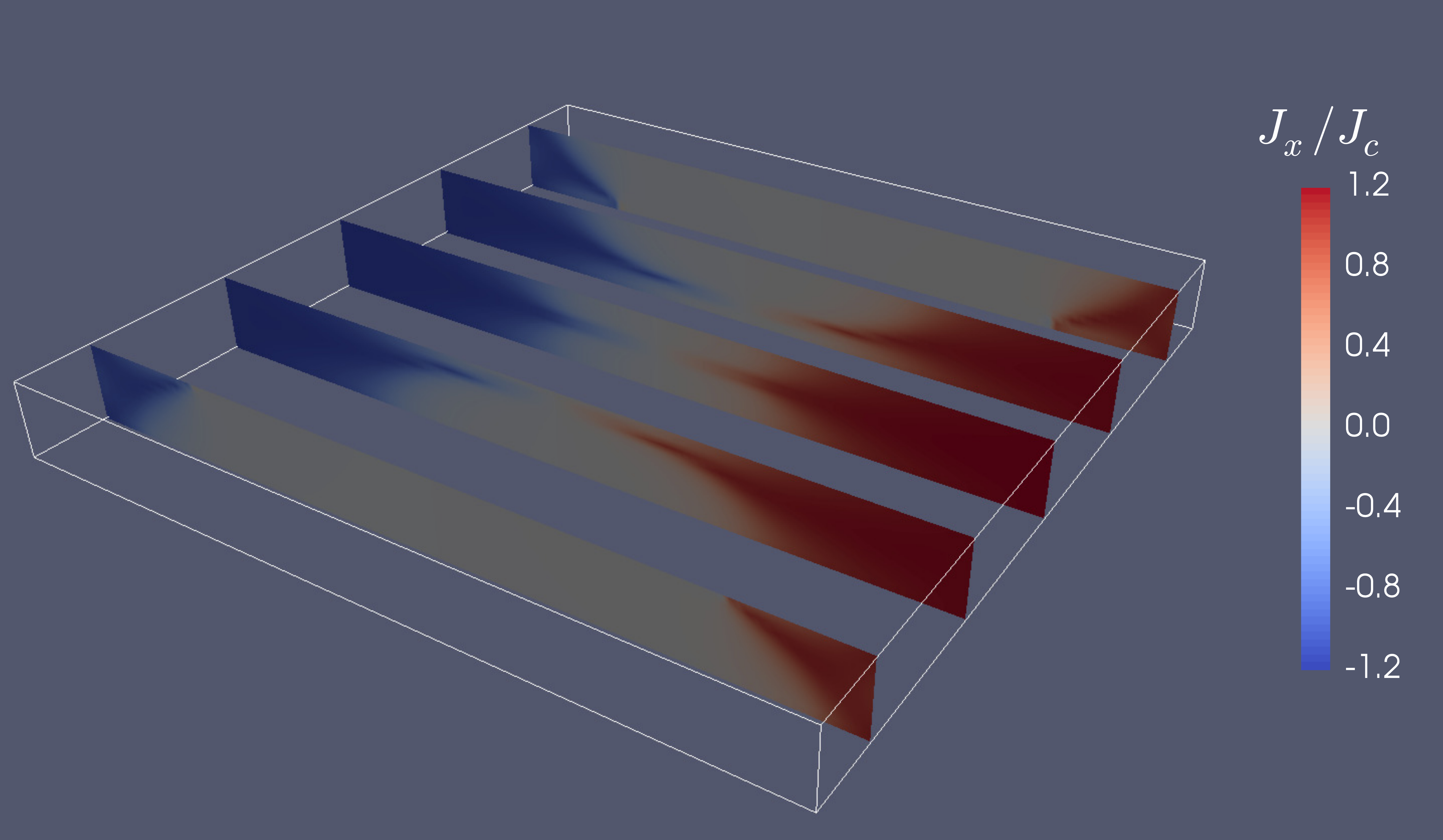}
        \caption{$J_x$ in cross-sections perpendicular to x-axis.}
    \end{subfigure}
        \begin{subfigure}[b]{0.49\textwidth}
        \includegraphics[width=\textwidth]{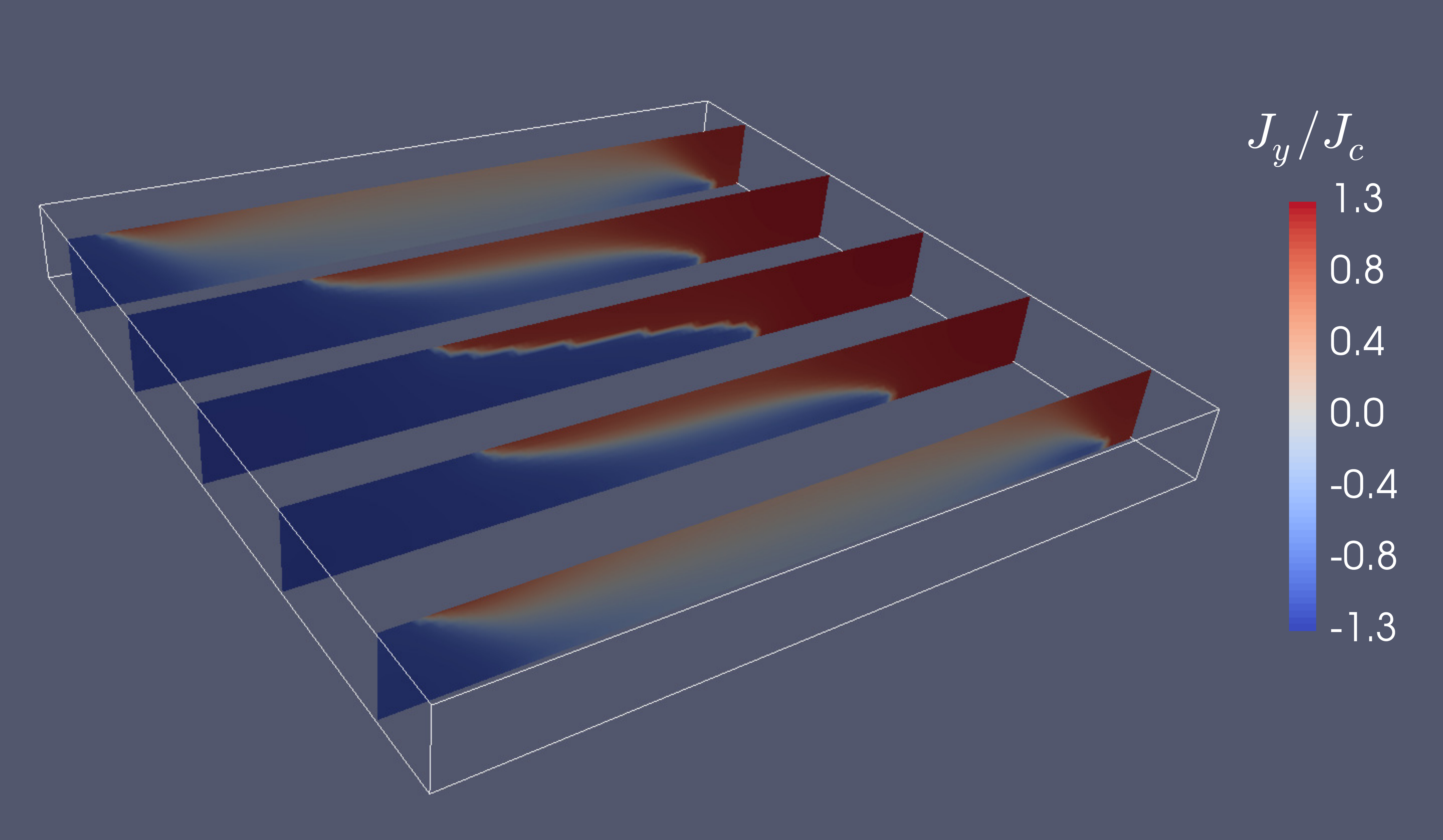}
        \caption{$J_y$ in cross-sections perpendicular to y-axis.}
    \end{subfigure}

    \begin{subfigure}[b]{0.49\textwidth}
        \includegraphics[width=\textwidth]{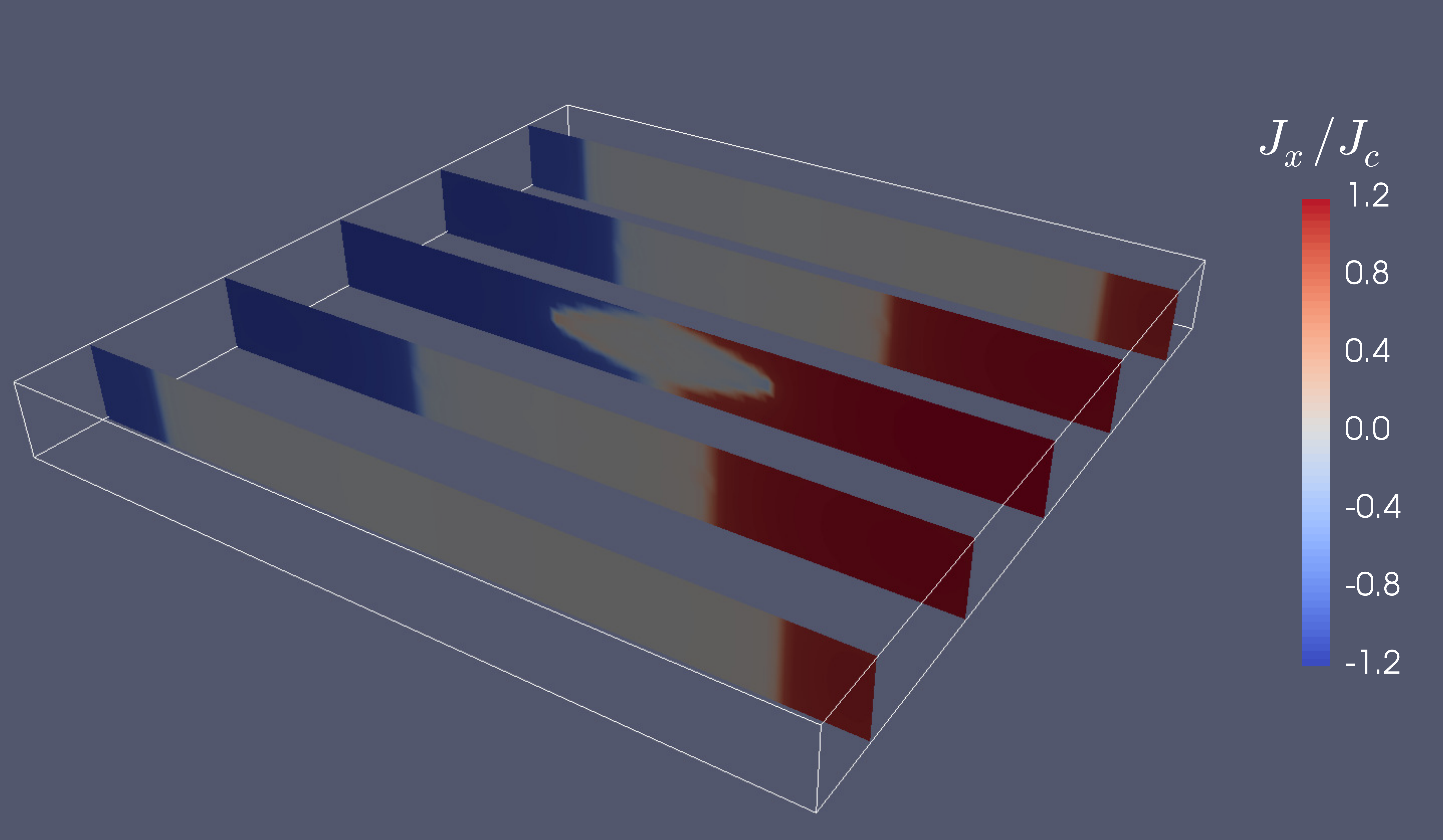}
        \caption{$J_x$ in cross-sections perpendicular to x-axis.}
    \end{subfigure}
    \begin{subfigure}[b]{0.49\textwidth}
        \includegraphics[width=\textwidth]{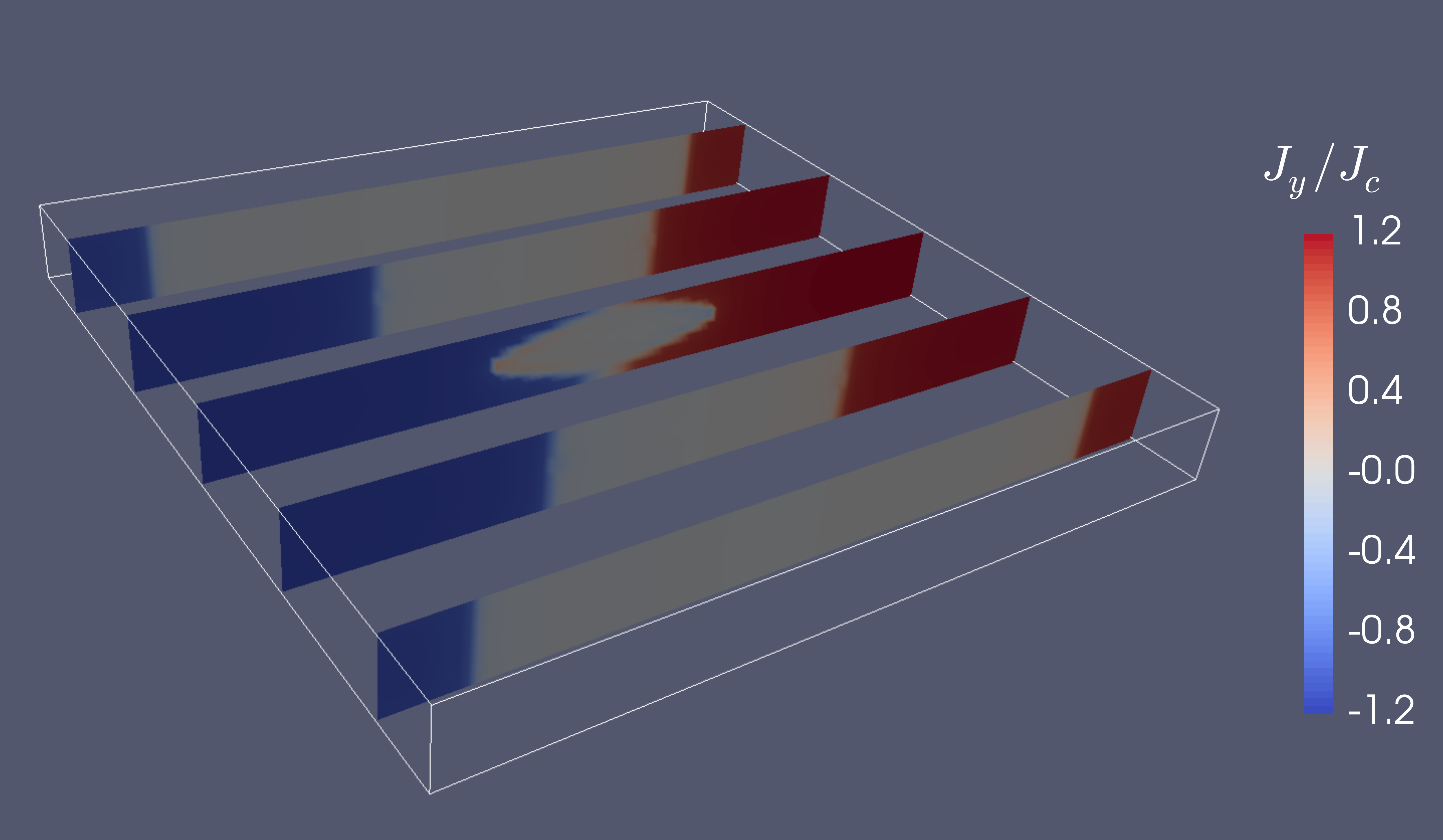}
        \caption{$J_y$ in cross section perpendicular to y-axis.}
    \end{subfigure}
    \caption{Distribution of normalized current densities for the bulk (top) and the stack (bottom). }
    \label{fig-J_distr}
\end{figure}

\begin{figure}[h!]
    \centering
    \begin{subfigure}[b]{0.49\textwidth}
        \includegraphics[width=\textwidth]{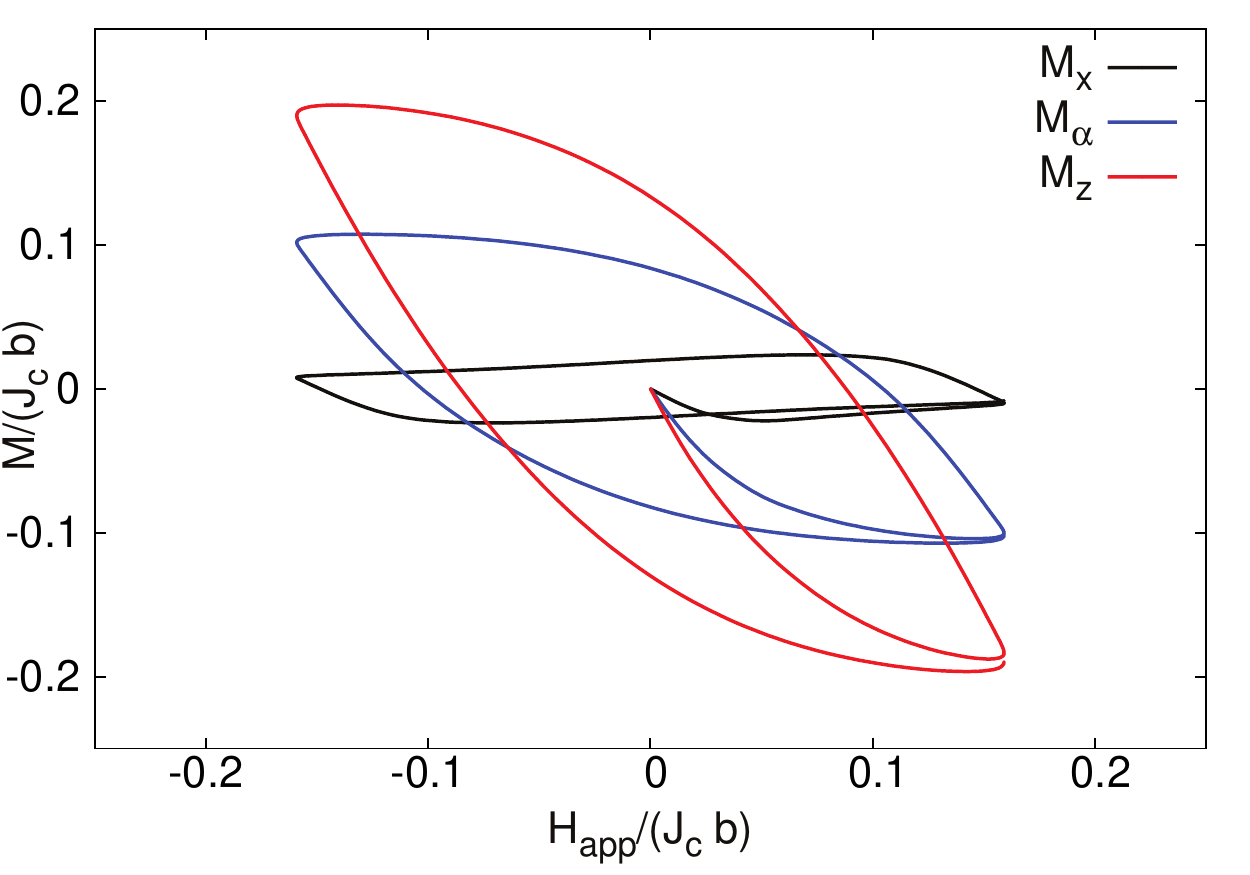}
        \caption{Bulk.}
    \end{subfigure}
        \begin{subfigure}[b]{0.49\textwidth}
        \includegraphics[width=\textwidth]{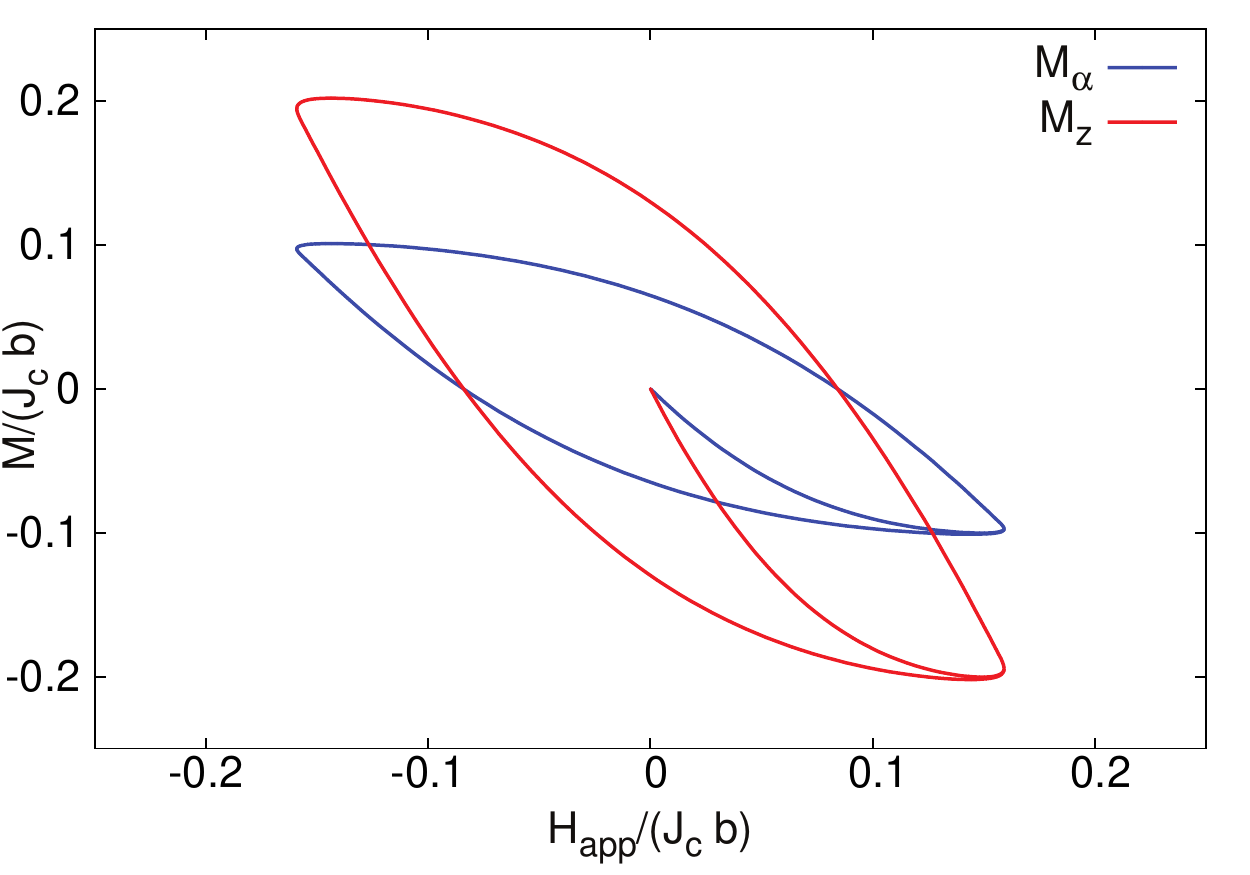}
        \caption{Stack.}
    \end{subfigure}
    \caption{Magnetization loops for the models. Magnitudes are adimensionalized with critical current density and HTS device size.}
    \label{fig-magnetization_loops}
\end{figure}

\begin{figure}[h!]
    \centering
        \includegraphics[width=0.45\textwidth]{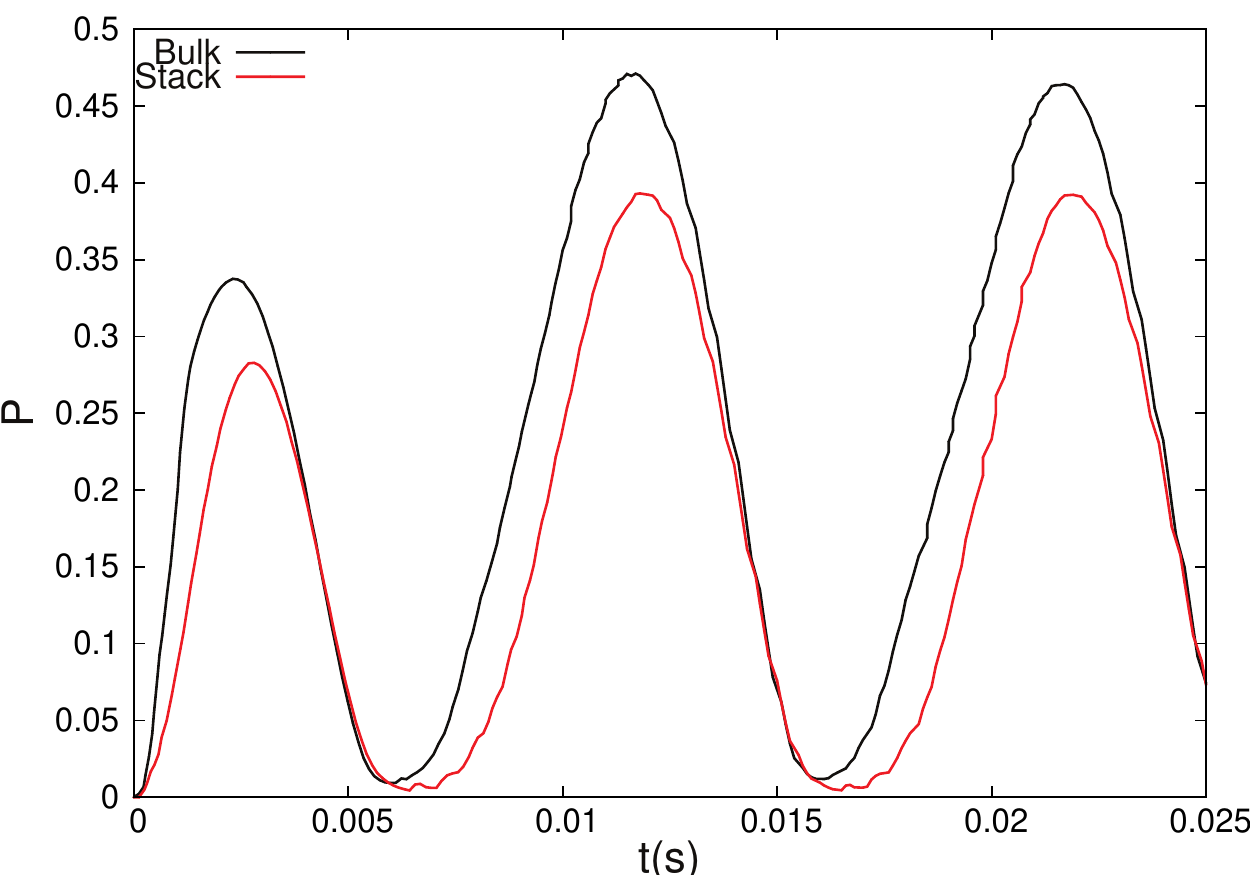}
    \caption{Instantaneous power dissipation in the bulk and the stack.}
    \label{fig-AC_loss}
\end{figure}

The estimation of AC losses is essential to assess the performance of superconducting devices (see \cite{grilli_computation_2013}, where a review of the field up to 2013 is presented). Therefore, any numerical tool modelling HTS devices must estimate AC losses accurately. Tab.~\ref{tab-comparison} shows a direct comparison between computed AC losses from the benchmark and our numerical results, computed with Mesh 2 (see Tab.~\ref{tab-meshes}) and lowest order N\'ed\'elec elements. In particular, presented AC losses are computed via two different methods. The magnitude, for a full cycle, can be calculated by integrating the instantaneous power dissipation $\J\cdot \E$ in the superconductor, i.e., 
\begin{align}\label{eq-Q_je}
Q_{\rm JE} = 2 \int_{\frac{T}{2}}^{T}\int_{{\Omega_{\rm hts}}} \J \cdot \E =  2 \int_{\frac{T}{2}}^{T}\int_{{\Omega_{\rm hts}}} \rho_{\rm hts} {\left\Vert \J \right\Vert}^2,
\end{align}
where the AC loss for half cycle is computed and doubled. Alternatively, the magnitude can be obtained with the magnetization loop in the direction of the applied external field $\H_{\alpha}$ as 
\begin{align}\label{eq-Q_mh}
Q_{\rm MH} = -\mu_0 \oint_{\H_\alpha} M_\alpha \left\Vert \Omega_{\rm hts} \right\Vert
\end{align}
for the full cycle, calculated in the time interval from peak to peak of the applied field. The numerical results are in excellent agreement with those presented in the benchmark, demonstrated by the integral quantities (\ref{eq-Q_je}) and (\ref{eq-Q_mh}), see Tab.~\ref{tab-comparison}. 

\begin{table}
\centering
\begin{tabular}{crrrr}
\hline
           & $Q_{\rm JE}$ bulk &  $Q_{\rm MH}$ bulk  & $Q_{\rm JE}$ stack & $Q_{\rm MH}$ stack \\ \hline 
Reference  &     4.59           &    4.62   &             3.47         & 3.45                      \\ 
Computed   &     4.64           &    4.62   &             3.48         & 3.46                      \\ \hline 
\end{tabular}
\caption{Comparison of AC losses (in mJ) in the bulk and in the stack calculated with two different methods against benchmark results.}
\label{tab-comparison}
\end{table}

In order to show the accuracy of high order FE schemes, we intentionally choose a coarse mesh with $12 \times 12 \times 2$ elements in the HTS domain (Mesh 1). It is well known that better approximations will be found either increasing the order of the elements ($p$-refinement) or reducing the mesh size ($h$-refinement). The effect of applying $r$ consecutive uniform refinements in all cells results in a mesh with $8^r N_c$ elements, where $N_c$ is the number of elements in the coarsest mesh. We recall that $\mathcal{ND}_k$, where the discrete solution $\H$ lies, is the discrete curl-conforming space based on the polynomial space $\mathcal{Q}_{k-1,k,k}\times\mathcal{Q}_{k,k-1,k}\times \mathcal{Q}_{k,k,k-1}$ (see Sect.~\ref{sec-fe_approx}). The current density is obtained as $\J=\nabla \times \H$, thus belonging to $\mathcal{D}_k:=\{\mathcal{Q}_{k,k-1,k-1}\times\mathcal{Q}_{k-1,k,k-1}\times \mathcal{Q}_{k-1,k-1,k}\}$. Fig.~\ref{fig-Jdir_accuracy} depicts how the solution is converging with the order of the elements and a fixed mesh (Mesh 1), whereas Fig.~\ref{fig-Jdir_mesh_accuracy} shows the behaviour of the solution for an uniformly size-refined mesh and lowest order elements; current densities are adimensionalized with the critical current density in all cases. Plots are taken for $t=5$~ms, i.e., first peak value of the applied external field, and over the line $ \{ x=0$, $z=0 \}$. The bottom axis represents the size of the parallelepiped, its center being located at $y=5$~mm in the plots. Out of this length, the adimensionalized current density is negligible since it is out of the limits of the superconducting domain. Note that current density profiles are in general discontinuous across elements. Furthermore, as $\J \in \mathcal{D}_k$, its $i$-th component, for $i\in\{1,2,3\}$, is a polynomial of degree $k$ with respect to $x_i$, and $k-1$ otherwise.

Convergence to a solution  is shown with respect to $p$-refinement  (Fig.~\ref{fig-Jdir_accuracy}) and $h$-refinement (Fig.~\ref{fig-Jdir_mesh_accuracy}). It is clear from the plots that, for coarse meshes, the current density is much better captured with high order elements. Fig.~\ref{fig-Jacc_Q2_r1} shows a direct comparison between quadratic FEs in the coarsest mesh and linear FEs in the mesh after one level of refinement, i.e., $r=1$. Note that both discretizations involve the same number of DoFs. More accurate solutions, i.e., closer to the solution for the most accurate simulation in Fig.~\ref{fig-Jacc_Q3_r2}, are obtained for quadratic FEs. It is expected, since, for smooth solutions, which is the case for the current profiles, $p$-refinement leads to exponential convergence rates.

Fig.~\ref{fig-Jacc_Q3_r2} shows a comparison between the coarsest mesh with third order FEs and first order elements in the mesh after two levels of refinement. The simulation with third order FEs involves a lower number of DoFs (136,038 vs 312,008). Again, $p$-refinement is more effective than $h$-refinement, achieving better results with a substantially lower number of DoFs. Summarizing, $p$ and $h$-refinement converge to the same solution, but $p$-refinement is more effective for the smooth solutions at hand (see Figs. \ref{fig-Jacc_Q2_r1} and \ref{fig-Jacc_Q3_r2}).

\begin{figure}[t!]
    \centering
    \begin{subfigure}[t]{0.32\textwidth}
        \includegraphics[width=\textwidth]{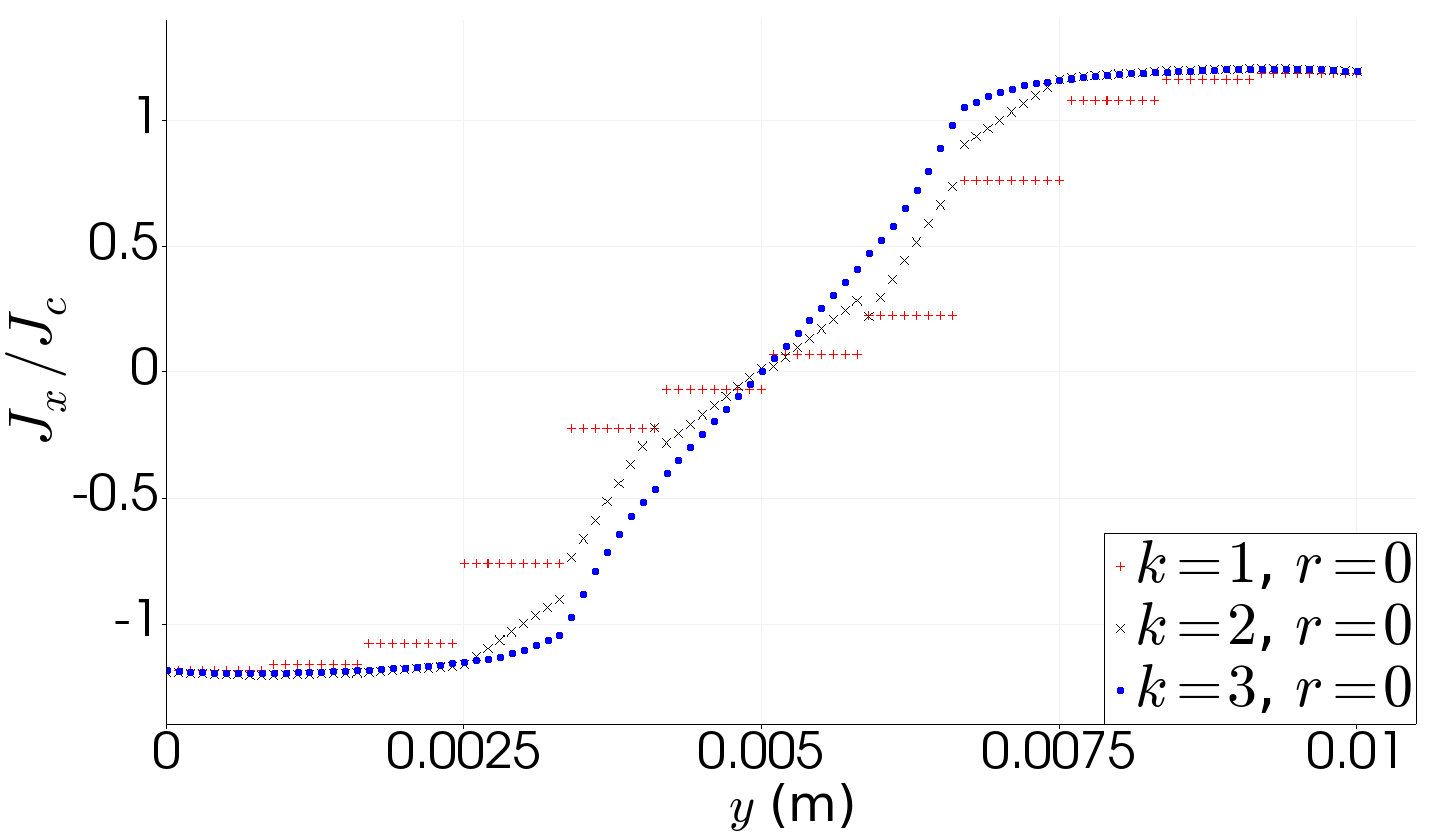}
        \caption{$J_x$ profiles.}
        \label{fig-Jx_order}
    \end{subfigure}
            \begin{subfigure}[t]{0.32\textwidth}
        \includegraphics[width=\textwidth]{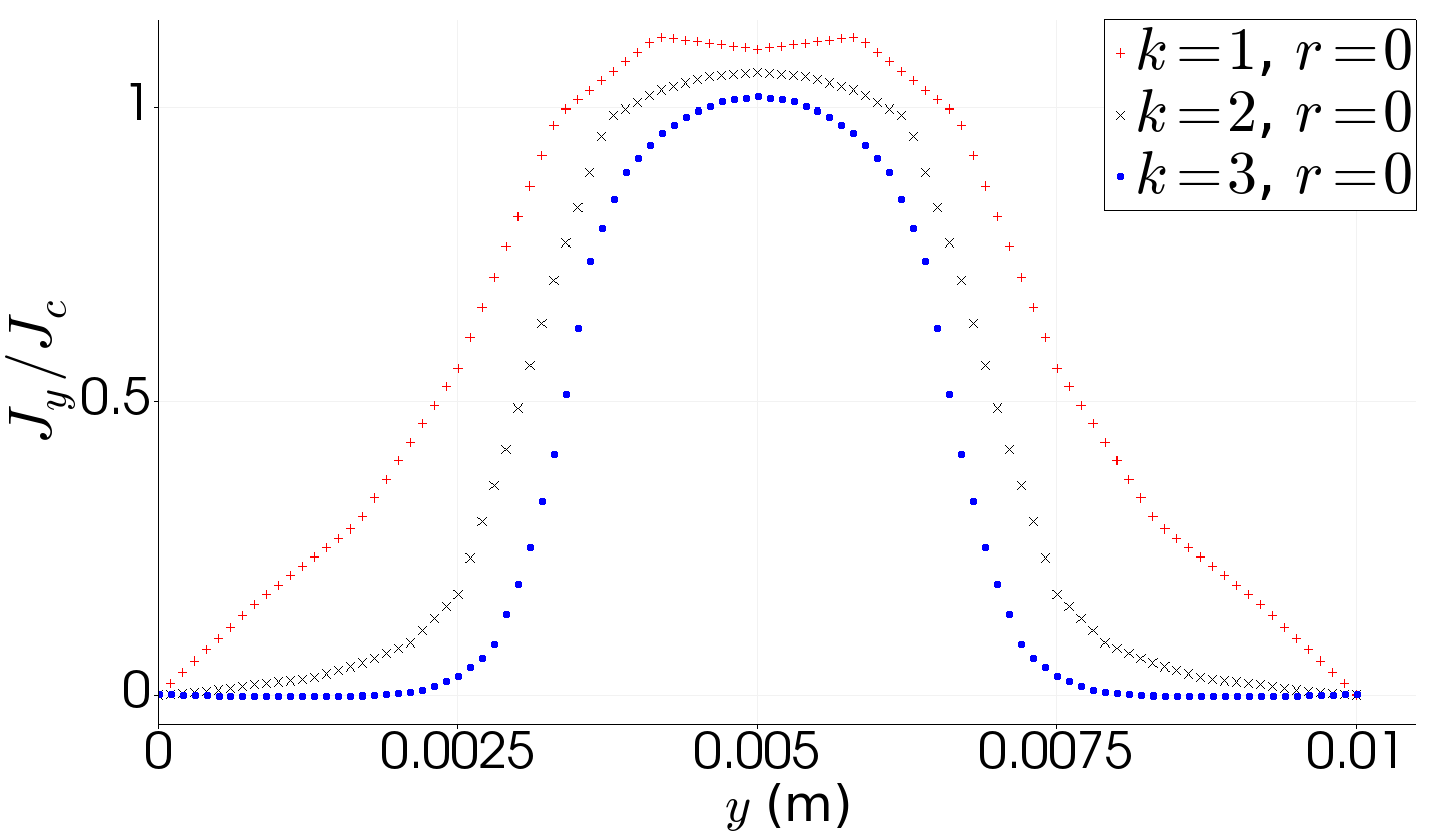}
        \caption{$J_y$ profiles.}
        \label{fig-Jy_order} 
    \end{subfigure}
        \begin{subfigure}[t]{0.32\textwidth}
        \includegraphics[width=\textwidth]{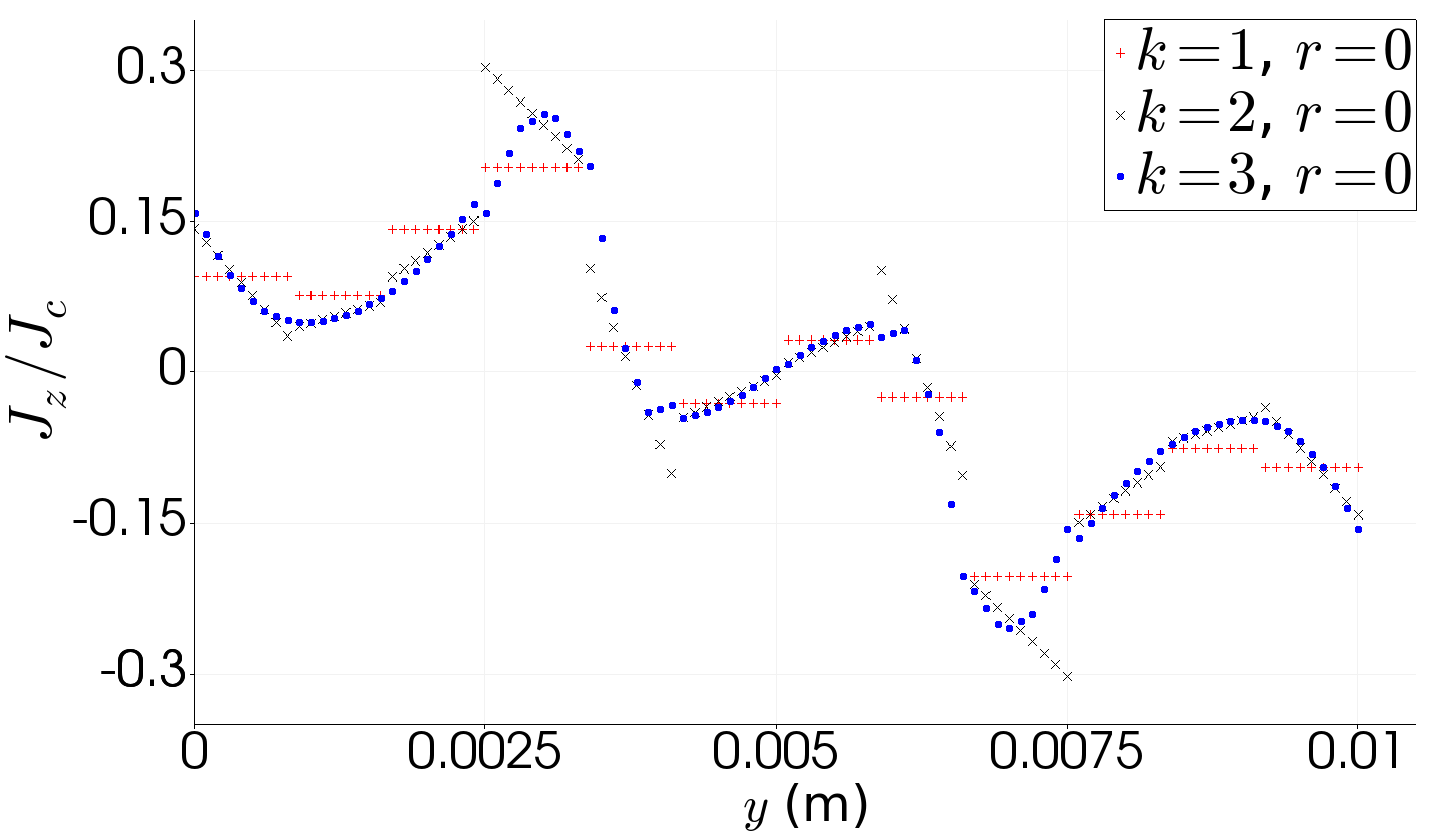}
        \caption{$J_z$ profiles.}
        \label{fig-Jz_order} 
    \end{subfigure}
    \caption{Adimensionalized $\J$ components over a line in the y-axis direction that passes through $z=0$ with a fixed mesh and different element orders.}
    \label{fig-Jdir_accuracy}
\end{figure}

\begin{figure}[t!]
    \centering
    \begin{subfigure}[t]{0.32\textwidth}
        \includegraphics[width=\textwidth]{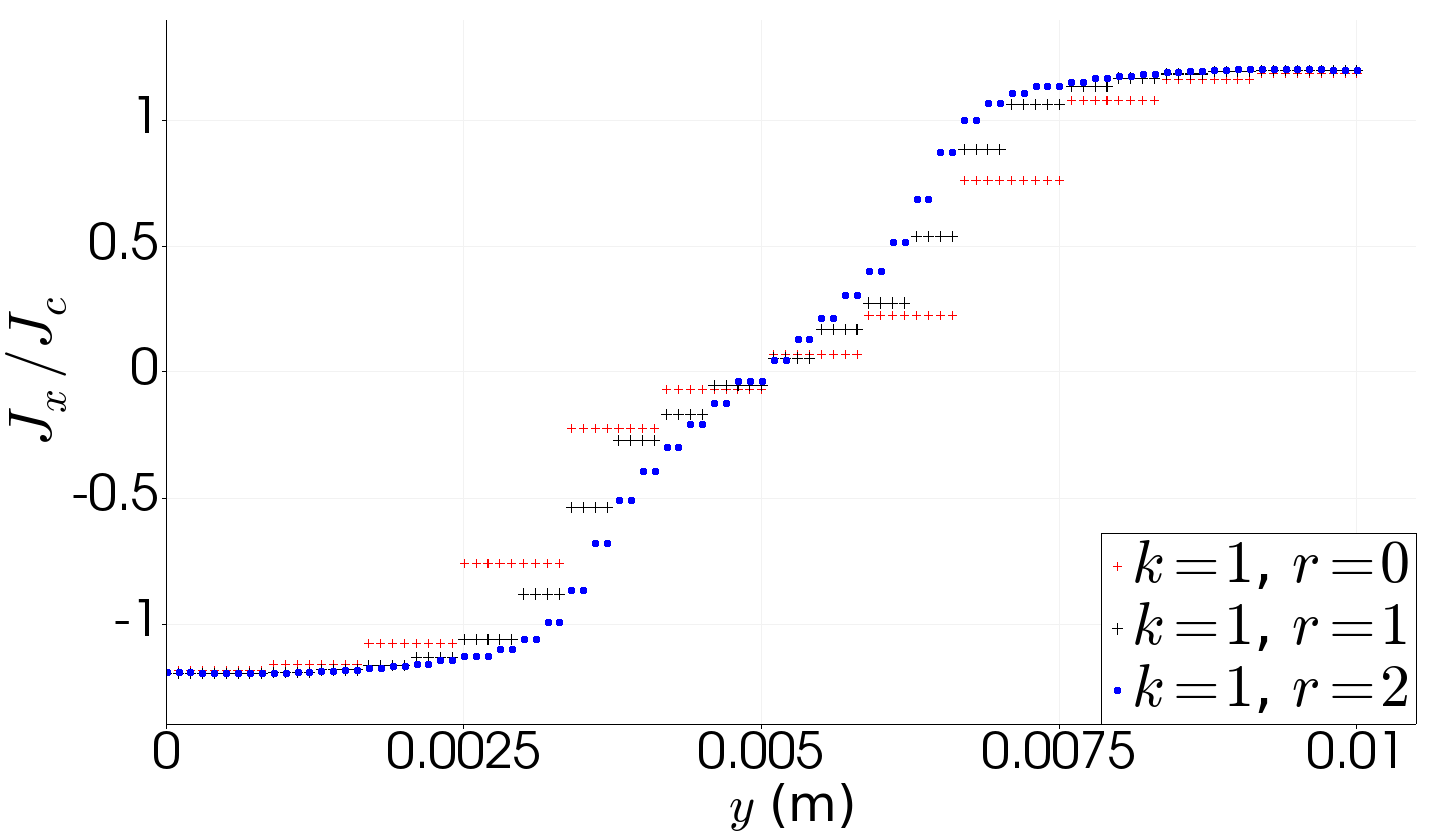}
        \caption{$J_x$ profiles.}
        \label{fig-Jx_mesh}
    \end{subfigure}
            \begin{subfigure}[t]{0.32\textwidth}
        \includegraphics[width=\textwidth]{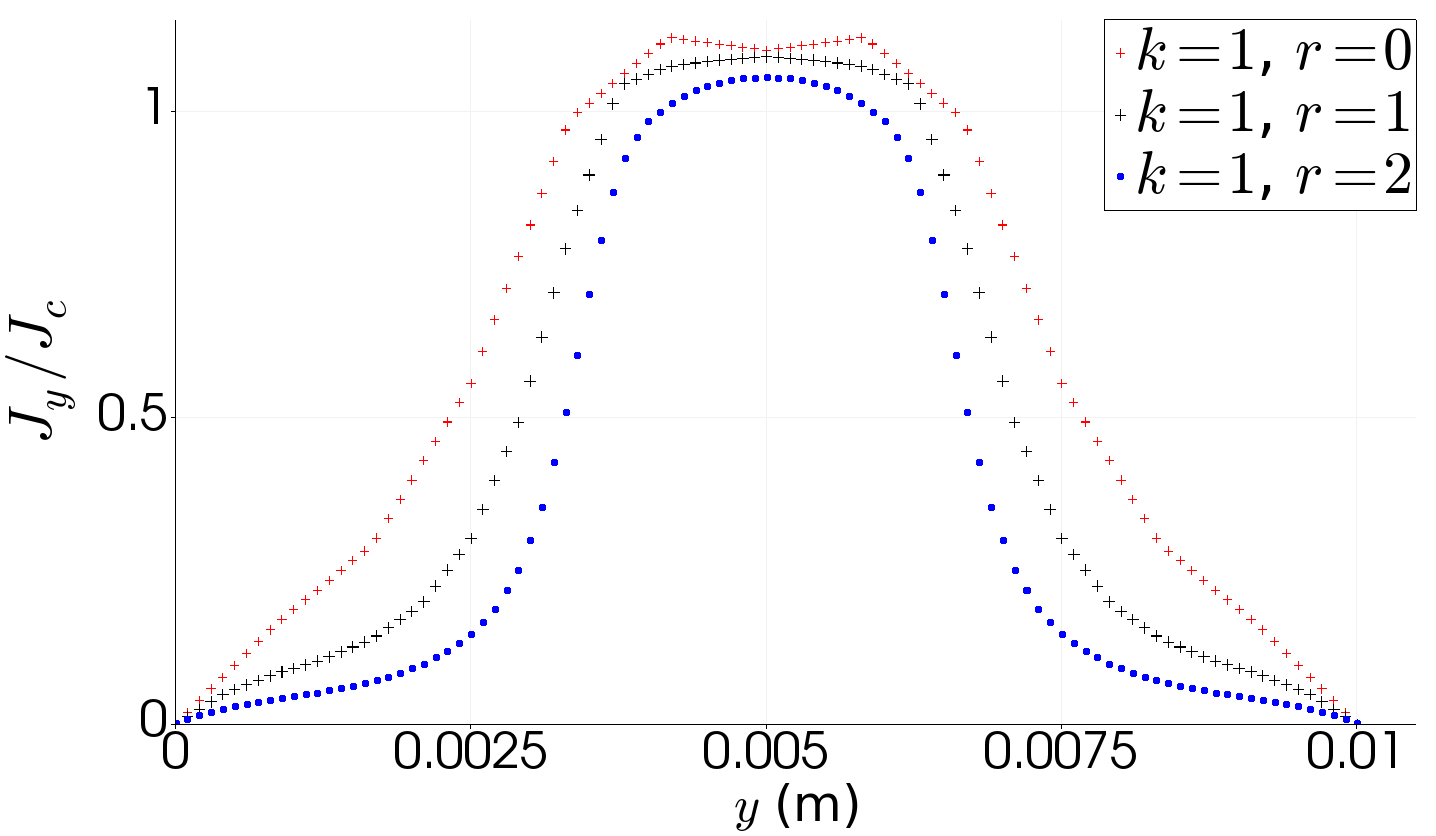}
        \caption{$J_y$ profiles.}
        \label{fig-Jy_mesh} 
    \end{subfigure}
        \begin{subfigure}[t]{0.32\textwidth}
        \includegraphics[width=\textwidth]{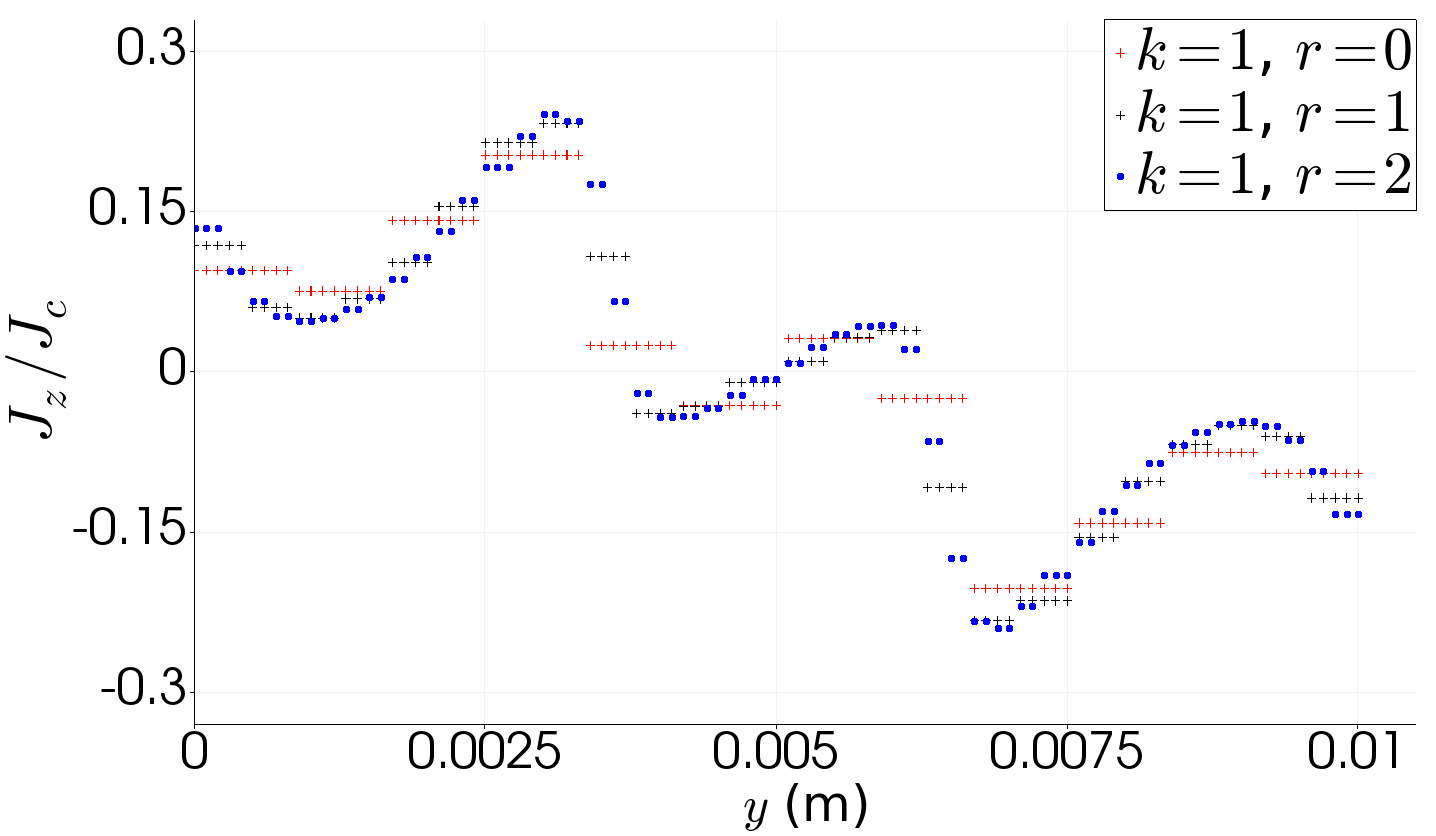}
        \caption{$J_z$ profiles.}
        \label{fig-Jz_mesh} 
    \end{subfigure}
    \caption{Adimensionalized $\J$ components over a line in the y-axis direction that passes through $z=0$ with a fixed element order and different uniformly refined meshes.}
    \label{fig-Jdir_mesh_accuracy}
\end{figure}

\begin{figure}[t!]
    \centering
    \begin{subfigure}[t]{0.49\textwidth}
        \includegraphics[width=\textwidth]{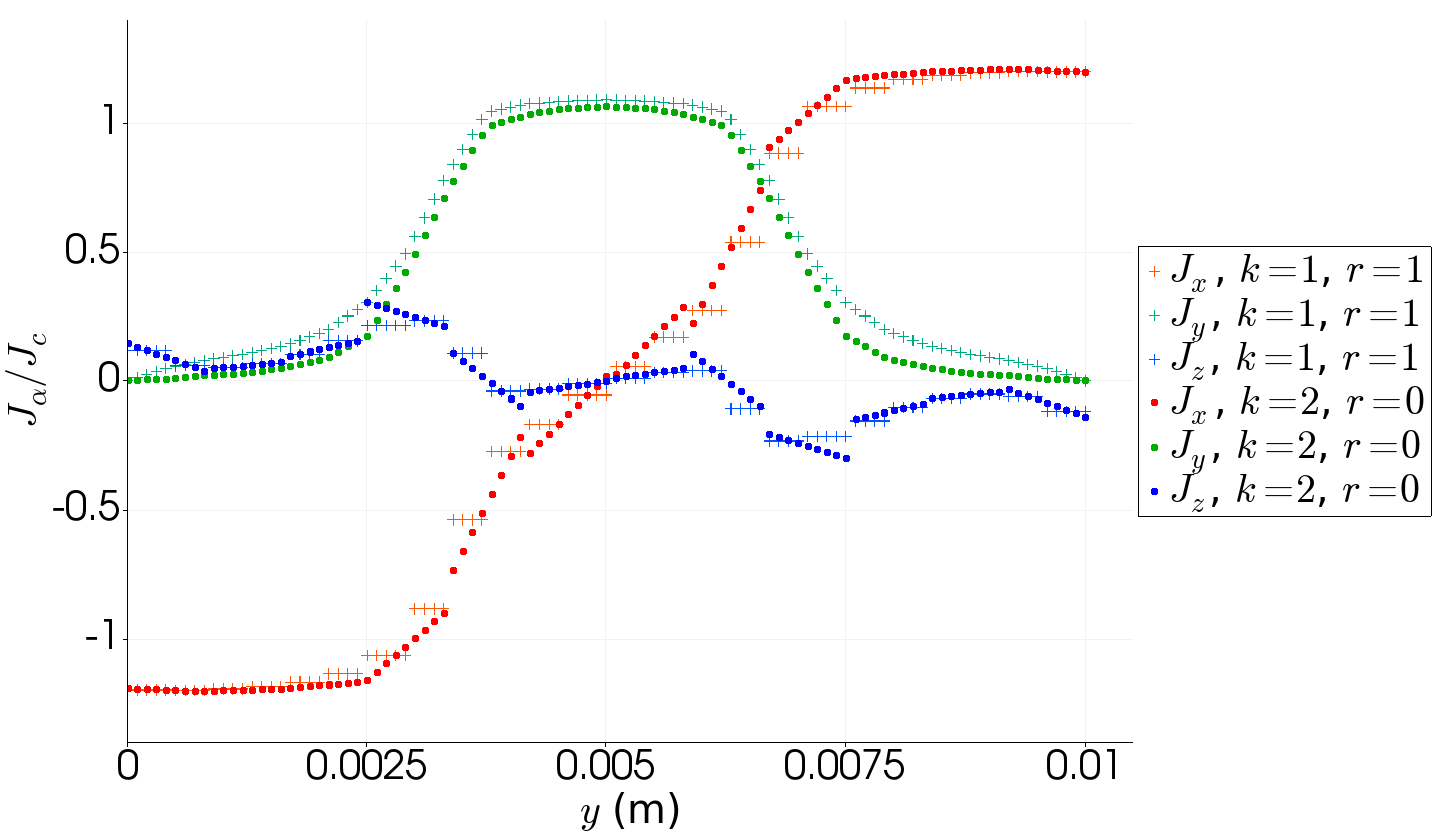}
        \caption{Comparison with same \#DoFs.}
        \label{fig-Jacc_Q2_r1}
    \end{subfigure}
        \begin{subfigure}[t]{0.49\textwidth}
        \includegraphics[width=\textwidth]{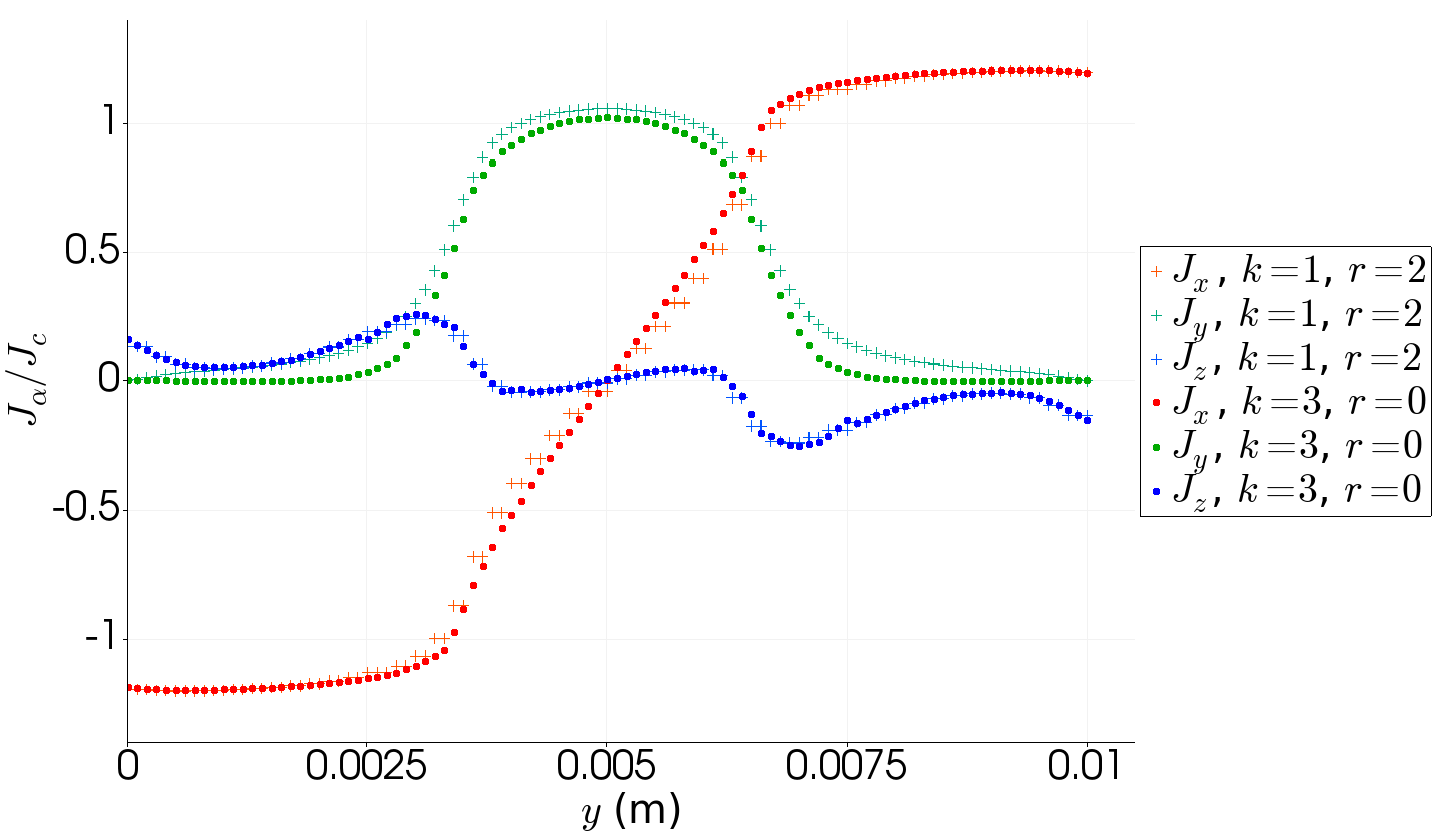}
        \caption{Higher order contains less \#DoFs.}
        \label{fig-Jacc_Q3_r2}
    \end{subfigure}
    \caption{$\J$ components over a line  in the y-axis direction  that passes through $z=0$ for different element orders and meshes.}
    \label{fig-J_comparison}
\end{figure}

Tabs.~\ref{tab-computing_times_mesh1} and \ref{tab-computing_times_mesh2} show total parallel execution times for the 3D benchmark with Meshes 2 and 3, respectively (see Tab.~\ref{tab-meshes}). This total time includes the time spent in \emph{every} single step of the simulation pipeline, including mesh generation. The BDDC-preconditioned CG iterative solver (see Sect.~\ref{subsec:parallel_solver}) was set up such that the coarse-grid problem is mapped and solved on a single node of the MN-IV supercomputer by means of Intel MKL PARDISO on 48 threads (cores). Hence, the simulations have actually been run in 48 + $P$ cores, $P$ being the number of subdomains in which the global domain is partitioned.  These tables show two key parameters for measuring strong scalability: the parallel speed-up, defined as the ratio between the parallel execution time $t_P$ when using $P$ processors and the sequential execution time $t_1$, i.e.,  $S_p=t_{P}/t_{1}$, and the parallel efficiency, defined as $E_p=S_p/p$. (Thus, the closer $S_p$ and $E_p$ are to $P$ and $1$, resp., the better.)  Note that the nonlinear convergence rate for a given problem does not depend on the linear solver being used (up to linear solver tolerance). Therefore, all the runs that involve the same mesh share the same the nonlinear convergence history (hence time stepping). The difference is found in the solution of the linearized problems, since the performance of the preconditioner (Sect. \ref{subsec:parallel_solver}) does depend on the partition being used. Aiming to show the impact of the preconditioner for each partition, Tabs.~\ref{tab-computing_times_mesh1} and \ref{tab-computing_times_mesh2} show the number of preconditioned Krylov iterations needed to attain convergence; the presented number of iterations is an arithmetic mean value of the number of iterations until convergence of all linearized problems taking place during the simulation, i.e., for every time step and nonlinear iteration.

\begin{table}
\centering
\begin{tabular}{crrrrrc}
\hline
\bf{P}   & \thead{ \bf{Wall clock} \\ \bf{time}}      &   $\mathbf{S_p}$   & $\mathbf{E_p}$ & \thead{ \bf{\#DoFs} \\ \bf{per part} } & \thead{ \bf{\#DoFs} \\ \bf{coarse} \\\bf{solver} } &   \thead{\bf{\#Linear solver} \\ \bf{iterations}} \\ \hline 
Serial run                 &        1d 19h 23'         &    1.00      &  1.00 &  101,428 & -    &  -  \\
6                          &            8h 29'         &    5.23      &  0.87 &  19,091 & 286  & 22 \\
12                         &            4h 28'         &    9.95      &  0.82 &  9,854  & 520  & 29 \\
24                         &            2h 25'         &    18.26     &  0.76 &  5,143  & 990  & 30 \\
48                         &            1h 32'         &    28.81     &  0.60 &  2,775  & 1,852 & 34 \\
72                         &            1h 14'         &    35.79     &  0.49 &  1,928  & 2,507 & 35 \\
96                         &            1h 47'         &    22.87     &  0.23 &  1,489  & 3,037 & 42 \\ \hline
\end{tabular}
\caption{Computing times for problem solved using Mesh 2. Iteration and free DoF counters show average values. Simulation ends after 246 converged time steps involving 976 linearized problem solves.}
\label{tab-computing_times_mesh1}
\end{table}

\begin{table}
\centering
\begin{tabular}{crrrrrc}
\hline
\bf{P}   & \thead{ \bf{Wall clock} \\ \bf{time}}      &  $\mathbf{S_p}$    &   $\mathbf{E_p}$  & \thead{ \bf{\#DoFs} \\ \bf{per part} } & \thead{ \bf{\#DoFs} \\ \bf{coarse} \\\bf{solver} } &   \thead{\bf{\#Linear solver} \\ \bf{iterations}} \\ \hline 
Serial run*                &        23d 06h 14'    &    1.00      &  1.00   & 495,190 & -    & -    \\
6                          &         4d 22h 30'    &    4.71      &  0.78   & 88,807  & 474  & 28   \\
12                         &         2d 08h 23'    &    9.90      &  0.82   & 45,043  & 834  & 29   \\
24                         &         1d 08h 31'    &    17.18     &  0.71   & 23,001  & 1,550 & 29   \\
48                         &           17h 16'     &    32.64     &  0.68   & 11,938  & 2,846 & 34   \\
96                         &           12h 05'     &    46.23     &  0.48   & 6,214   & 4,924 & 42   \\
120                        &           10h 46'     &    51.86     &  0.43   & 5,087   & 6,253 & 44   \\
144                        &           10h 38'     &    52.53     &  0.36   & 4,316   & 7,446 & 47   \\ \hline
\end{tabular} 
\caption{Computing times for problem solved using Mesh 3. Iteration and free DoF counters show average values. Simulation ends after 459 converged time steps involving 2374 linearized problem solves. *Serial run time is computed with an extrapolation with the number of linearized problem solves after 3 days of computation due to limited computing time in the access to MN-IV.}
\label{tab-computing_times_mesh2}
\end{table}

Let us comment on the results in Tabs.~\ref{tab-computing_times_mesh1} and \ref{tab-computing_times_mesh2}. Clearly, the most salient property of the algorithms at hand is the remarkable reduction in time-to-solution in both cases. By exploiting parallel resources, the computational time is reduced by a factor of 35.8 and 52.5 for Mesh 2 (Tab.~\ref{tab-computing_times_mesh1}) and Mesh 3 (Tab.~\ref{tab-computing_times_mesh2}), respectively. As far as we know, such speed-ups have not been presented so far for FE HTS modelling. In practice, these speed-ups allow us to reduce the time-to-solution for a practical HTS simulation from days to hours. For the largest problem size, time-to-solution is reduced up to 144 parts. Above this core count, time-to-solution increases due to parallelism related overheads; more computationally intensive simulations (i.e., larger loads per processor) would be required to exploit additional computational resources efficiently. Although average iteration counts increase mildly with the number of processors, the size of the coarse problem keeps growing while the local problems become smaller. Therefore, at some point, the coarse solver, which only exploits a bounded number of cores (i.e., 48), dominates computing times, loosing parallel efficiency. Fortunately, there is a large room for improvement in this direction, e.g., a multilevel version of the preconditioner is expected to push forward the limits of the presented strong scalability results (see, e.g., \cite{badia_multilevel_2016}).

\section{Conclusions}\label{sec-conclusions}

In this article, we present a parallel, fully-distributed FE framework suitable for the solution of nonlinear problems modelling the electromagnetic behaviour of HTS devices. We have selected the widespread $H$-formulation as a demonstrator of the potential of the presented advanced numerical algorithms, which have been tailored for the problem at hand and combined through the simulation pipeline, even though the ingredients presented in this work could be applied to other formulations. For the mesh generation, we have considered an advanced AMR technique, which provides an aggressive coarsening in dielectric regions. The smart coarsening is restricted by the 2:1 balance, which allows for efficient parallel implementations. For the FE approximation, we choose edge (or N\'ed\'elec) elements of arbitrary order. These elements are favoured in electromagnetics simulations due to their sound mathematical structure. For the solution of the arising algebraic systems, we have presented the design of a tailored nonlinear parallel solver, which includes a linearization with a Newton-Raphson method (with exact Jacobian derivation) and advanced domain decomposition preconditioners for $\mathcal{H}$(\textbf{curl}) spaces on adaptive meshes and heterogeneous problems. Time integration was performed with the Backward Euler integrator and a variable time step, taking advantage of the convergence history of the nonlinear solver, and thus reducing time-to-solution. Finally, we have provided a detailed set of numerical experiments. First, a comparison with experimental data has shown an excellent agreement between experimental and numerical data. Second, a time-to-solution study reproducing a 3D benchmark has shown a remarkable reduction of computing times when exploiting parallel resources, and thus the capability of our algorithms to efficiently exploit HPC platforms. The work here presented has been implemented in the open source simulation software \FEMPAR{}, which can become a powerful tool for the HTS modelling community.
\begin{small}

\bibliographystyle{myabbrvnat}
\bibliography{art024}

\begin{thebibliography}{64}
\providecommand{\natexlab}[1]{#1}
\providecommand{\url}[1]{\texttt{#1}}
\expandafter\ifx\csname urlstyle\endcsname\relax
  \providecommand{\doi}[1]{doi: #1}\else
  \providecommand{\doi}{doi: \begingroup \urlstyle{rm}\Url}\fi

\bibitem[Bíró(1999)]{BIRO1999391}
O.~Bíró.
\newblock Edge element formulations of eddy current problems.
\newblock \emph{Computer Methods in Applied Mechanics and Engineering},
  169\penalty0 (3):\penalty0 391 -- 405, 1999.

\bibitem[Grilli(2004)]{grilli_numerical_2004}
F.~Grilli.
\newblock \emph{Numerical modelling of high temperature superconducting tapes
  and cables}.
\newblock PhD thesis, École Polytechnique Fédérale de Lausanne, Lausanne,
  2004.

\bibitem[Lousberg et~al.(2009)Lousberg, Ausloos, Geuzaine, Dular, Vanderbemden,
  and Vanderheyden]{lousberg_numerical_2009}
G.~P. Lousberg, M.~Ausloos, C.~Geuzaine, P.~Dular, P.~Vanderbemden, and
  B.~Vanderheyden.
\newblock Numerical simulation of the magnetization of high-temperature
  superconductors: a 3{D} finite element method using a single time-step
  iteration.
\newblock \emph{Superconductor Science and Technology}, 22\penalty0
  (5):\penalty0 055005, 2009.

\bibitem[Campbell(2009)]{campbell_direct_2009}
A.~Campbell.
\newblock A direct method for obtaining the critical state in two and three
  dimensions.
\newblock \emph{Superconductor Science and Technology}, 22\penalty0
  (3):\penalty0 034005, 2009.

\bibitem[Grilli et~al.(2005)Grilli, Stavrev, Floch, Costa-Bouzo, Vinot,
  Klutsch, Meunier, Tixador, and Dutoit]{grilli_finite_2005}
F.~Grilli, S.~Stavrev, Y.~L. Floch, M.~Costa-Bouzo, E.~Vinot, I.~Klutsch,
  G.~Meunier, P.~Tixador, and B.~Dutoit.
\newblock Finite-element method modeling of superconductors: from 2-d to 3-d.
\newblock \emph{IEEE Transactions on Applied Superconductivity}, 15\penalty0
  (1):\penalty0 17--25, 2005.

\bibitem[Amemiya et~al.(2006)Amemiya, Sato, and Ito]{amemiya_magnetic_2006}
N.~Amemiya, S.~Sato, and T.~Ito.
\newblock Magnetic flux penetration into twisted multifilamentary coated
  superconductors subjected to ac transverse magnetic fields.
\newblock \emph{Journal of Applied Physics}, 100\penalty0 (12):\penalty0 3907,
  2006.

\bibitem[Stenvall and Tarhasaari(2010)]{stenvall_programming_2010}
A.~Stenvall and T.~Tarhasaari.
\newblock Programming finite element method based hysteresis loss computation
  software using non-linear superconductor resistivity and {$T-\varphi$}
  formulation.
\newblock \emph{Superconductor Science and Technology}, 23\penalty0
  (7):\penalty0 075010, 2010.

\bibitem[Pecher et~al.(2003)Pecher, McCulloch, Chapman, Prigozhin, and
  Elliott]{pecher_3d-modelling_2003}
R.~Pecher, M.~McCulloch, S.~Chapman, C.~Prigozhin, and L.~Elliott.
\newblock 3{D}-modelling of bulk type-{II} superconductors using unconstrained
  {H}-formulation.
\newblock In \emph{{Proceedings of the 6th European Conference on Applied
  Superconductivity}, EUCAS}, volume 181, 2003.

\bibitem[Escamez et~al.(2016)Escamez, Sirois, Lahtinen, Stenvall, Badel,
  Tixador, Ramdane, Meunier, Perrin-Bit, and Bruzek]{7422024}
G.~Escamez, F.~Sirois, V.~Lahtinen, A.~Stenvall, A.~Badel, P.~Tixador,
  B.~Ramdane, G.~Meunier, R.~Perrin-Bit, and C.~{\'E}. Bruzek.
\newblock 3-{D} {N}umerical modeling of {AC} losses in multifilamentary
  {M}g{B}2 wires.
\newblock \emph{IEEE Transactions on Applied Superconductivity}, 26\penalty0
  (3):\penalty0 1--7, 2016.

\bibitem[Grilli et~al.(2013)Grilli, Brambilla, Sirois, and
  Memiaghe]{grilli_development_2013}
F.~Grilli, R.~Brambilla, F.~Sirois, and S.~Memiaghe.
\newblock Development of a three-dimensional finite-element model for
  high-temperature superconductors based on the {H}-formulation.
\newblock \emph{Cryogenics}, 53:\penalty0 142--147, 2013.

\bibitem[Zhang and Coombs(2012)]{zhang_3d_2012}
M.~Zhang and T.~A. Coombs.
\newblock 3{D} modeling of high-${T_c}$ superconductors by finite element
  software.
\newblock \emph{Superconductor Science and Technology}, 25\penalty0
  (1):\penalty0 015009, 2012.

\bibitem[Stenvall et~al.(2014)Stenvall, Lahtinen, and
  Lyly]{Stenvall_hformulation_2014}
A.~Stenvall, V.~Lahtinen, and M.~Lyly.
\newblock An {H}-formulation-based three-dimensional hysteresis loss modelling
  tool in a simulation including time varying applied field and transport
  current: the fundamental problem and its solution.
\newblock \emph{Superconductor Science and Technology}, 27\penalty0
  (10):\penalty0 104004, 2014.

\bibitem[Lahtinen et~al.(2015)Lahtinen, Stenvall, Sirois, and
  Pellikka]{lahtinen_homology_2015}
V.~Lahtinen, A.~Stenvall, F.~Sirois, and M.~Pellikka.
\newblock A finite element simulation tool for predicting hysteresis losses in
  superconductors using an {H}-oriented formulation with cohomology basis
  functions.
\newblock \emph{Journal of Superconductivity and Novel Magnetism}, 28\penalty0
  (8):\penalty0 2345--2354, 2015.

\bibitem[Bossavit(1994)]{Bossavit_Numerical_1994}
A.~Bossavit.
\newblock Numerical modelling of superconductors in three dimensions: a model
  and a finite element method.
\newblock \emph{IEEE Transactions on Magnetics}, 30\penalty0 (5):\penalty0
  3363--3366, 1994.

\bibitem[Elliott and Kashima(2007)]{elliott_finite-element_2007}
C.~M. Elliott and Y.~Kashima.
\newblock A finite-element analysis of critical-state models for type-{II}
  superconductivity in 3{D}.
\newblock \emph{IMA Journal of Numerical Analysis}, 27\penalty0 (2):\penalty0
  293--331, 2007.

\bibitem[Badía-Majós and López(2012)]{badia_electromagnetics}
A.~Badía-Majós and C.~López.
\newblock Electromagnetics close beyond the critical state: thermodynamic
  prospect.
\newblock \emph{Superconductor Science and Technology}, 25\penalty0
  (10):\penalty0 104004, 2012.

\bibitem[Pardo and Kapolka(2017)]{pardo_3d_2017}
E.~Pardo and M.~Kapolka.
\newblock 3{D} computation of non-linear eddy currents: {Variational} method
  and superconducting cubic bulk.
\newblock \emph{Journal of Computational Physics}, 344:\penalty0 339--363,
  2017.

\bibitem[Prigozhin(1996)]{PRIGOZHIN1996190}
L.~Prigozhin.
\newblock The bean model in superconductivity: Variational formulation and
  numerical solution.
\newblock \emph{Journal of Computational Physics}, 129\penalty0 (1):\penalty0
  190 -- 200, 1996.

\bibitem[Pardo et~al.(2015)Pardo, {\v S}ouc, and Frolek]{pardo_electro_2015}
E.~Pardo, J.~{\v S}ouc, and L.~Frolek.
\newblock Electromagnetic modelling of superconductors with a smooth
  current–voltage relation: variational principle and coils from a few turns
  to large magnets.
\newblock \emph{Superconductor Science and Technology}, 28\penalty0
  (4):\penalty0 044003, 2015.

\bibitem[Prigozhin(1998)]{prigozhin_solution_1998}
L.~Prigozhin.
\newblock Solution of {Thin} {Film} {Magnetization} {Problems} in {Type}-{II}
  {Superconductivity}.
\newblock \emph{Journal of Computational Physics}, 144:\penalty0 180--193,
  1998.

\bibitem[Navau et~al.(2008)Navau, Sanchez, Del-Valle, and
  Chen]{navau_alternating_2008}
C.~Navau, A.~Sanchez, N.~Del-Valle, and D.~Chen.
\newblock Alternating current susceptibility calculations for thin-film
  superconductors with regions of different critical-current densities.
\newblock \emph{Journal of Applied Physics}, 103\penalty0 (113907), 2008.

\bibitem[van Nugteren(2016)]{van_nugteren_high_2016}
J.~van Nugteren.
\newblock \emph{High temperature superconductor accelerator magnets}.
\newblock PhD thesis, University of Twente, Enschede, 2016.

\bibitem[N\'ed\'elec(1980)]{nedelec_mixed_1980}
J.~N\'ed\'elec.
\newblock Mixed finite elements in {$R^3$}.
\newblock \emph{Numer. Math.}, 35:\penalty0 315--341, 1980.

\bibitem[Mur(1994)]{mur_edge_1994}
G.~Mur.
\newblock Edge elements, their advantages and their disadvantages.
\newblock \emph{IEEE transactions on magnetics}, 30\penalty0 (5):\penalty0
  3552--3557, 1994.

\bibitem[Badia and Codina(2012)]{badia_nodal-based_2012}
S.~Badia and R.~Codina.
\newblock A {Nodal}-based {Finite} {Element} {Approximation} of the {Maxwell}
  {Problem} {Suitable} for {Singular} {Solutions}.
\newblock \emph{SIAM Journal on Numerical Analysis}, 50\penalty0 (2):\penalty0
  398--417, 2012.

\bibitem[Granados et~al.(2016)Granados, Gonçalves~Sotelo, Carrera, and
  Lopez-Lopez]{granados_h-formulation_2016}
X.~Granados, G.~Gonçalves~Sotelo, M.~Carrera, and J.~Lopez-Lopez.
\newblock {H}-{Formulation} {FEM} {Modeling} of the {Current} {Distribution} in
  2g {HTS} {Tapes} and {Its} {Experimental} {Validation} {Using} {Hall} {Probe}
  {Mapping}.
\newblock \emph{IEEE Transactions on Applied Superconductivity}, 26\penalty0
  (8), 2016.

\bibitem[Brandt(1995)]{brandt_1995}
E.~H. Brandt.
\newblock Square and rectangular thin superconductors in a transverse magnetic
  field.
\newblock \emph{Phys. Rev. Lett.}, 74:\penalty0 3025--3028, 1995.

\bibitem[Bouzo et~al.(2004)Bouzo, Grilli, and Yang]{grilli_integral}
M.~C. Bouzo, F.~Grilli, and Y.~Yang.
\newblock Modelling of coupling between superconductors of finite length using
  an integral formulation.
\newblock \emph{Superconductor Science and Technology}, 17\penalty0
  (10):\penalty0 1103, 2004.

\bibitem[van Nugteren et~al.(2016)van Nugteren, van Nugteren, Gao, Bottura,
  Dhallé, Goldacker, Kario, ten Kate, Kirby, Krooshoop, de~Rijk, Rossi,
  Senatore, Wessel, Yagotintsev, and Yang]{nugteren_2016}
J.~van Nugteren, B.~van Nugteren, P.~Gao, L.~Bottura, M.~Dhallé, W.~Goldacker,
  A.~Kario, H.~ten Kate, G.~Kirby, E.~Krooshoop, G.~de~Rijk, L.~Rossi,
  C.~Senatore, S.~Wessel, K.~Yagotintsev, and Y.~Yang.
\newblock Measurement and numerical evaluation of ac losses in a rebco roebel
  cable at 4.5 k.
\newblock \emph{IEEE Transactions on Applied Superconductivity}, 26\penalty0
  (3):\penalty0 1--7, 2016.

\bibitem[Burstedde et~al.(2011)Burstedde, Wilcox, and
  Ghattas]{BursteddeWilcoxGhattas11}
C.~Burstedde, L.~C. Wilcox, and O.~Ghattas.
\newblock {\texttt{p4est}}: Scalable algorithms for parallel adaptive mesh
  refinement on forests of octrees.
\newblock \emph{SIAM Journal on Scientific Computing}, 33\penalty0
  (3):\penalty0 1103--1133, 2011.

\bibitem[Saad(2003)]{Saad_book}
Y.~Saad.
\newblock \emph{Iterative Methods for Sparse Linear Systems}.
\newblock Society for Industrial and Applied Mathematics, 2nd edition, 2003.

\bibitem[Dohrmann(2003)]{dohrmann_2003}
C.~R. Dohrmann.
\newblock A preconditioner for substructuring based on constrained energy
  minimization.
\newblock \emph{SIAM Journal on Scientific Computing}, 25\penalty0
  (1):\penalty0 246--258, 2003.

\bibitem[Badia et~al.(2014)Badia, Martín, and Principe]{badia_highly_2014}
S.~Badia, A.~F. Martín, and J.~Principe.
\newblock A {Highly} {Scalable} {Parallel} {Implementation} of {Balancing}
  {Domain} {Decomposition} by {Constraints}.
\newblock \emph{SIAM Journal on Scientific Computing}, 36\penalty0
  (2):\penalty0 C190--C218, 2014.

\bibitem[Badia et~al.(2016)Badia, Martín, and Principe]{badia_multilevel_2016}
S.~Badia, A.~F. Martín, and J.~Principe.
\newblock Multilevel {Balancing} {Domain} {Decomposition} at {Extreme}
  {Scales}.
\newblock \emph{SIAM Journal on Scientific Computing}, pages C22--C52, 2016.

\bibitem[Toselli(2006)]{toselli_dual-primal_2006}
A.~Toselli.
\newblock Dual-primal {FETI} algorithms for edge finite-element approximations
  in 3{D}.
\newblock \emph{IMA Journal of Numerical Analysis}, 26\penalty0 (1):\penalty0
  96--130, 2006.

\bibitem[Badia et~al.(2017)Badia, Mart\'in, and
  Nguyen]{badia_physics_based_2017}
S.~Badia, A.~F. Mart\'in, and H.~Nguyen.
\newblock Physics-based balancing domain decomposition by constraints for
  heterogeneous problems.
\newblock \emph{hal}, 26\penalty0 (-01337968v3), 2017.

\bibitem[Hong et~al.(2006)Hong, Campbell, and Coombs]{hong_numerical_2006}
Z.~Hong, A.~M. Campbell, and T.~Coombs.
\newblock Numerical solution of critical state in superconductivity by finite
  element software.
\newblock \emph{Superconductor Science and Technology}, 19\penalty0 (12), 2006.

\bibitem[Ainslie et~al.(2017)Ainslie, Hu, Zermeno, and
  Grilli]{ainslie_numerical_2016}
M.~D. Ainslie, D.~Hu, V.~M.~R. Zermeno, and F.~Grilli.
\newblock Numerical simulation of the performance of high-temperature
  superconducting coils.
\newblock \emph{Journal of Superconductivity and Novel Magnetism}, 30\penalty0
  (7):\penalty0 1987--1992, 2017.

\bibitem[Pardo and Kapolka(2017)]{pardo_3d_magnetization_2017}
E.~Pardo and M.~Kapolka.
\newblock 3{D} magnetization currents, magnetization loop, and saturation field
  in superconducting rectangular prisms.
\newblock \emph{Superconductor Science and Technology}, 30\penalty0
  (6):\penalty0 064007, 2017.

\bibitem[Farinon et~al.(2014)Farinon, Iannone, Fabbricatore, and
  Gambardella]{farinon_2d_2014}
S.~Farinon, G.~Iannone, P.~Fabbricatore, and U.~Gambardella.
\newblock 2{D} and 3{D} numerical modeling of experimental magnetization cycles
  in disks and spheres.
\newblock \emph{Superconductor Science and Technology}, 27\penalty0
  (10):\penalty0 104005, 2014.

\bibitem[Badia et~al.(2018{\natexlab{a}})Badia, Mart{\'i}n, and
  Principe]{badia-fempar}
S.~Badia, A.~F. Mart{\'i}n, and J.~Principe.
\newblock \texttt{FEMPAR}: An object-oriented parallel finite element
  framework.
\newblock \emph{Archives of Computational Methods in Engineering}, 25\penalty0
  (2):\penalty0 195--271, 2018{\natexlab{a}}.

\bibitem[Badia et~al.(2018{\natexlab{b}})Badia, Mart\'in, and
  Principe]{fempar-web-page}
S.~Badia, A.~F. Mart\'in, and J.~Principe.
\newblock \texttt{FEMPAR} {W}eb page.
\newblock \url{http://www.fempar.org}, 2018{\natexlab{b}}.

\bibitem[Kapolka et~al.(2018)Kapolka, Zermeno, Zou, Morandi, Ribani, Pardo, and
  Grilli]{kapolka_3d_2017}
M.~Kapolka, V.~M.~R. Zermeno, S.~Zou, A.~Morandi, P.~L. Ribani, E.~Pardo, and
  F.~Grilli.
\newblock Three-dimensional modeling of the magnetization of superconducting
  rectangular-based bulks and tape stacks.
\newblock \emph{IEEE Transactions on Applied Superconductivity}, 28\penalty0
  (4):\penalty0 1--6, 2018.

\bibitem[Bean(1964)]{bean_magnetization_1964}
C.~Bean.
\newblock Magnetization of {High}-{Field} superconductors.
\newblock \emph{Reviews of modern physics}, 36\penalty0 (1):\penalty0 31--39,
  1964.

\bibitem[Kim et~al.(1963)Kim, Hempstead, and Strnad]{kim_magnetization_1963}
Y.~Kim, C.~Hempstead, and A.~Strnad.
\newblock Magnetization and critical supercurrents.
\newblock \emph{Physical Reviews Letters}, 129\penalty0 (528), 1963.

\bibitem[Monk(2003)]{monk_finite_2003}
P.~Monk.
\newblock \emph{Finite {Element} {Methods} for {Maxwell}'s {Equations}}.
\newblock Oxford Science Publications, 2003.

\bibitem[N\'ed\'elec(1986)]{nedelec_new_mixed_1986}
J.~N\'ed\'elec.
\newblock A new family of mixed finite elements in {$R^3$}.
\newblock \emph{Numer. Math.}, 50:\penalty0 57--81, 1986.

\bibitem[Tu et~al.(2005)Tu, O'Hallaron, and Ghattas]{Tu_2005}
T.~Tu, D.~R. O'Hallaron, and O.~Ghattas.
\newblock Scalable parallel octree meshing for terascale applications.
\newblock In \emph{Proceedings of the ACM/IEEE SC 2005 Conference,
  Supercomputing, 2005}, pages 4--4, 2005.

\bibitem[Kůs and Šístek(2017)]{coupling_kus_2017}
P.~Kůs and J.~Šístek.
\newblock Coupling parallel adaptive mesh refinement with a nonoverlapping
  domain decomposition solver.
\newblock \emph{Advances in Engineering Software}, 110:\penalty0 34 -- 54,
  2017.

\bibitem[Brune et~al.(2013)Brune, Knepley, Smith, and Tu]{brune_composing_2013}
P.~Brune, M.~Knepley, B.~Smith, and X.~Tu.
\newblock Composing scalable nonlinear algebraic solvers.
\newblock \emph{Argonne National Laboratory, Preprint ANL/MCS-P2010-0112},
  2013.

\bibitem[Balay et~al.(2018)Balay, Abhyankar, Adams, Brown, Brune, Buschelman,
  Dalcin, Eijkhout, Gropp, Kaushik, Knepley, May, McInnes, Rupp, Smith,
  Zampini, Zhang, and Zhang]{petsc-web-page}
S.~Balay, S.~Abhyankar, M.~F. Adams, J.~Brown, P.~Brune, K.~Buschelman,
  L.~Dalcin, V.~Eijkhout, W.~D. Gropp, D.~Kaushik, M.~G. Knepley, D.~A. May,
  L.~C. McInnes, K.~Rupp, B.~F. Smith, S.~Zampini, H.~Zhang, and H.~Zhang.
\newblock {PETS}c {W}eb page.
\newblock \url{http://www.mcs.anl.gov/petsc}, 2018.

\bibitem[Dohrmann and Widlund(2016)]{dohrmann_bddc_2016}
C.~Dohrmann and O.~Widlund.
\newblock A {BDDC} {Algorithm} with {Deluxe} {Scaling} for
  {Three}-{Dimensional} {H}(curl) {Problems}.
\newblock \emph{Communications on Pure and Applied Mathematics}, 69\penalty0
  (4):\penalty0 745--770, 2016.

\bibitem[int()]{intel_pardiso}
Intel {{MKL PARDISO}} - {{Parallel Direct Sparse Solver Interface}}.
\newblock \url{https://software.intel.com/en-us/articles/intel-mkl-pardiso}.

\bibitem[MNI(2018)]{MNIV}
{M}arenostrum {IV} website.
\newblock \url{https://www.bsc.es/marenostrum/marenostrum}, 2018.

\bibitem[Norris(1971)]{norris_calculation_1971}
W.~T. Norris.
\newblock Calculation of hysteresis losses in hard superconductors:
  polygonal-section conductors.
\newblock \emph{Journal of Physics D: Applied Physics}, 4\penalty0
  (9):\penalty0 1358, 1971.

\bibitem[Nibbio and Stavrev(2001)]{nibbio_effect_2001}
N.~Nibbio and S.~Stavrev.
\newblock Effect of the geometry of hts on ac loss by using finite element
  method simulation with b-dependent e-j power law.
\newblock \emph{IEEE Transactions on Applied Superconductivity}, 11\penalty0
  (1):\penalty0 2627--2630, 2001.

\bibitem[Pardo et~al.(2004)Pardo, Chen, Sanchez, and Navau]{pardo_alter_2004}
E.~Pardo, D.-X. Chen, A.~Sanchez, and C.~Navau.
\newblock Alternating current loss in rectangular superconducting bars with a
  constant critical-current density.
\newblock \emph{Superconductor Science and Technology}, 17\penalty0
  (1):\penalty0 83, 2004.

\bibitem[Dutoit et~al.(2005)Dutoit, Duron, Stavrev, and
  Grilli]{dutoit_measuring_2005}
B.~Dutoit, J.~Duron, S.~Stavrev, and F.~Grilli.
\newblock Dynamic field mapping for obtaining the current distribution in
  high-temperature superconducting tapes.
\newblock \emph{IEEE Transactions on Applied Superconductivity}, 15\penalty0
  (2):\penalty0 3644--3647, 2005.

\bibitem[Fukui et~al.(2003)Fukui, Tonsho, Nakayama, Yamaguchi, Torii, Ueda, and
  Takao]{FUKUI2003224}
S.~Fukui, H.~Tonsho, H.~Nakayama, M.~Yamaguchi, S.~Torii, K.~Ueda, and
  T.~Takao.
\newblock Numerical evaluation of measured ac loss in hts tape in ac magnetic
  field carrying ac transport current.
\newblock \emph{Physica C: Superconductivity}, 392-396\penalty0 (Part
  1):\penalty0 224 -- 228, 2003.
\newblock Proceedings of the 15th International Symposium on Superconductivity
  (ISS 2002): Advances in Superconductivity XV. Part I.

\bibitem[Sim et~al.(2010)Sim, Kim, Lee, Cho, and Ko]{kim_estimation_2010}
K.~Sim, S.~Kim, S.~Lee, J.~Cho, and T.~K. Ko.
\newblock The estimation of the current distribution on the hts cable by
  measuring the circumferential magnetic field.
\newblock \emph{IEEE Transactions on Applied Superconductivity}, 20\penalty0
  (3):\penalty0 1981--1984, 2010.

\bibitem[Brandt(1996)]{brandt_superconductors_1996}
E.~H. Brandt.
\newblock Superconductors of finite thickness in a perpendicular magnetic
  field: Strips and slabs.
\newblock \emph{Phys. Rev. B}, 54:\penalty0 4246--4264, 1996.

\bibitem[Amemiya et~al.(1997)Amemiya, Miyamoto, Banno, and
  Tsukamoto]{amemiya_numerical_1997}
N.~Amemiya, K.~Miyamoto, N.~Banno, and O.~Tsukamoto.
\newblock Numerical analysis of ac losses in high t/sub c/ superconductors
  based on e-j characteristics represented with n-value.
\newblock \emph{IEEE Transactions on Applied Superconductivity}, 7\penalty0
  (2):\penalty0 2110--2113, 1997.

\bibitem[Pardo et~al.(2004)Pardo, Chen, Sanchez, and
  Navau]{pardo_the_transverse_2004}
E.~Pardo, D.-X. Chen, A.~Sanchez, and C.~Navau.
\newblock The transverse critical-state susceptibility of rectangular bars.
\newblock \emph{Superconductor Science and Technology}, 17\penalty0
  (3):\penalty0 537, 2004.

\bibitem[Grilli et~al.(2013)Grilli, Pardo, Stenvall, Nguyen, Yuana, and
  Gömöry]{grilli_computation_2013}
F.~Grilli, E.~Pardo, A.~Stenvall, D.~N. Nguyen, W.~Yuana, and F.~Gömöry.
\newblock Computation of {Losses} in {HTS} {Under} the {Action} of {Varying}
  {Magnetic} {Fields} and {Currents}.
\newblock \emph{IEEE Transactions on Applied Superconductivity}, 24\penalty0
  (1), 2013.

\end{thebibliography}

\end{small}

\end{document}